\documentclass[journal,draftcls,onecolumn]{IEEEtran}
\usepackage{epsfig}
\usepackage{times}
\usepackage{float}
\usepackage{afterpage}
\usepackage{amsmath}
\usepackage{amstext}
\usepackage{amssymb,bm}
\usepackage{latexsym}
\usepackage{color}
\usepackage{graphicx}
\usepackage{amsmath}
\usepackage{amsthm}
\usepackage{pstricks}
\usepackage{subfigure}
\usepackage{booktabs}
\usepackage{array}
\usepackage{makecell}
\usepackage{diagbox}
\usepackage{enumerate}
\usepackage{algorithm}
\usepackage{algorithmic}
\usepackage[normalem]{ulem}
\usepackage{url}
\usepackage{setspace}
\usepackage{enumitem}
\usepackage{lineno}

\usepackage[noadjust]{cite}

\usepackage{xcolor}

\newtheorem{theorem}{Theorem}

\newtheorem{proposition}{Proposition}
\newtheorem{definition}{Definition}

\allowdisplaybreaks

\usepackage{mathtools}  
\newcommand*{\prob}{\mathsf{P}}  

\begin{document}

\title{Achievable Error Exponents of One-Way and Two-Way AWGN Channels}



\author{
    \IEEEauthorblockN{Kenneth Palacio-Baus\IEEEauthorrefmark{1}\IEEEauthorrefmark{2}, Natasha Devroye\IEEEauthorrefmark{1}} \\
    \IEEEauthorblockA{\IEEEauthorrefmark{1}University of Illinois at Chicago, \{kpalac2, devroye\}@uic.edu}\thanks{This work was presented in part at the 2018 International Symposium on Information Theory.} \\
    \IEEEauthorblockA{\IEEEauthorrefmark{2}University of Cuenca, Ecuador}
}



\maketitle

\thispagestyle{plain}
\pagestyle{plain}

%
\IEEEpeerreviewmaketitle

\begin{abstract}
Achievable error exponents for the one-way with noisy feedback and two-way AWGN channels are derived for the transmission of a finite number of messages $M$ using fixed block length $n$, under the almost sure (AS) and the expected block (EXP) power constraints.  
	In the one-way setting under noisy AWGN feedback, it is shown that under the AS constraint and when the feedback link is much stronger than the direct link, active feedback leads to a larger gain over the non-feedback error exponent than passive feedback.  Under the EXP constraint, a previously known error exponent for the transmission of two messages is generalized to any arbitrary but finite number of messages $M$.
	In the two-way setting, where each user has its own message to send in addition to (possibly) aiding in the transmission of feedback for the opposite direction, error exponent regions are defined and derived for the first time for the AWGN two-way channel under both AS and EXP power constraints. It is shown that feedback or interaction may lead to error exponent gains in one direction, possibly at the expense of a decrease in the error exponents attained in the other direction. The relationship between $M$ and $n$ supported by our achievability strategies is explored.

	
\end{abstract}

\section{Introduction}
\label{sec:Introd}



The reliability function \cite{Fano:book,Gallager:book,Gallager1965}, or error exponent, of a one-way channel characterizes the rate of decay of the probability of error when communicating one of $2^{nR}$ messages as 
\begin{equation} \label{eq:ErrExp}
		E(R) = \lim_{n \to \infty} -\frac{1}{n} \log \mathsf{P}_e^{(n)},
\end{equation}
where $\mathsf{P}_e^{(n)}$ is the smallest probability of error that can be achieved by a code of rate $R$ with block length $n$. 
Error exponents have been the subject of intense interest in both the absence \cite{Shannon1959,Gallager1965} 
and presence 
of feedback (to be reviewed later). 
If feedback is available, the transmitter is given access to a (possibly noisy, possibly encoded) function of the received output, that may dramatically increase the error exponents of one-way channels relative to when feedback is absent.  
Error exponents in the presence of ideal, noiseless feedback were considered by Berlekamp in \cite{berlekamp_block_1964} for general discrete memoryless channels, and by Pinsker \cite{pinsker1968probability} for the AWGN case. 
It was shown that while feedback cannot increase the capacity of non-anticipatory channels, 
it may greatly improve the error exponents achieved. Shannon pointed out this fact for the reliability of discrete memoryless channels with perfect feedback \cite{Shannon1956}. This was first demonstrated for the AWGN channel in \cite{SchalkwijkNKailath1966}, in which the probability of error decays double-exponentially in  the blocklength $n$, and later by \cite{Zigangirov1970SuperEXP} who demonstrated that a decay rate equal to any number of exponential levels is possible.
One natural question is whether this increase in error exponents may be attributed to the  feedback being noiseless. In recent years, it has been shown that even noisy feedback is useful in improving error exponents (though less dramatically),  with a limited number of available results in the one-way setting.  In this article we continue this line of work and study error exponents of one-way additive white Gaussian noise (AWGN) channels with noisy AWGN feedback for the transmission of a finite number of messages (zero-rate). 

We also extend the study of error exponents with noisy feedback to the two-way channel, in which two terminals exchange independent messages. In this bidirectional model, each transmitter's encoding function output at time $k$ is a function not only of the message, but also of the past available channel outputs. This transmitter's encoding function is said to be  \emph{adaptative} or \emph{interactive}, a term used to emphasize how in two-way networks, each terminal's channel input may adapt to its received channel outputs. 
The capacity region of the two-way AWGN channel (with independent noise across the terminals) is known, and is a rectangular region where both users may simultaneously attain their interference-free AWGN capacity. That is, adaptation at the transmitters is useless from a capacity perspective. The question we ask and answer here is whether adaptation may improve the error exponents of this two-way AWGN channel, for the transmission of a  finite number of messages. This is the first study of two-way error exponents, to the best of our knowledge, and it is of initial interest as the exact error exponent for the one-way channel with noisy feedback is still open in this finite message regime. Error exponents for positive rates are left for future work, and may extend existing work such as \cite{SchalkwijkNKailath1966,sahai_boosting_2005,Guo2013:PartialSeqFB,BenYishai2015:ImprovER,Yamamoto2009posRateBSC,Burnashev2010posRate}. We also initially study error exponents for fixed block length codes rather than variable block length codes, as studied in \cite{burnashev1976data,GallagerNakibogluVariable2008,SatoYamamoto2010VLC} and the references therein for discrete memoryless channels.
We focus on AWGN channels, and will review results that address the problem for the transmission of a finite number of messages for such channels {\cite{Kim:PassiveNoisyFB,Kim:ActiveNoisyFB,Kim:ThreeCWPeak2013,Burnashev2012AWGN-ZR,YamamotoNoisyFB2014} in the coming sections. 



The achievable error exponents of AWGN channels are sensitive to the type of power constraint imposed on the channel inputs. In this work, we consider two constraints, defined as the 
\begin{itemize}
\item Almost sure (AS) power constraint: 
\begin{equation} \label{eq:ASpowerConstraint}
\sum^{n}_{k=1} X_{i,k}^2 \leq n P_i
\end{equation}
\item Expected block (EXP) power constraint: 
\begin{equation} \label{eq:EXPpowerConstraint}
E\left[ \sum^{n}_{k=1} X_{i,k}^2 \right] \leq n P_i.
\end{equation}
\end{itemize}
In \eqref{eq:ASpowerConstraint} and \eqref{eq:EXPpowerConstraint}, $X_{i,k}$ corresponds to the $k$-th  channel input of user $i$, and $P_i$ to the power available at the $i$-th terminal. 
The EXP constraint is less stringent than the AS constraint, and allows very high amplitude transmissions to occur with exponentially small probability. These rare events may correspond to decoding errors, and transmissions may be used to correct such errors, thereby increasing achievable error exponents.  Burnashev and Yamamoto \cite{Yamamoto2009posRateBSC} comment that this ``trick'' can not be used for general discrete memoryless channels,  but is useful for channels characterized by additive noise.

%


\subsection{Contributions}
\label{sec:Contrib}

We present results on the one-way and two-way AWGN channel error exponents under both the AS and EXP power constraint in two separate sections. Each section includes the problem statement, past related work and our new achievable error exponent (regions). Our contributions are as follows:
\begin{enumerate}
\item Theorem  \ref{th:OneWayActiveASM3}, proven in Section \ref{sec:ActiveM3OneWay}, demonstrates how under the AS power constraint, the transmission of a finite number of messages $M$, and a feedback link stronger than the forward link, the use of new proposed {\it active} feedback scheme in a one-way AWGN channel with noisy AWGN feedback leads to a higher error exponent gain over the non-feedback transmission exponent than that reported for {\it passive} feedback by Xiang and Kim in \cite{Kim:ThreeCWPeak2013}. 

\item Theorem \ref{th:OneWayEXPM2}, proven  in Section \ref{sec:ProofOneWayEXPM}, generalizes an achievable error exponent under the EXP power constraint presented by Kim, Lapidoth and Weissman in \cite{Kim:ActiveNoisyFB} for the transmission of two messages to any finite number of messages $M$, again for the one-way AWGN channel with AWGN noisy feedback. The generalization is based on the use of a simplex code, and is an active feedback scheme.

\item   Theorems \ref{Prop2} and \ref{Prop3} demonstrate new achievable error exponent regions (using passive and active feedback respectively) for the two-way AWGN channel under the AS power constraint,  provided that one channel's signal to noise ratio (SNR)  is better than the other. 
This non-symmetric SNR scenario is of interest since for the one-way AWGN channel under the AS constraint, only a feedback link significantly stronger that the forward has been shown to lead to error exponent gains over feedback-free transmissions. These results follow as a direct application of Theorem \ref{th:OneWayActiveASM3} for active feedback and the results reported in \cite{Kim:ThreeCWPeak2013} for passive feedback.

\item Theorem \ref{th:EXPMachievTW}, proven in  Section \ref{sec:M2EXP}, uses the approach employed in  Theorem \ref{th:OneWayEXPM2} to demonstrate an achievable error exponent region for the two-way channel under EXP power constraints in both directions.
Note that the non-symmetric SNR condition required for the AS constraint is not necessary here. 
\end{enumerate}

The achievable error exponents for the transmission of a finite number of messages $M$ presented here will all be based on the use of 
a simplex code for the non-feedback transmissions present in our schemes.  The use of simplex codes enable the use of a geometric approach for upper bounding the probability of error, that can be visualized for a small number of messages ($M=3$) and can be extended to any finite $M\geq 3$. 
Under the EXP constraint, some feedback and retransmission signals that occur with exponentially small probability employ very high amplitude signals. These rarely occurring transmissions may be used to ensure an exceedingly small probability of error. 

\subsection{Article outline}
\label{sec:OutlineIntro}

The remainder of this article is organized as follows. Section \ref{sec:ResultsOneWay} presents a summary of previous results followed by our findings for the one-way AWGN channel at zero-rate: Theorems  \ref{th:OneWayActiveASM3} and  \ref{th:OneWayEXPM2}, which are proven in Sections \ref{sec:ActiveM3OneWay} and \ref{sec:ProofOneWayEXPM} respectively.  The two-way AWGN channel is studied in Section \ref{sec:ResultsTwoWayII}, with the main results presented in Theorems \ref{Prop2} and \ref{Prop3}, that follow from the use of one-way achievability schemes under the AS constraint, and Theorem \ref{th:EXPMachievTW}, which addresses the case of EXP constraint and is proven in Section \ref{sec:M2EXP}. {Section \ref{sec:HowLargeIsM} address the relation between the number of messages $M$ and the block length $n$ for our proposed schemes.} Numerical simulations are presented in Section \ref{sec:Simul}. Finally, Section \ref{sec:Conclu} presents our conclusions and a discussion of open problems.  

{\bf Notation.} 
We use $\prob_w(x)$ to indicate the probability of any event $x$ conditioned on the transmission of message $W={w}$, i.e. the probability of error given that $W=1$ has been sent as: $\prob(\text{error} \mid w=1) = \prob_1(\text{error})$. We indicate the length of a sequence using a superscript, i.e. $x^{\lambda_1 n}_i$ denotes that sequence $x$ lasts for $\lambda_1 n$ channel uses. Subscript $i$ indicates the terminal that generated the sequence. Error exponents are denoted by $E$, accompanied by a superscript to denote the forward direction power constraint, and a subscript to denote the feedback link power constraint, if applicable.  We use $\mathsf{E}\left[ \cdot \right]$ to denote the expectation operator. Random variables are indicated by upper case letters, taking on instance in lower cases from alphabets in calligraphic font (random variable $X$ takes on $x\in {\cal X}$). We use $a_n \doteq b_n$ to indicate that $\frac{1}{n} \ln (\frac{a_n}{b_n}) \to 0$ as $n\to \infty$. We use the terms \textit{terminal} and \textit{user} interchangeably to refer to the devices involved in the communication process.

\section{Error exponents for the one-way AWGN channel at zero-rate}
\label{sec:ResultsOneWay}


This section defines error exponents for one-way channels with feedback, and presents existing results on achievable error exponents for the one-way AWGN channel at zero-rate under different power constraints and types of feedback. We also introduce new achievable error exponents for the transmission of a finite number of messages $M$ (and this is what we will refer to as zero-rate from now on), under the AS power constraint with active feedback (Theorem \ref{th:OneWayActiveASM3}) and under the EXP power constraint (Theorem \ref{th:OneWayEXPM2}).

\subsection{Definitions for the One-Way AWGN channel}
\label{sec:DefsOneWay}

In the one-way AWGN channel, terminal 1 (transmitter) wishes to transmit a message $W$, selected uniformly from a set of $M$  equally likely messages $\mathcal{W}=\{1,2,...,M\}$  to terminal 2 (receiver) using a code of fixed block length $n$. 
Forward and backward directions are characterized by independent AWGN channels. The AWGN noises are of zero mean and of variances $\sigma^2$ and $\sigma_{\text{FB}}^2$ respectively, both identically distributed and independent across users and channel uses. 
Figure  \ref{fig:OneWayGaussianActive} shows the one-way AWGN channel with active noisy AWGN feedback, characterized by Equations \eqref{eq:OneWayGEnModelY} and \eqref{eq:OneWayGEnModelZ} for the $k$-th channel use: 
\begin{align} \label{eq:OneWayGEnModelY}
		Y_k &= X_k + N_{k}, \; N_{k} \sim \mathcal{N}(0,\sigma^2) \;\;\forall k \in {1,...,n} \\
		Z_k &= U_k + N_{\text{FB}_k}, \;\; N_{\text{FB}_k}\sim \mathcal{N}(0,\sigma_{\text{FB}}^2)  \;\; \forall k \in {1,...,n}. \label{eq:OneWayGEnModelZ}
\end{align}
\begin{figure}[H]
	\centering
	\includegraphics[width=0.5\textwidth]{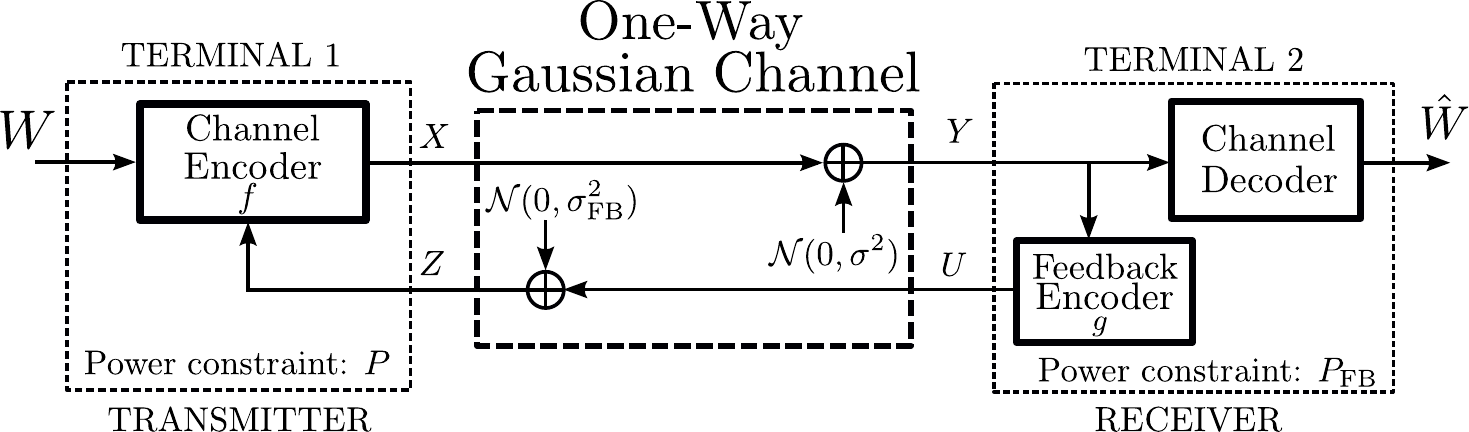}
	\caption{One-way AWGN channel with active feedback.}
	\label{fig:OneWayGaussianActive}
\end{figure}
The model above captures noiseless feedback by taking $\sigma_{\text{FB}}^2\to 0$, and the absence of feedback by taking $\sigma_{\text{FB}}^2\to \infty$.  Channel inputs  for each direction are subject to either an $X \in \{\text{AS},\text{EXP} \}$ power constraint according to Equations \eqref{eq:ASpowerConstraint} and \eqref{eq:EXPpowerConstraint}. In the following we will make definitions for a system with feedback;  definitions in the absence of feedback should be clear from context and omission of the corresponding feedback-related terms.

Let $\mathcal{X,Y,U,Z}$ all be the set of reals  and $\left({P},{\sigma^2},{P_{\text{FB}}},{\sigma_{\text{FB}}^2},n\right)$ be a block length-$n$ code for the transmission of $M$ messages consisting of $n$ forward and feedback encoding functions,
\begin{align} 
f_k &: \mathcal{W} \times \mathcal{Z}^{k-1} \to \mathcal{X}, \;\; k=1,...,n,\label{eq:ForwardENC} \\
g_k &: \mathcal{Y}^{k} \to \mathcal{U}, \;\; k=1,...,n, \label{eq:BackwardENC}
\end{align}
 leading to channel inputs $X_k = f_k\left( W, Z^{k-1} \right)$ and $U_k = g_k\left(Y^k\right)$, and a decoding rule $\phi: \mathcal{Y}^n \to \{ 1,...,M\}$ that determines the best estimate of the transmitted message $W$, denoted by $\hat{W}$. 


Let $\prob^X_{e}\left(M, {P},{\sigma^2}, {P_{\text{FB}}},{\sigma_{\text{FB}}^2}, n\right) := \frac{1}{M} \sum_{w=1}^M \text{Pr}(\phi(y^n) \neq w \mid W=w \text{ sent})$ denote the probability of error attained by a particular $\left({P},{\sigma^2},{P_{\text{FB}}},{\sigma_{\text{FB}}^2},n\right)$ code under an $X \in \{\text{AS}, \text{EXP}\}$ power constraint. 
We define the achievable error exponent for the one-way AWGN channel with feedback under the $X$ power constraint as:
\begin{equation} \label{eq:ErrExpDefOneWay}
		E^X_{\text{FB}}(M,P,\sigma^2,P_{\text{FB}},\sigma^2_{\text{FB}}) := \liminf_{n\to \infty} -\frac{1}{n} \log \prob_e^X \left(M, {P},{\sigma^2}, {P_{\text{FB}}},{\sigma_{\text{FB}}^2}, n\right),
\end{equation}
where the subscript $\text{FB}$ indicates the presence of feedback and will be omitted for the non-feedback case.  This notation follows that in \cite{Kim:ActiveNoisyFB} for the transmission of two messages with active noisy feedback.


\subsection{Error exponents for the one-way AWGN channel without feedback and with perfect output feedback}

Results presented in this section correspond to the cases of feedback-free and perfect output feedback transmissions for a finite number of transmitted messages.  These are introduced as references since they lower and upper bound what can be achievable with noisy feedback, respectively. 
The feedback-free scenario ($\sigma_{\text{FB}} \to \infty$) was studied by Shannon in \cite{Shannon1959}, who showed that under the AS power constraint and the transmission of $M$ messages, the best achievable error exponent is that of Equation \eqref{Eq:NoFBErrExp}   and attainable using a simplex code: 
\begin{equation} \label{Eq:NoFBErrExp}
E^{\text{AS}}\left( M,{P},{\sigma^2} \right) = \frac{P}{\sigma^2} \frac{M}{4(M-1)}.
\end{equation}
For $M=2$ and $M=3$, Equation \eqref{Eq:NoFBErrExp} becomes  $E^{\text{AS}}\left( 2,{P},{\sigma^2} \right) = \frac{P}{2\sigma^2} $, and $E^{\text{AS}}\left( 3,{P},{\sigma^2} \right) = \frac{3}{8} \frac{P}{\sigma^2} $. Note that for large $M$, $E^{\text{AS}}\left( M,{P},{\sigma^2} \right)  \approx \frac{P}{4\sigma^2}$.


The perfect-feedback scenario ($\sigma_{\text{FB}}^2 \to 0$) 
was studied by Pinsker in \cite{pinsker1968probability}, who showed that for the transmission of $M\geq 3$ messages under the AS power constraint, the error exponent of the non-feedback AWGN channel shown in \eqref{Eq:NoFBErrExp}, can be improved up to: 
\begin{equation} \label{eq:ErrExpPerfectFB}
 E^{\text{AS}}_{\text{FB}}\left( M,{P},{\sigma^2}, P_{\text{FB}}, \sigma_{\text{FB}}^2 =0 \right) =  \frac{P}{2 \sigma^2}.
\end{equation}
{Note that for  $M=2$, this coincides with the achievable exponent in \eqref{Eq:NoFBErrExp} and hence} no further improvements over the non-feedback error exponent are possible subject to the AS power constraint, even using perfect feedback \cite{SheppBinary1969,YamamotoNoiselessFB2016}. 
Pinsker's result in \eqref{eq:ErrExpPerfectFB} constitutes an upper bound on the error exponent under the AS constraint for the one-way AWGN channel with noisy feedback.

\subsection{Error exponents for the One-Way AWGN channel with noisy feedback: past work}
\label{sec:RecentResultsOneWay}

When feedback is over a noisy channel, it is relevant to distinguish between active and passive feedback, roughly defined as follows:
\begin{itemize}
	\item In \textbf{passive feedback}, the forward channel output observed at the receiver is directly sent to the source and no encoding function is used. The transmitter sees a noisy version of the signal obtained by the receiver. 
	
	\item In  \textbf{active feedback}, the forward channel output is encoded using 
a function $g$ that takes as argument the sequence of all channel outputs available at the receiver, i.e. $U_k = g(Y^{k})$, and this is returned to the
	 transmitter at the $k$-th channel use.  Active feedback renders the returned transmission more robust against noise. 
	 Note that active feedback may mimic passive feedback by taking the encoding function $g$ in Figure  \ref{fig:OneWayGaussianActive} as a one-to-one mapping of signal $Y$ for each channel use $k$. 
\end{itemize}  
 


Yamamoto and Burnashev addressed the problem for the AWGN channel and noisy feedback under the AS power constraint for the transmission of a non-exponentially growing number of messages $M$ (zero-rate) using fixed block length encoding \cite{Burnashev2012AWGN-ZR,Yamamoto2012,Burnashev2014noisyFB,YamamotoNoisyFB2014}.  Their work extends previous techniques used to demonstrate achievable error exponent results on the Binary Symmetric Channel (BSC) with noisy feedback \cite{burnashev_bsc_2008,burnashev_zerorate_2008}, to the AWGN channel. 
Under the AS power constraint and for a large (non-exponentially growing with $n$) number of messages $M$, such that $\ln M= o(n), \; n\to \infty$ and $\sigma^2_{\text{FB}}\to 0$, the following error exponent is achievable as in \cite{Yamamoto2012}:
\begin{equation} \label{eq:YamBurn2012}
		E^{\text{AS}}_{\text{FB}_1}(M,P,\sigma, \sigma^2_{\text{FB}} ) \geq \frac{PM}{4\sigma^2(M-1)}\left[ 1 + \frac{1}{2+\sqrt{5}}-\frac{1}{2M} + o(1) \right].
\end{equation}
The number in the subscript of $E^{\text{AS}}_{\text{FB}_1}$ stands for the use of one \textit{switching moment}, or a change in the forward encoding function. Also, when $(\sigma^2_{\text{FB}}\to \infty)$, this becomes:
\begin{equation} 
		E^{\text{AS}}_{\text{FB}_1}(M,P,\sigma) \geq \frac{PM}{4\sigma^2(M-1)}\left[ 1 + \frac{1}{56\sigma^2_{\text{FB}}} + O(\sigma^{-4}_{\text{FB}}) \right] > E(M,P,\sigma, \sigma^2_{\text{FB}} = \infty) = \frac{PM}{4\sigma^2(M-1)}.
\end{equation}
{The result for very small feedback noise \eqref{eq:YamBurn2012} was improved in \cite{Burnashev2014noisyFB}.  Then, for $M\to \infty$ with $\ln M = o(n)$ as $n\to \infty$, an error exponent as that of Equation \eqref{eq:YamBurn2014AWGN} is attainable after one switching moment. Note that for a very small noise variance in the feedback link, this yields a larger improvement than that of Equation \eqref{eq:YamBurn2012} for very large $M$.}
\begin{equation} \label{eq:YamBurn2014AWGN}
		E^{\text{AS}}_1\left(P,\sigma^2, \sigma^2_{\text{FB}} \right) = \frac{P}{3\sigma^2} \left(1- \sigma^2_{\text{FB}}\right)
\end{equation}

\bigskip

Kim, Lapidoth and Weissman {\cite{kim:2006reliability,kim:BoundsEEAWGNNFB2006}} addressed the error exponents for the AWGN channel with feedback for the transmission of a small number of messages. They presented bounds on error exponents for active and passive feedback for the transmission of $M=2$ messages \cite{Kim:PassiveNoisyFB, Kim:ActiveNoisyFB} over the one-way AWGN channel with noisy AWGN feedback. 
 Results for $M\geq 3$ messages using passive feedback were presented by Xiang and Kim in \cite{Kim:ThreeCWPeak2013}. We summarize these results next. 
 Several of their techniques have been useful in the extensions to general $M$ and to the two-way channel presented here. 

\subsubsection{Achievable error exponents for the transmission of two messages}
\label{sec:ErrExpM2}
In \cite{Kim:PassiveNoisyFB} Kim et al. presented the following achievable error exponent for the transmission of two messages under the EXP constraint and passive feedback: 
\begin{equation} \label{eq:TwoPassKIM}
E^{\text{EXP}}_{\text{FB}}\left( 2,P, \sigma^2=1, \sigma_{\text{FB}}^2 \right) \geq  \frac{P}{2 \sigma_{\text{FB}}^2}.
\end{equation}
The use of active noisy feedback in the transmission of two messages is presented in \cite{Kim:ActiveNoisyFB}, considering the channel shown in Figure \ref{fig:OneWayGaussianActive} under the EXP power constraint for the forward channel and both AS and EXP power constraints for the feedback link. The achievable error exponent expressions using active feedback are respectively:
\begin{itemize}[leftmargin=0.5cm]
\item \textbf{AS power constraint}: $\sum^{n}_{k=1} g^2_k(Y^{k}) \leq n P_{\text{FB}}$
\begin{align} \label{eq:ExpectedActiveAS2}
	E_{\text{FB}^{\text{AS}}}^{\text{EXP}}\left(2, {P},{\sigma^2}, {P_{\text{FB}}},{\sigma^2_{\text{FB}}}  \right) &\geq \frac{P}{2\sigma^2}+ \frac{2 P_{\text{FB}}}{\sigma^2_{\text{FB}}}
\end{align}

	\item \textbf{EXP power constraint}: $\mathsf{E}\left[ \sum^{n}_{k=1} g^2_k(Y^{k}) \right] \leq n P_{\text{FB}}$
\begin{align} 
	E_{\text{FB}^{\text{EXP}}}^{\text{EXP}}\left(2, {P},{\sigma^2}, {P_{\text{FB}}},{\sigma^2_{\text{FB}} } \right) &\geq 2 \left(\frac{P}{\sigma^2}+ \frac{ P_{\text{FB}}}{\sigma^2_{\text{FB}}} \right). \label{eq:ExpectedActiveEB1}
\end{align}
\end{itemize}

\subsubsection{Achievable error exponents for $M\geq3$ messages under the AS power constraint and \textbf{passive feedback}}
\label{sec:OneWayM3AS}

Xiang and Kim \cite{Kim:ThreeCWPeak2013} studied the reliability function for the transmission of $M\geq3$ messages under the Peak Energy constraint (PE): $\mathsf{P} [ \sum_{i=1}^n f_k^2(w,\tilde{Y}^{k-1}) \leq nP ] = 1, \forall\; W=w \in \{1,2,...,M\}$ (which satisfies the AS constraint as well) on the forward channel, and passive feedback.  A block diagram of this scheme is shown in Figure \ref{fig:BLKDGMcombinedOWay} (left), which reflects how in passive feedback, the signal at the receiver is immediately returned to the transmitter at each channel use. Figure \ref{fig:BLKDGMcombinedOWay} (right) shows the scheme we propose for active feedback presented in Section \ref{sec:OneWayM3ASactive}.
\begin{figure}[H]
	\centering
		\includegraphics[width=1\textwidth]{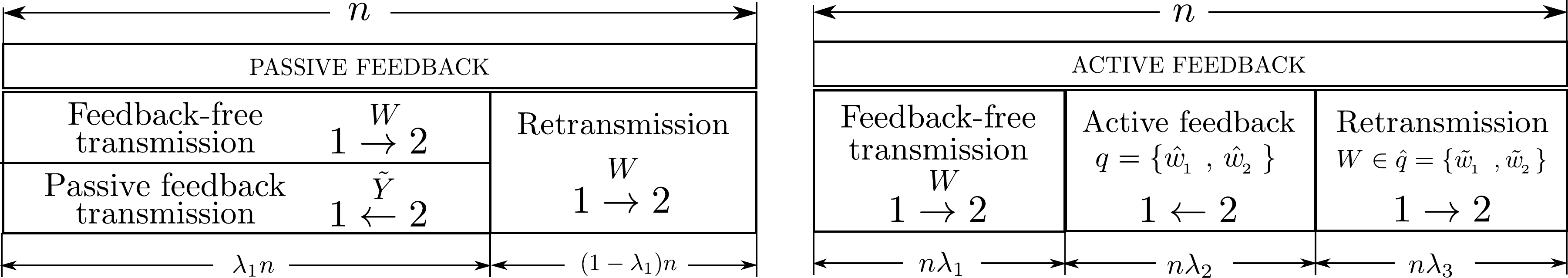}
		\caption{Block diagram for the achievability scheme under the AS power constraint: passive feedback (left), and active feedback (right).}
	\label{fig:BLKDGMcombinedOWay}
\end{figure}
The achievable error exponent derived in \cite{Kim:ThreeCWPeak2013} is expressed as the minimum of three terms; the first two related to the forward and feedback transmissions, and the last  resulting from a decoding rule that considers both:
\begin{align} \label{eq:PassiveFBM3}
		E_{\text{FB}^{\text{AS}}}^{\text{AS}}\left(M=3,{P},{\sigma^2}, P_{\text{FB}},{\sigma^2_{\text{FB}}}  , s\right) \geq \min{ \left\{\frac{ P}{\sigma^2} \frac{\lambda_1 (s^2-2s+4)}{8} , \frac{P_{\text{FB}}}{\sigma_{\text{FB}}^2}  \frac{3s^2 \lambda_1 }{32} , \frac{P}{\sigma^2} \frac{\left( 1 - \frac{\lambda_1}{4}\right) }{2} \right\} }.
\end{align}
Above, $\lambda_1 \in [0,1]$ is used to characterize the duration of the transmission ($ n \lambda_1$ channel uses) and retransmission  ($n(1-\lambda_1 )$ channel uses) stages, and $s \in (0,1)$ the size of a protection region that determines whether a retransmission is necessary, see \cite[Figs. 5-6]{Kim:ThreeCWPeak2013}. 
Since the first and third terms in \eqref{eq:PassiveFBM3} are equal for $\lambda_1 = \frac{4}{s^2-2s+5}$, the above may be rewritten in a way that explicitly shows the contribution of the passive feedback stage as:
\begin{align} \label{eq:PassiveFBM3simplified}
E_{\text{FB}^{\text{AS}}}^{\text{AS}}\left(M=3,{P},{\sigma^2},P_{\text{FB}},\sigma_{\text{FB}} , s \right) \geq \min{ \left\{\frac{ P}{2\sigma^2} \left( \frac{s^2-2s+4}{s^2-2s+5} \right) ,  \underbrace{ \frac{3}{8} \left( \frac{P_{\text{FB}}}{\sigma_{\text{FB}}^2} \frac{s^2}{s^2-2s+5} \right)}_{\text{passive feedback}}  \right\} }.
\end{align}
 
Comparing the above result with \eqref{Eq:NoFBErrExp} for $M=3$ indicates that the first argument of the $\min$ function yields a gain over the feedback-free error exponent for any $s\in (0,1)$ (with a maximum of $\frac{2}{5} \frac{P}{\sigma^2 }$ for $s\to0$). Once $s$ is chosen,  therefore ensuring that the first argument does provide an error exponent larger than that in the absence of feedback, the improvement is maintained if the second argument is equal to or greater than the first. This leads to:
\begin{equation} \label{eq:MinNoiseFBM3}
\frac{P_{\text{FB}}}{\sigma_{\text{FB}}^2} \geq \frac{P}{\sigma^2} \frac{s^2-2s+5}{s^2},
\end{equation}
i.e., if $s$ is very small, the feedback link's SNR needs to be remarkably larger than the forward link's. Note that the largest error exponent achievable by \eqref{eq:PassiveFBM3simplified} requires choosing $s$ very close to zero, and requires a very high SNR in the feedback link.  
In contrast, choosing $s$ close to $1$ leads to the smallest error exponent attainable by the first argument of \eqref{eq:PassiveFBM3simplified} (which coincides with the feedback-free exponent of Equation \eqref{Eq:NoFBErrExp}). This error exponent is achievable as long as the feedback SNR satisfies $\frac{P_{\text{FB}}}{\sigma_{\text{FB}}^2} > 4 \frac{P}{\sigma^2}$, otherwise, the scheme would not even achieve the feedback-free error exponent.
The error exponent of \cite[Theorem 1 ]{Kim:ThreeCWPeak2013} shown in Equation \eqref{eq:PassiveFBM3} can be generalized for all finite $M$ (see \cite[Appendix]{Kim:ThreeCWPeak2013}), which may be manipulated into the following form: 
\small 
\begin{equation} \label{eq:ThreeMessagesM}
	E_{\text{FB}}^{\text{PE}}\left(M,{P},{\sigma^2},s\right) \geq  \min\left\{ M \frac{P}{2\sigma^2} \left( \frac{s^2 - 2s +4}{M(s^2 - 2s +4)+3(M-2)} \right)   ,   \underbrace{\frac{P_{\text{FB}}}{\sigma^2_{\text{FB}}}  \frac{3M s^2}{8}\left( \frac{1}{M(s^2 - 2s +4)+3(M-2)} \right)}_{\text{passive feedback}}  \right\}.
\end{equation}
\normalsize
As in the case of $M=3$, Equation \eqref{eq:ThreeMessagesM} results from finding an optimal $\lambda_1^*(s) = \left(  \frac{M}{6(M-1)} (s^2-2s+4)+\frac{M-2}{2(M-1)} \right)^{-1}$ that equates the first and third terms of the $\min$ function.
Note that setting $M=3$ in \eqref{eq:ThreeMessagesM} leads to \eqref{eq:PassiveFBM3simplified}.  

\subsection{Error exponents for the one-way AWGN channel with noisy feedback: contributions}
\label{sec:RecentResultsOneWayContributions}

In this section we present new achievable error exponents for the one-way AWGN channel for the transmission of a finite number of messages under the AS and EXP power constraints, with active, noisy feedback. 

\subsubsection{Achievable error exponents for $M\geq 3$ messages under the AS power constraint and \textbf{active feedback}}
\label{sec:OneWayM3ASactive}

The achievability scheme presented in Figure \ref{fig:BLKDGMcombinedOWay} (left) and proposed in \cite{Kim:ThreeCWPeak2013} can be slightly modified to employ active feedback as shown in Figure  \ref{fig:BLKDGMcombinedOWay} (right) for a finite number of messages $M$. 
In contrast to the two stage block diagram of passive feedback, active feedback involves three non-overlapping stages. First, a feedback-free transmission based on a simplex code of $M$ messages is used to send $W$ during $\lambda_1 n$ channel uses. Once the first stage is complete, the receiver opts for immediate decoding only if the transmitted signal has been received within a protection region (similar to that defined  for passive feedback in \cite{Kim:ThreeCWPeak2013}) and if so, it ignores all future transmissions until the next message is sent. If the received signal is outside the protection region, the active feedback stage takes place for $\lambda_2 n$ channel uses. In this stage, the receiver uses a simplex code of ${M \choose 2}$ messages to inform the transmitter of the most likely pair of codewords it has determined: $q= \{\hat{w}_1 , \hat{w}_2 \}$, where $\hat{w}_1, \hat{w}_2  \in \{1,2,...,M\}$ represent the two closest (minimum distance) messages to the received signal $y^{\lambda_1 n}$. 
The transmitter decodes message $q$ as $\hat{q} = \{ \tilde{w}_1 , \tilde{w}_2\}$, and uses this to generate a binary retransmission signal aimed to help the receiver making the right decision based on the true message $W=w$. The final retransmission stage lasts for $\lambda_3 n$ channel uses ($\lambda_3 = 1 - (\lambda_1 + \lambda_2)$ ), and corresponds to the transmission of two antipodal signaling codewords that are generated depending on whether the true message $w$ is equal to the first or second element in $\hat{q}$. If the true message is not in $\hat{q}$, this is counted as an error, and nothing is transmitted to the receiver. This approach leads to the first of our main results:
\begin{theorem} \label{th:OneWayActiveASM3}
An achievable error exponent for the transmission of a finite number of messages $M\geq3$ over a one-way AWGN channel with active noisy feedback under the AS power constraint is given by:  
\begin{equation} \label{eq:ASactiveM}
	E_{\text{FB}^{\text{AS}}}^{\text{AS}}\left(M,{P},{\sigma^2},P_{\text{FB}},\sigma^2_{\text{FB}},s\right) \geq  \min\left\{ M \frac{P}{2\sigma^2} \left( \frac{s^2 - 2s +4}{M(s^2 - 2s +4)+3(M-2)} \right)   ,   \underbrace{ \frac{P_{\text{FB}}}{\sigma^2_{\text{FB}}}  \frac{{M \choose 2}}{4\left({M \choose 2}-1\right)}}_{\text{active feedback}}    \right\},
\end{equation}
\normalsize
for parameter  $s\in [0,1]$. 
\end{theorem}
In Equation \eqref{eq:ASactiveM}, the second argument of the $\min$ operator corresponds to the active feedback contribution to the probability of error given by the use of a simplex code -Equation \eqref{Eq:NoFBErrExp}- and the transmission of ${M \choose 2}$ messages corresponding to all possible unordered codeword pairs, which as in the case of three messages, are labeled in lexicographic order for the sake of identification.  
A comparison of Equations \eqref{eq:ASactiveM} and \eqref{eq:ThreeMessagesM} shows that the second argument of the $\min$ function under active feedback does not depend on $s$.  
Active feedback uses a feedback-free transmission of messages based on a simplex code, and not on the transmission of the exact location of the received signal $y^{\lambda_1 n}$ as it is in passive feedback.  
In the latter, the probability of error is based on the geometry defined for the protection regions determined by parameter $s$.  
Section \ref{sec:ActiveM3OneWay} presents the complete proof of this result. First, we  present the illustrative case of $M=3$ and later generalize this for any finite $M$. 
\subsubsection{Achievable error exponents for the transmission of $M$ messages under the EXP power constraint and active noisy feedback}
\label{sec:EXPGenM}

This section presents our second contribution that results from a direct generalization of the work of Kim, Lapidoth and Weissman \cite{Kim:ActiveNoisyFB} for $M=2$ messages 
to an arbitrary but finite number $M$. This is presented in Theorem \ref{th:OneWayEXPM2}, whose proof is presented in Section \ref{sec:ProofOneWayEXPM}.
\begin{theorem} \label{th:OneWayEXPM2}
An achievable error exponent for the transmission of a finite number of messages $M$ over a one-way AWGN channel with active noisy feedback under the EXP power constraint in both the forward and feedback directions is given by:
\begin{align} \label{eq:Theorem2EXPMow}
	 E^{\text{EXP}}_{\text{FB}^{\text{EXP}}}\left( M,{P},{\sigma^2}, {P_{\text{FB}}},{\sigma_{\text{FB}}^2} \right)	\geq \frac{M}{M-1} \left( \frac{P}{\sigma^2} +  \frac{ P_{\text{FB}}}{\sigma_{\text{FB}}^2}    \right).
\end{align}
\end{theorem}
Note that for $M=2$, \eqref{eq:Theorem2EXPMow} yields the result of \cite[Equation (15) ]{Kim:ActiveNoisyFB} and shown in Equation \eqref{eq:ExpectedActiveEB1}. Also, observe that the error exponent lower bound is proportional to the summation of the SNRs of the forward and backward directions. 

\section{Error exponents for the Two-Way AWGN channel} 
\label{sec:ResultsTwoWayII}

The two-way channel was first introduced by Shannon in \cite{Shannon:1961} and further studied by Han in \cite{Han:1984}. The capacity region of the two-way AWGN channel (with independent noise across the terminals) is known, and is a rectangular region where both users may simultaneously attain their interference-free AWGN capacity. The rates achievable for this channel can not be increased by interaction or adaptation between the two terminals. The question pursued here is whether the same is true of error exponents -- may they be improved through the use of adaptation/interaction between the terminals in a two-way setting, where feedback and messages must share the same resources. 
 Two-way error exponents have not been studied in the past, to the best of our knowledge. 
In the two-way AWGN channel each terminal may intuitively perform two types of tasks: 1) transmission of their own message, and 2) transmission of feedback information for the other terminal. 
Since each terminal may use part of its available power to cooperate with the opposite direction, we have found that gains resulting from interaction may come at the price of a reduction of the error exponent of the terminal providing feedback -- there is a tradeoff between the achievable error exponents in the two directions (at least under the presented schemes).\footnote{Recall that we are operating at zero-rate, so the tradeoff with rate is not captured in this simplified, yet still challenging setting. However, the error exponent region does generally depend on $M$, the number of messages being transmitted {in each direction. Extensions to having a different number of messages be sent in each direction is left for future work.}} 

\medskip

\subsection{Two-way AWGN channel model description and definitions}
\label{sec:ProbStatTWOway}

The two-way AWGN channel is depicted in Figure  \ref{fig:TwoWayGaussian}, comprising two users denoted as terminal $i$ for $i=\{1,2\}$.  Terminal $i$ transmits message $W_i$, uniformly selected from $ {\cal W}_i:= \{1,2,\cdots, M\}$ to terminal $(3-i)$ using a fixed block length code of size $n$. 
\begin{figure}[htb]
	\centering
		\includegraphics[width=0.6\textwidth]{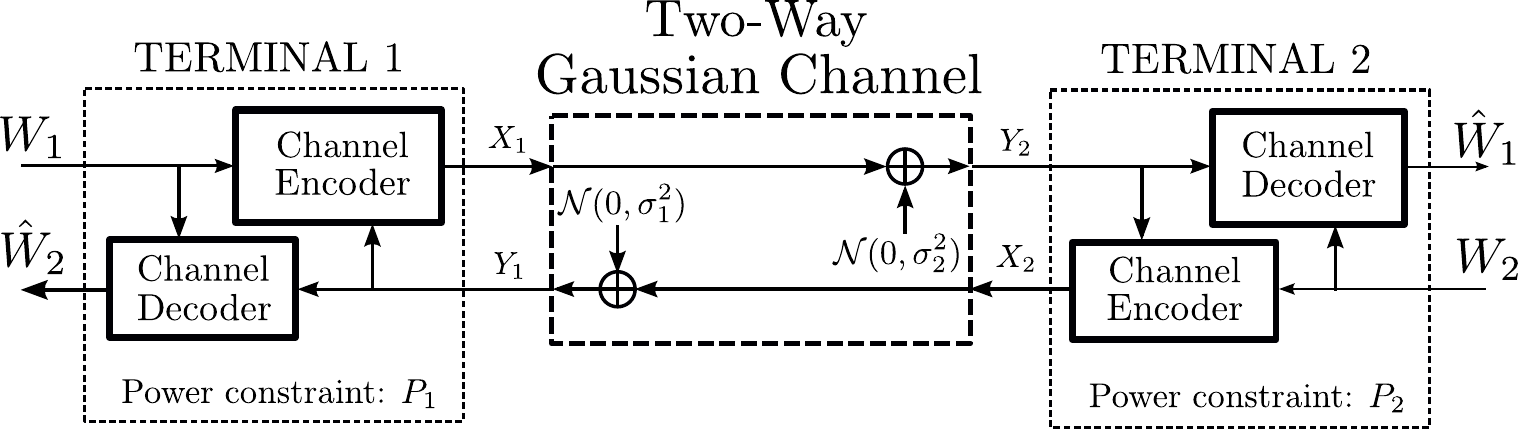}
		\caption{Two-way AWGN channel.}
	\label{fig:TwoWayGaussian}
\end{figure}
The general two-way AWGN channel model is presented in Equation \eqref{eq:TWChModel1}, which characterizes the channel output received at the $i$-th terminal 
\begin{align} 
	Y_i &= X_i + a_i X_{3-i} + N_i,  \label{eq:TWChModel1}
\end{align}
where, for $i=\{1,2\}$,  $a_i$ is a constant,  {$X_i\in\mathbb{R}$ corresponds to channel inputs satisfying the input block power constraint $P_i$, $Y_i \in \mathbb{R}$} to channel outputs and $N_i \sim {\cal N}(0,\sigma_i^2)$ to  zero-mean Gaussian noise processes, each independent and identically distributed across channel uses. 
The model described by \eqref{eq:TWChModel1} can be simplified by noting that each terminal can subtract its own transmission $X_i$, thus, the two-way AWGN channel may equivalently be represented as in  \eqref{eq:TWChModelSimple1}, where each link is modeled as an independent and interference free AWGN channel with a noise variance $\sigma^2_i$ perceived at the $i$-th terminal's receiver, with $a_i=1$ for simplicity:
\begin{align} \label{eq:TWChModelSimple1}
	Y_i &= X_{3-i} + N_i  ; \;\; N_i \sim {\cal N}(0,\sigma_i^2)
\end{align}
The capacity region of this channel is rectangular \cite{SatoH1977,Han:1984}, with each user being able to transmit at rates up to its interference-free AWGN capacity  (denoted by $C_{12}$ and  $C_{21}$ for each direction respectively). 

Here, we characterize error exponents for the zero rate operational rate pair, i.e. for $(R_{12} , R_{21}) = (0,0)$, and for different ratios between the SNRs in the two directions.
Let $\frac{P_i}{\sigma^2_{3-i}}$ be the signal-to-noise ratio for link $i$. Then we will consider both symmetric $\left(\frac{P_1}{\sigma^2_{2}} = \frac{P_2}{\sigma^2_{1}} \right)$ and non-symmetric $\left(\frac{P_1}{\sigma^2_{2}} \neq \frac{P_2}{\sigma^2_{1}}\right) $ channels. 
The non-symmetric case is of particular importance under the AS power constraint, since as indicated in \cite{Kim:ThreeCWPeak2013,Yamamoto2012,YamamotoNoisyFB2014} only a feedback channel stronger that the forward direction is able to attain gains over feedback-free error exponents on the one-way AWGN channel. This condition becomes even more critical for the two-way channel, since the feedback link is also used for messages transmission. Moreover, symmetric SNR channels seem to be unable to achieve an error exponent region greater than those achieved by independent transmissions under the AS constraint.


 
In Section \ref{sec:ResultsOneWay}, we reviewed how error exponents of the one-way AWGN channel can be improved by the inclusion of (even noisy) feedback. 
In all cases, the receiver's resources are dedicated solely to feedback and helping the message transmission. 
If the receiver transmits its own messages as well, the one-way channel with noisy feedback transforms into the two-way channel and results from the former can be applied in the latter.
Since messages flow in two directions, an error exponent pair must be considered now. 

Let $\left({P_1},{\sigma_2^2},{P_2},{\sigma_1^2},n\right)$ be a block length-$n$ code, consisting of two encoding and two decoding rules as depicted in Figure  \ref{fig:TwoWayGaussian}. Each terminal's encoding rule consists of a set of $n$ functions, defined for the $k$-th channel use as:
\begin{equation}
 x_{i,k} :\{1,2,...,M\}  \times Y_{i}^{k-1} \rightarrow \mathcal{X}_{i}, \;\; \text{for } k=1,...n,
\end{equation}
leading to the $k$-th channel inputs for terminal $i$: $X_{i,k} = x_{i,k} \left(W_i,Y_{i}^{k-1}\right)$. 
Decoding rules are denoted by $\phi_{i}$, and estimate the received message based on the sequence $Y_{i}^n$  as:
\begin{equation}
\phi_i:\mathcal{Y}^n_{i} \times {\cal W}_i \rightarrow {\cal W}_{3-i}, \;\;\text{for } i=1,2.
\end{equation}

Let $P^X_{e_{12}}\left(M, {P_1},{\sigma_2^2}, {P_2},{\sigma_1^2}, n\right) =\sum_{w_1, w_2} \text{Pr}(\phi_2(y_2^n, w_2) \neq w_{1}|w_{1}, w_{2} \text{ sent})$ and similarly $P^X_{e_{21}}\left(M, {P_1},{\sigma_2^2}, {P_2},{\sigma_1^2}, n\right)$ denote the probability of error in the forward and backward directions simultaneously achieved by a particular $({P_1},{\sigma_2^2},{P_2},{\sigma_1^2},n)$ code under $X\in \{\text{AS}, \text{EXP}\}$ power constraints. 

\begin{definition}
A pair of error exponents $\left(E_{12}, E_{21} \right)^X$ is called achievable for 
the transmission of a finite number of messages $M$, under the $X$ power constraint for the two-way AWGN channel, if there exists a $\left({P_1},{\sigma_2^2},{P_2},{\sigma_1^2},n\right)$ code such that for large $n$, simultaneously 
 		\begin{align}
		-	\frac{1}{n} \log{\prob^X_{e_{12}}}\left(M, {P_1},{\sigma_2^2}, {P_2},{\sigma_1^2}, n\right) &> E^X_{12}\left(M,{P_1},{\sigma^2_2}, {P_2},{\sigma^2_1} \right),  	  \text{ and }\\
		-	\frac{1}{n} \log{\prob^X_{e_{21}}}\left(M, {P_1},{\sigma_2^2}, {P_2},{\sigma_1^2}, n\right) &> E^X_{21}\left(M,{P_1},{\sigma^2_2}, {P_2},{\sigma^2_1} \right).
		\end{align}
\end{definition}

\begin{definition}
The error exponent region for the two-way AWGN channel for the transmission of $M$ messages corresponds to the union over all achievable error exponent pairs $(E_{12},E_{21})^X$, where we will often drop the arguments of $E_{ij}$ for simplicity and sometimes we may refer to $(E_{12},E_{21})^{\text{AS}}$ as $(E_{12}^{\text{AS}},E_{21}^{\text{AS}})$. 
\end{definition}



Next we present and extend the results initially demonstrated in \cite{PalacioDevroyeISIT2018}, with proofs in the upcoming sections. 

\subsection{Achievable error exponent for non-interactive terminals}
\label{sec:TwoWayResultsProps}
While in the two-way setting, both terminals may adapt their current inputs to past received channel outputs, they need not do so, and may ignore (for the purpose of generating their channel inputs) the received outputs altogether. 
The following proposition establishes the achievable error exponent region when the terminals do not interact:
\begin{proposition} 
\label{Prop1}
The achievable error exponent region for non-interactive terminals is formed by the union over all simultaneously achieved error exponents pairs under both, AS and EXP power constraints for the transmission of $M$ messages: 
\begin{align} \label{eq:prop1}
		E_{12} &\geq \frac{P_1}{\sigma^2_2} \frac{M}{4(M-1)}, \\
		E_{21} &\geq \frac{P_2}{\sigma^2_1} \frac{M}{4(M-1)}.
\end{align}
\end{proposition}
The above equations follow directly from applying Shannon's result of \cite{Shannon1959}, given in Equation \eqref{Eq:NoFBErrExp}, to each communicating pair, and using a simplex code in each direction. In this scheme, both terminals are concerned about their own message transmission only, and do not allocate any resources to help the other direction. 

\subsection{Achievable error exponent region under interactive terminals}
\label{sec:InteractiveTwoWay}

In this section we analyze the error exponent achieved by interactive transmission protocols, and have identified  certain scenarios (one direction much better than the other) for which under the AS constraint, interaction leads to an improvement over the feedback-free error exponent region characterized by Proposition \ref{Prop1}. This improvement comes at the cost of an error exponent decrease linked to the amount of power a terminal is unable to use for its own message transmission, as it was allocated to serve the opposite direction through feedback.

As described for the one-way AWGN channel, the power constraint imposed on the input block codewords and the number of messages to be transmitted, along with channels' SNRs  determine whether error exponent gains are feasible for one or both communication directions {through interaction}. 
Consider for example the AS constraint and the transmission of two messages, and observe that interaction can not improve the two-way error exponent as even noiseless feedback is unable to improve the non-feedback error exponent of a binary transmission for a one-way AWGN channel under AS constraint \cite{pinsker1968probability}. 
Therefore, even in the ideal case that a noiseless feedback link is available for each direction (and not used to transmit the true messages), the highest attainable error exponent for each direction coincides with the non-feedback one, which suggests that non-interactive transmissions suffice for $M=2$.  
In general, the same upper bound applies for the transmission of $M$ messages (Pinsker's upper bound is independent of $M$) and error exponent gains over non-feedback are  possible for $M>2$ in both directions.

\medskip
\subsubsection{Achievable error exponents region under the AS power constraint}
\label{sec:TwoWayAS}

We study the transmission of $M\geq 3$ messages over an AWGN two-way channel for which  $\frac{P_1}{\sigma_2^2} < \frac{P_2}{\sigma_1^2}$, as no gains over Proposition \ref{Prop1} appear (with current known achievability schemes) to be attainable for symmetric SNRs, as it was the case for the 
one-way AWGN channel under the AS constraint as well
{\cite{Kim:ThreeCWPeak2013,Yamamoto2012,Burnashev2014noisyFB}}.
{Assuming that the $1\to 2$ communication direction is noisier than the $1\leftarrow 2$ direction, consider a two-way achievability scheme that takes advantage of this asymmetry: the stronger link can be used during a fraction of the block length $n$ to transmit message $W_2$ without the help of feedback using a simplex code, and for another fraction of time, to improve the error exponent of the weaker direction by providing passive or active feedback in the transmission of message $W_1$. Figure \ref{fig:BLOCK2wayAS} shows these two approaches, for passive and active feedback respectively. The two schemes differ in when the transmission of message $W_2$ occurs. In the case $W_1$ is transmitted with passive feedback, message $W_2$ is sent during the first $\lambda n$ channel uses (for $\lambda \in [0,1]$) over the stronger link while the weaker channel remains idle, since Terminal 1 may initiate its own transmission (helped by terminal 2) only once the $1\leftarrow 2$ message transmission concludes. In the remaining $(1-\lambda)n$ channel uses, message $W_1$ is transmitted employing passive feedback, as described in Section \ref{sec:ResultsOneWay} for the one-way AWGN. Since terminal 2 has already used part of its available power $P_2$ for the forward transmission, only the remainder can be used to serve the other direction through feedback. Note that in passive feedback, a signal received at Terminal 2 at the $i$-th channel use is immediately fed back to the Terminal 1 without delay, and that both directions are busy at the same time.} 

\begin{figure}[H]
	\centering
		\includegraphics[width=0.9\textwidth]{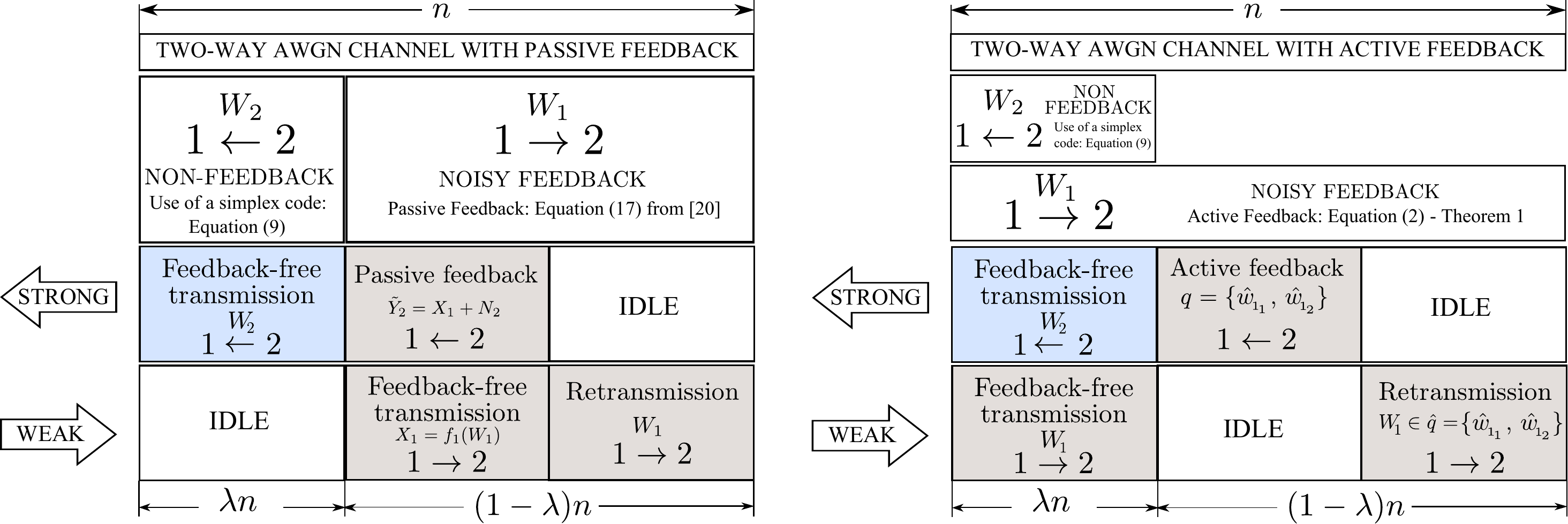}
		\caption{Two-Way AWGN channel achievability block diagram under AS power constraint.}
	\label{fig:BLOCK2wayAS}
\end{figure}
{The case of active feedback, shown in Figure \ref{fig:BLOCK2wayAS} (right), differs from passive feedback in that the transmission of $W_2$, and the first stage of the active feedback supported transmission of $W_1$  may occur simultaneously. Note that this is possible since both directions are independent, and also, because the active feedback stage can only start once the first stage transmission is concluded, which leaves room for the transmission of $W_2$.}

\begin{theorem}
\label{Prop2}
An achievable error exponent region for the transmission of $M\geq3$ messages over a two-way AWGN channel with non-symmetric SNR $\frac{P_1}{\sigma_2^2} < \frac{P_2}{\sigma_1^2}$, under the AS power constraint and \textbf{passive feedback} is the union over all error exponent pairs $(E_{12},E_{21})^{\text{AS}}$ over parameters $\lambda \in [0,1]$, and $s \in (0,1)$ satisfying:  
\begin{align}
 E_{21}^{\text{AS}} &\geq  \frac{M}{4(M-1)} \lambda \frac{P_2}{\sigma^2_1}, \label{eq:nonFBpassiveM}\\
 E_{12}^{\text{AS}} &\geq \min\left\{ M \frac{P_1}{2\sigma^2_2} \left( \frac{s^2 - 2s +4}{M(s^2 - 2s +4)+3(M-2)} \right)   ,   \underbrace{\frac{P_{2}}{\sigma^2_{1}}  \frac{3M s^2}{8}\left( \frac{1-\lambda}{M(s^2 - 2s +4)+3(M-2)} \right)}_{\text{passive feedback}}  \right\}. \label{eq:PassiveMTW} 
\end{align}
\end{theorem}
Equation \eqref{eq:nonFBpassiveM} follows from \eqref{Eq:NoFBErrExp}, the use of a non-feedback transmission for message $W_2$ in the first $\lambda n$ channel uses using a simplex code of $M$ symbols. Equation \eqref{eq:PassiveMTW} follows from direct application of the noisy feedback aided scheme presented in \cite[Section II-B]{Kim:ThreeCWPeak2013} and reviewed in Section \ref{sec:OneWayM3AS}, which is used for the transmission of message $W_1$ in the remaining $(1-\lambda)n$ channel uses. 
Note that \eqref{eq:PassiveMTW} is presented in the form of \eqref{eq:PassiveFBM3simplified} since it explicitly shows the error exponent contribution of the forward and feedback transmissions. 
\begin{figure}[htb]
	\centering
		\includegraphics[width=0.5\textwidth]{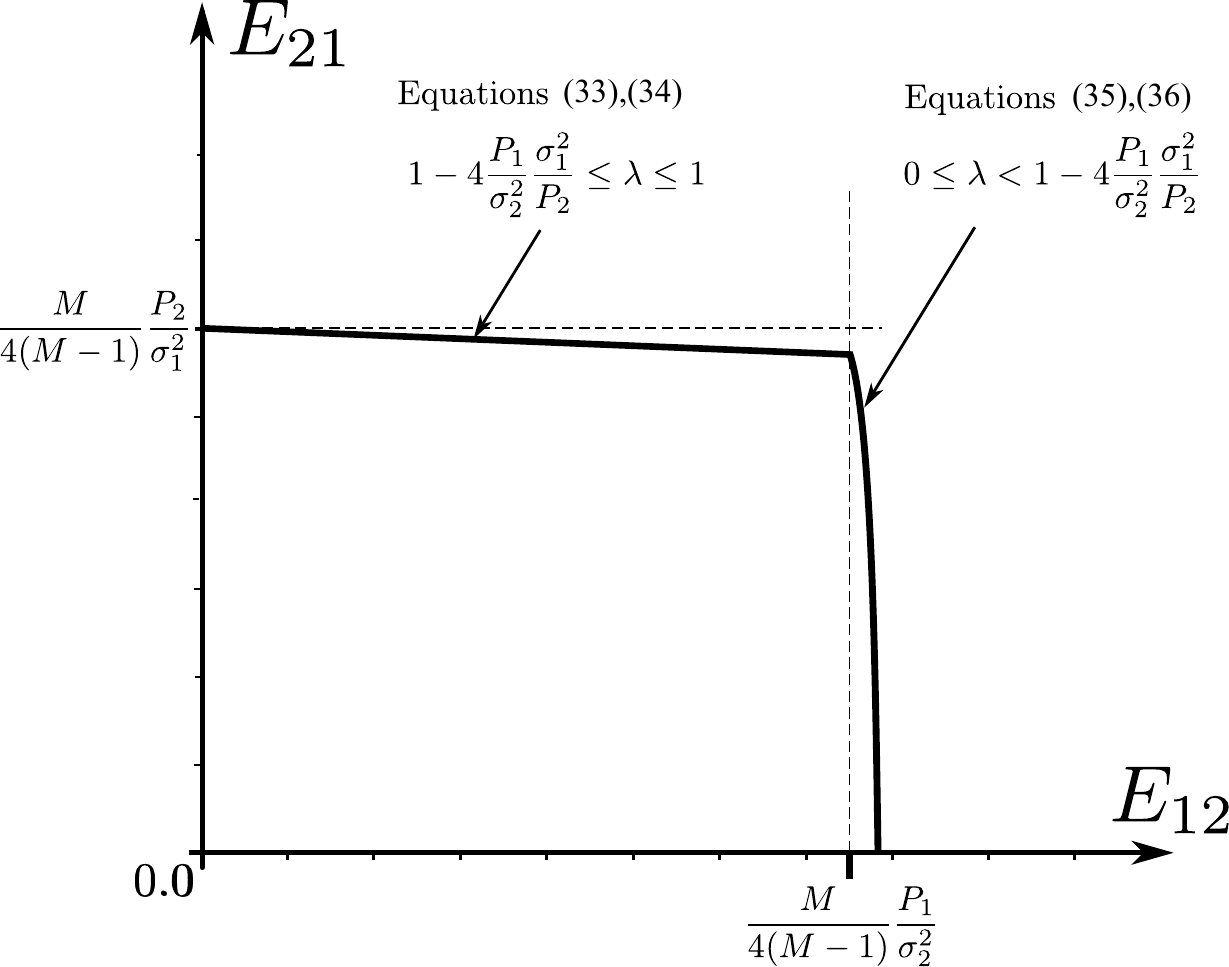}
		\caption{Achievable error exponent region for the two-way AWGN channel with passive feedback under the AS power constraint, as parametrized by $\lambda \in [0,1]$ for the transmission of $M$ messages. The upper-right corner point is attained at $\lambda= 1- 4\frac{P_1}{\sigma_2^2} \frac{\sigma_1^2}{P_2} $. }
	\label{fig:M3soloSIMPLE2SNR160Analy}
\end{figure}
The achievable error exponent region for the two-way AWGN channel with passive feedback under the AS power constraint is presented in Figure \ref{fig:M3soloSIMPLE2SNR160Analy}, parametrized by $\lambda$ as: 
\begin{enumerate}
	\item For $ \left( 1- 4\frac{P_1}{\sigma_2^2} \frac{\sigma_1^2}{P_2} \right) \leq \lambda \leq 1 $:
	\begin{align}
	  E_{12}  &\geq  \left( \frac{1-\lambda}{4} \right)\frac{M}{4(M-1)} \frac{P_2}{\sigma_1^2}, \\
			E_{21} &\geq \lambda \frac{M}{4(M-1)} \frac{P_2}{\sigma_1^2}.
\end{align}
	\item For $ 0 \leq \lambda < \left( 1- 4\frac{P_1}{\sigma_2^2} \frac{\sigma_1^2}{P_2} \right)$:
\begin{align}
		E_{12} &\geq \frac{1}{2} \frac{P_1}{\sigma_2^2} \frac{s^2-2s+4}{s^2-2s+5} ,				\\
		E_{21} &\geq \frac{M}{4(M-1)} \frac{P_2}{\sigma_1^2} \lambda,
\end{align}
where $s =  \frac{  \frac{P_1}{\sigma_2^2}  }{  \frac{P_1}{\sigma_2^2}  - \frac{3}{4} \frac{P_{2}}{\sigma_1^2} (1-\lambda) }  - \frac{\sqrt{ 3 \frac{P_1}{\sigma_2^2} \frac{P_{2}}{\sigma_{1}^2} (1-\lambda) - 3 \left(\frac{P_1}{\sigma_2^2}\right)^2  }}{  \frac{P_1}{\sigma_2^2}  - \frac{3}{4} \frac{P_{2}}{\sigma_{1}^2} (1-\lambda) } $.

\end{enumerate}

An achievable error exponent region using active feedback, under asymmetric SNRs and the AS power constraint in each direction is presented in Theorem \ref{Prop3}:
\begin{theorem}	
\label{Prop3}
An achievable error exponent region for the transmission of $M\geq3$ messages over a two-way AWGN channel with non-symmetric SNR $\frac{P_1}{\sigma_2^2} < \frac{P_2}{\sigma_1^2}$, under an AS power constraint and \textbf{active feedback} is the union over all error exponent pairs $(E_{12},E_{21})^{\text{AS}}$ over parameters $\lambda\in [0,1]$ and $s \in (0,1)$ satisfying:    
\begin{align} 
    E_{21}^{\text{AS}} &\geq \frac{M}{4(M-1)} \lambda \frac{P_2}{\sigma^2_1} \label{eq:nonFBM3ii}\\
		E_{12}^{\text{AS}} &\geq \min\left\{ M \frac{P_1}{2\sigma^2_2} \left( \frac{s^2 - 2s +4}{M(s^2 - 2s +4)+3(M-2)} \right)   ,   \underbrace{ \frac{P_{2}}{\sigma^2_{1}}  \frac{{M \choose 2} (1-\lambda)}{4\left[{M \choose 2}-1\right]}}_{\text{active feedback}}    \right\} . \label{eq:ErrExpAFB12}
\end{align}	
\end{theorem}
As in Theorem \ref{Prop2}, Equation \eqref{eq:nonFBM3ii} follows from using \eqref{Eq:NoFBErrExp} for $\lambda n$ channel uses and a simplex code of $M$ symbols for the transmission of $W_2$.  Equation \eqref{eq:ErrExpAFB12} results from the use of encoded feedback.  Specifically, message $W_1$ is transmitted in $(1-\lambda)n$ channel uses employing the active noisy feedback-aided scheme for the one-way AWGN channel communication presented in Section \ref{sec:OneWayM3ASactive},  in Theorem \ref{th:OneWayActiveASM3}, Equation \eqref{eq:ASactiveM}.
 
The error exponent region for the case of active feedback is shown in Figure \ref{fig:M3soloSIMPLE2SNR160AnalyACTIVE}. A portion of this  region can be characterized for $0 \leq E_{12} \leq  \frac{P_1}{\sigma_2^2} \frac{2M}{7M - 6)} $ as the line
\begin{align}
	E_{21} \geq  				
      \frac{M}{4(M-1)} \frac{P_2}{\sigma_1^2} - \frac{M}{M-1} \frac{{M \choose 2}-1}{{M \choose 2}} E_{12}.
\end{align}

\begin{figure}[htb]
	\centering
		\includegraphics[width=0.5\textwidth]{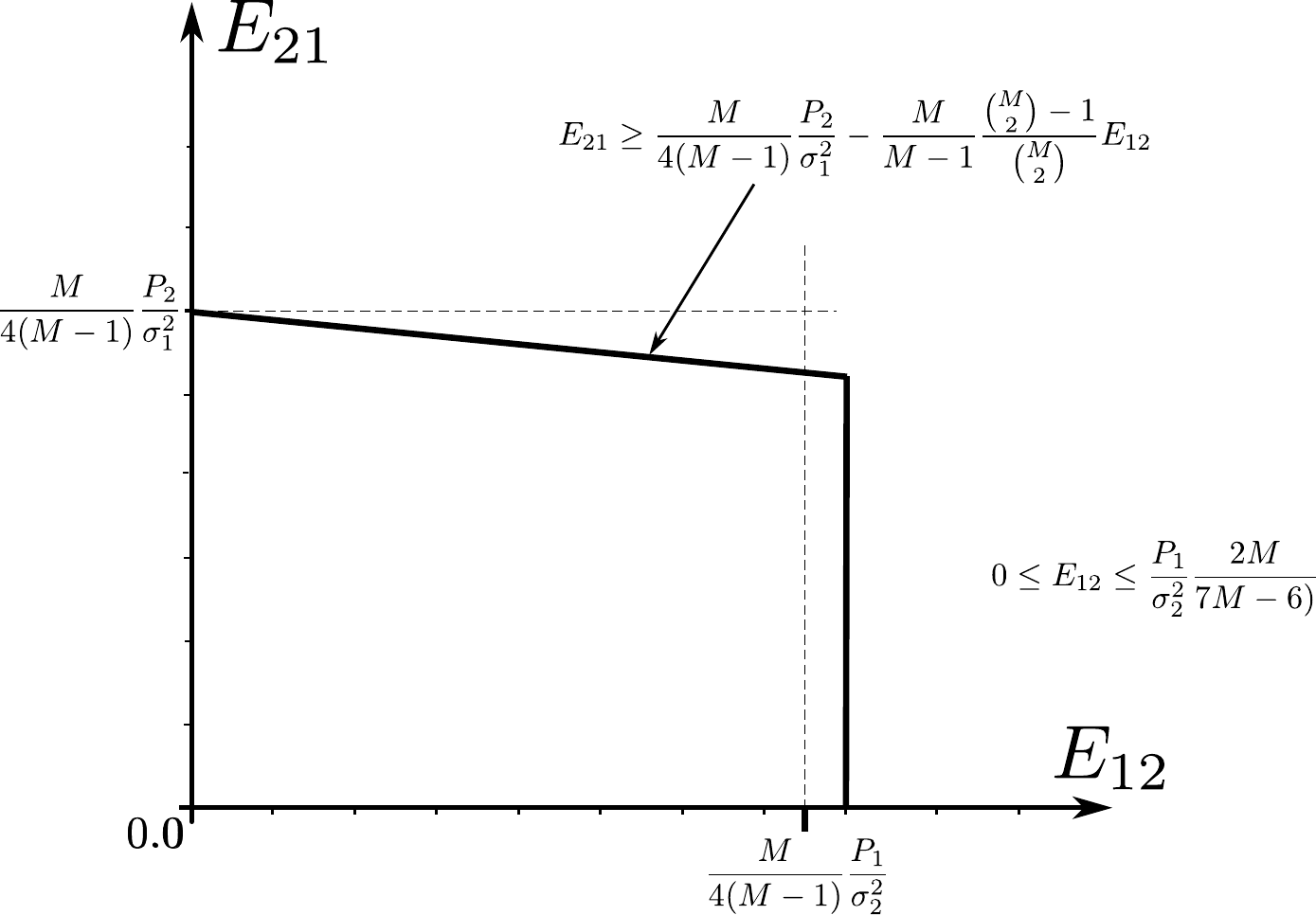}
		\caption{Achievable error exponent region for the two-way AWGN channel with active feedback under the AS power constraint  and the transmission of $M$ messages.}
	\label{fig:M3soloSIMPLE2SNR160AnalyACTIVE}
\end{figure}

The scheme shown in the  block diagram of Figure \ref{fig:BLOCK2wayAS}  yields the largest error exponent region derived so far for interactive terminals under non-symmetric SNRs under the AS power constraint. 
It appears that error exponent gains over feedback free transmission are possible only for the weaker direction, which is not surprising given the results for the one-way channel with noisy feedback under an AS power constraint. 

 Next, we consider the EXP power constraint, which holds for all relative SNR conditions. 
As described in Section \ref{sec:Introd}, the EXP power constraint permits
 very high amplitude transmissions associated to very rarely occurring events, which are used to correct detected decoding errors. In the following subsections, we show that this unique feature can be leveraged for the two-way AWGN channel as well. 

\bigskip
\subsubsection{Achievable error exponents region under the EXP power constraint}
\label{sec:TwoWayEXP}
 
This achievability scheme is based on the use of the building block component introduced by Kim et al. in \cite{Kim:ActiveNoisyFB} for both directions. We modified this component, originally designed for the one-way AWGN channel, to operate for a general number of messages as presented in Theorem \ref{th:OneWayEXPM2} and shown in Section \ref{sec:ProofOneWayEXPM}. Here, we utilize it for the simultaneous transmission of $M$ messages in opposite directions. 
Note that we can employ a similar approach as that used for the one-way transmission, specifically: an initial transmission of the true message followed by a feedback transmission that returns the message estimated by the receiver, followed (possibly) by a high amplitude retransmission to correct decoding errors if necessary. In the two-way channel, these three stages can occur simultaneously, with the restriction that each transmitter must satisfy the power constraints.
This scheme results in the region of Theorem \ref{th:EXPMachievTW}, whose proof is presented in Section \ref{sec:M2EXP}: 
\begin{theorem}
\label{th:EXPMachievTW}
An achievable error exponent region for the two-way AWGN channel and the transmission of a finite number of messages $M$ under the EXP power constraint for both directions is given by the union over all error exponent pairs $(E_{12},E_{21})^{\text{EXP}}$  over $\lambda \in (0,1)$ for which:  
\begin{align} \label{eq:E12expMTW}
	E_{12}^{\text{EXP}} &\geq \frac{M}{M-1} \left(\lambda K_1 \frac{ P_1}{\sigma_2^2} + (1-\lambda) J_2 \frac{P_2}{\sigma_1^2} \right)\\
	E_{21}^{\text{EXP}} &\geq \frac{M}{M-1} \left( \lambda K_2\frac{ P_2}{\sigma_1^2} + (1-\lambda) J_1 \frac{ P_1}{\sigma_2^2} \right), \label{eq:E21expMTW}
\end{align}
where $K_1, K_2 \in [0, \frac{1}{\lambda}]$ and $ J_1, J_2 \in [0,\frac{1}{1-\lambda}]$  such that $\lambda K_i + (1-\lambda)J_i \leq 1$ for $i = 1,2$. 		
\end{theorem}

An achievable error exponent sum-rate can be found by adding Equations \eqref{eq:E12expMTW} and \eqref{eq:E21expMTW}.
\begin{align} \label{eq:E12expMTWsum}
	E_{12}^{\text{EXP}} + E_{21}^{\text{EXP}} &\geq \frac{M}{M-1} \left(\lambda K_1 \frac{ P_1}{\sigma_2^2} + (1-\lambda) J_2 \frac{P_2}{\sigma_1^2} \right) + \frac{M}{M-1} \left( \lambda K_2\frac{ P_2}{\sigma_1^2} + (1-\lambda) J_1 \frac{ P_1}{\sigma_2^2} \right)\\
		&= \frac{M}{M-1} \left( \lambda K_1 \frac{ P_1}{\sigma_2^2} + (1-\lambda) J_2 \frac{P_2}{\sigma_1^2} + \lambda K_2\frac{ P_2}{\sigma_1^2} + (1-\lambda) J_1 \frac{ P_1}{\sigma_2^2}   \right)\\
		&= \frac{M}{M-1} \left(  \frac{ P_1}{\sigma_2^2} \underbrace{\left( \lambda K_1 +(1-\lambda) J_1\right)}_{\leq 1 \text{ by Eq.} \eqref{eqn:PwrUsr1} }   +  \frac{P_2}{\sigma_1^2}  \underbrace{ \left( \lambda K_2 +(1-\lambda) J_2\right)}_{\leq 1 \text{ by Eq.} \eqref{eqn:PwrUsr1}} \right)   \\
		&= \frac{M}{M-1} \left(  \frac{ P_1}{\sigma_2^2} +  \frac{P_2}{\sigma_1^2} \right), \label{eq:EEREXPM}
\end{align}
where the equality in \eqref{eq:EEREXPM} follows from \eqref{eqn:PwrUsr1} and taking  $ \lambda K_i +(1-\lambda) J_i =1$ (to use all the power). 
Figure \ref{fig:M2allsimpleAnalytic} shows this error exponent region. 
Observe that this region can be viewed as  time-sharing two one-way schemes, each characterized by the exponent in Theorem \ref{th:OneWayEXPM2}, in which each direction operates for a fraction of time aided by the other terminal. 
The axis-crossing points are obtained when the whole block length $n$ and all power is dedicated to the transmission of one direction only.

\begin{figure}[H]
	\centering
		\includegraphics[width=0.5\textwidth]{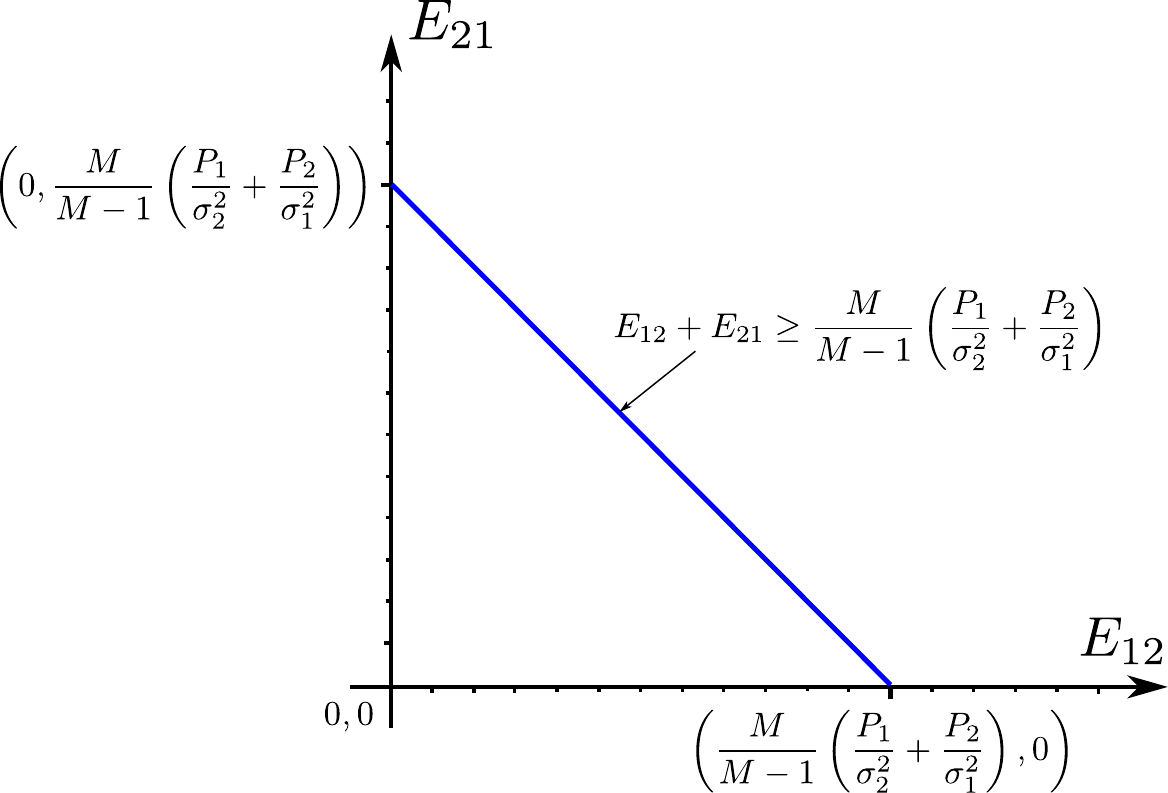}
		\caption{Achievable error exponent region for the transmission of $M$ messages under the EXP power constraint.}
	\label{fig:M2allsimpleAnalytic}
\end{figure}

Theorem \ref{th:EXPMachievTW} concludes the statement of our main results. The remainder of the paper consists of the proofs and numerical evaluations of these regions, so that their performance can be visually compared.



Under the EXP power constraint and $M=2$, outer bounds on the error exponent region for the two-way AWGN channel have been presented in \cite{PalacioDevroyeISIT2018}, which follow directly from using the one-way channel outer bound presented in \cite{Kim:ActiveNoisyFB} for each direction. Tighter outer bounds for $M\geq2$ have been left for future work.

\section{Achievability scheme for the One-Way AWGN channel for $M\geq3$ under AS power constraint and \textbf{active feedback}: Proof of Theorem \ref{th:OneWayActiveASM3}}
\label{sec:ActiveM3OneWay}


In this section, the proof of Theorem \ref{th:OneWayActiveASM3}, Equation  \eqref{eq:ASactiveM}  follows by  generalizing the geometric technique used for $M=3$ messages introduced by Xiang and Kim \cite{Kim:ThreeCWPeak2013} for passive feedback and modified towards the use of encoded feedback. 
We have reused most of their notation in order to highlight the differences when active feedback is used. 
We first derive the achievable error exponent expression for the transmission of $M=3$ messages, then generalize it to arbitrary but finite $M$.

Some of the results presented here rely on the use of a simplex code denoted by $\mathcal{C}\left( \Omega_M, E_\omega \right)$ for the transmission of $M$ symbols $ \omega \in \Omega := \{ \omega_1, \omega_2, \omega_3, ..., \omega_M \}$ from the $i$-th transmitter, using symbol energy $E_\omega$ and leading to codewords $X^{j}_i(\omega)$ of length $j$. Equation \eqref{eq:simplexCode} shows the definition of the simplex code $\mathcal{C}\left( \Omega_3, E_\omega \right)$.
\begin{equation} \label{eq:simplexCode}
		X^{j}_i(\omega) = 		
		\begin{cases*}
      \sqrt{E_\omega}\cdot (0,1,0,...,0),															  & if   $\omega = \omega_1$ \\
			\sqrt{E_\omega}\cdot (\frac{-\sqrt{3}}{2},-\frac{1}{2},0,...,0),	& if $\omega = \omega_2$ \\
			\sqrt{E_\omega}\cdot (\frac{+\sqrt{3}}{2},-\frac{1}{2},0,...,0),	& if $\omega = \omega_3$
    \end{cases*}.
\end{equation} 

\subsection{Achievable error exponents for the transmission of three messages under the AS power constraint and active feedback}
\label{sec:ActiveM3geom}

Consider transmitting one of three equally likely messages from ${\cal W} = \{1,2,3 \}$ over  the one-way AWGN channel with active feedback of Figure \ref{fig:OneWayGaussianActive}  and assume the feedback link is strictly better than the forward channel, specifically: $\frac{P_{\text{FB}}}{\sigma^2_{\text{FB}}} > \frac{P}{\sigma^2}$.  
The active feedback achievability scheme block diagram under the AS power constraint is shown in Figure  \ref{fig:BLKDGMcombinedOWay} (right).   Transmission of $W$ is performed in $n$ channel uses through three stages: transmission, active feedback and retransmission, and corresponds to a modified version of the scheme proposed in \cite{Kim:ThreeCWPeak2013} that uses active feedback instead of passive. The three stages are analyzed next. 

\textbf{1. Transmission}: The first stage occurs during $  \lambda_1 n$ channel uses, where message $W=w$ is transmitted without feedback as codeword $X^{\lambda_1 n}(w) $ using the simplex code $\mathcal{C}\left({\cal W}, \lambda_1 n P \right)$  defined in \eqref{eq:simplexCode}.  
At the receiver, the signal $y^{\lambda_1 n}$ is decoded using protection regions $B_w$,  
one for each transmitted codeword as shown in Figure \ref{fig:ProtRegionB} and defined in \eqref{eq:ProtRegsBw}, as in \cite{Kim:ThreeCWPeak2013}.  All received signals inside region $B_{w}$ are immediately declared as the codeword $w$ (and no feedback is necessary). 
These regions are parametrized by $s\in[0,1]$ following \cite[Equation 6]{Kim:ThreeCWPeak2013} for a parameter $t \in [0, \frac{\sqrt{3} - 1}{2}]$ which is geometrically coupled to $s$.  Further reading regarding this idea can be found in \cite{Kim:ThreeCWPeak2013}.	
\begin{align} \label{eq:ProtRegsBw}
 B_w &= \left\{ y^{\lambda_1 n}: || x^{\lambda_1 n}(w) - y^{\lambda_1 n}|| \leq || x^{\lambda_1 n}(w') - y^{\lambda_1 n}||   \text{ for } w'\neq w,  \right. \nonumber \\ 
&\left. \text{ and } \left| || x^{\lambda_1 n}(w') - y^{\lambda_1 n}|| -  ||x^{\lambda_1 n}(w'') - y^{\lambda_1 n}|| \right| \leq td',  \text{ for } w',w''\neq w  \right\}.
\end{align}

\begin{figure}[H]
	\centering
	\includegraphics[width=0.4\textwidth]{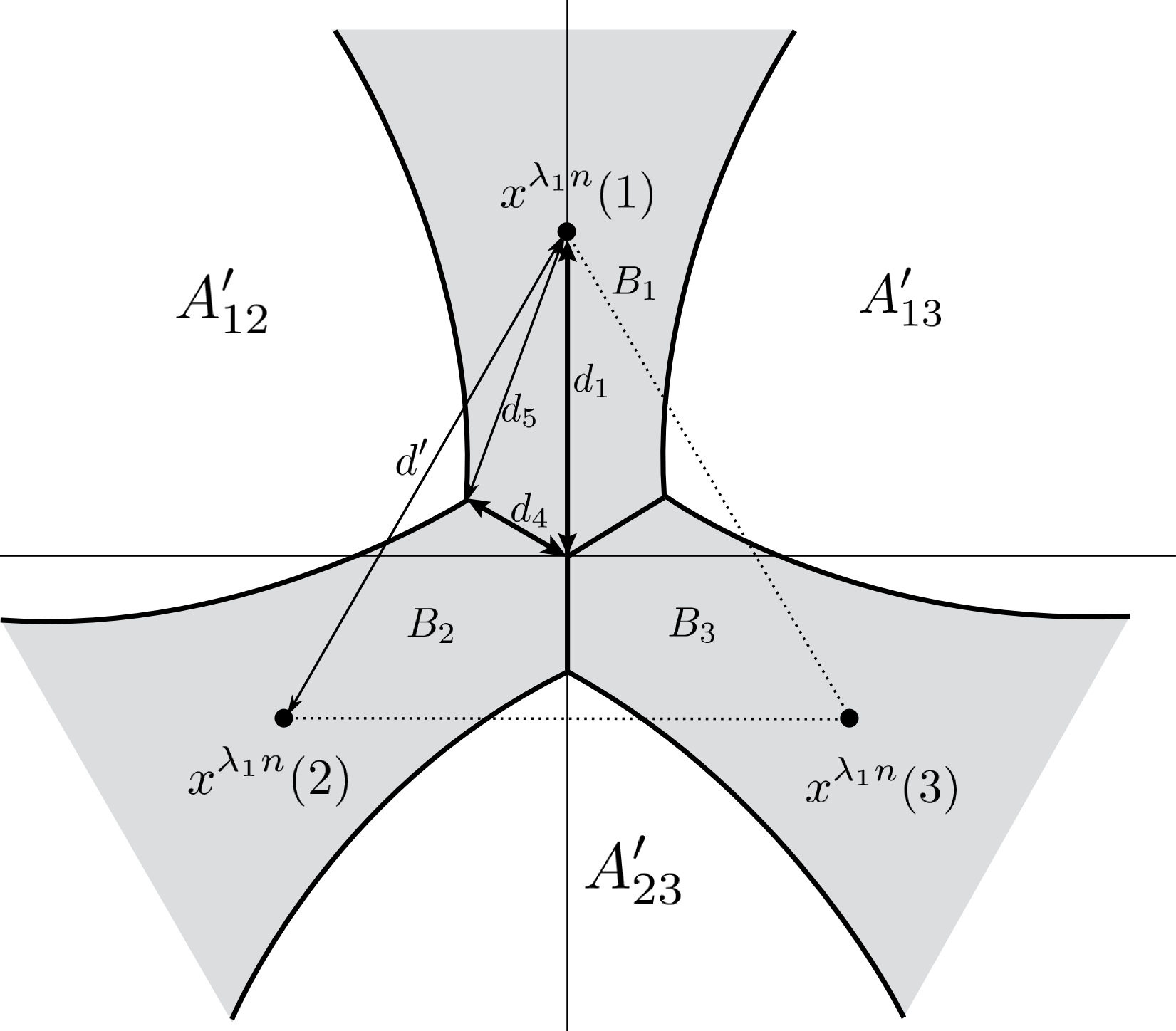}
	\caption{Protection regions defined in \cite{Kim:ThreeCWPeak2013} and used for decoding message $W$ after the transmission stage. Here, $d_1 = \sqrt{\lambda_1 n P}$, $d_4 = \frac{s}{2} d_1$, $d' = \sqrt{3 \lambda_1 n P}$, and $d_5 = \sqrt{\frac{\lambda_1 n P}{4\sigma^2}(s^2-2s+4)} $.}
	\label{fig:ProtRegionB}
\end{figure}
Figure  \ref{fig:ProtRegionB} also illustrates regions $A'_{12}, A'_{13}, A'_{23}$, that represent regions in which the receiver 
is undecided between two codewords. The definition of these regions can be found in \cite[Section II-B]{Kim:ThreeCWPeak2013}. Assuming $W=1$ is sent, the probability of error of this stage corresponds to the occurrence of event $\mathcal{E}_{T}$, defined in \eqref{eq:ET}, and representing a signal received in the wrong protection region, $B_2 \cup B_3$ or within an ambiguous region not including symbol $W=1$, i.e. $ y^{\lambda_1n} \in A'_{23}$:
\begin{equation} \label{eq:ET}
		\mathcal{E}_{T} = \{  y^{\lambda_1n} \in B_1 \cup B_2 \cap A'_{23}  \}.
\end{equation}
The achievable error exponent of this stage is derived from $\mathsf{P}({\mathcal{E}_T})$ as in 
 \cite[Equation 7]{Kim:ThreeCWPeak2013}:
\begin{equation} \label{eq:ErrExpT}
\lim_{n\rightarrow\infty} -\frac{1}{n}\log \left(\mathsf{P}({\cal E}_T)\right)	:=	E_{\mathcal{E}_T} \geq  \frac{\lambda_1 P}{8\sigma^2}(s^2-2s+4).
\end{equation}



	\textbf{2. Active feedback}:	
If the received signal is not within a protection region, we progress to the second stage, where
the receiver determines which two message / codeword form the most likely codeword pair $q= \{ \hat{w}_{1}, \hat{w}_{2} \}$, determined using minimum distance decoding based on $y^{\lambda_1n}$ as:
\begin{equation}
q = 
\begin{cases*}
\{1,2\},   & if $y^{\lambda_1 n} \in A'_{12} $ \\
\{2,3\},   & if $y^{\lambda_1 n} \in A'_{23} $ \\
\{1,3\},   & if $y^{\lambda_1 n} \in A'_{13} $ 
\end{cases*}.
\end{equation}	
Let ${\cal Q}$ be the set of $q$ values defined above.  The most likely unordered pair (labeled lexicographically)
is encoded and sent to the transmitter as $U^{\lambda_2 n}(q) $ using the simplex code $\mathcal{C}\left({\cal Q}, n P_{\text{FB}} \right)$ over the feedback link in $\lambda_2 n$ channel uses (active feedback). The transmitter estimates $q$ as $\hat{q}=\{\tilde{w}_{1}, \tilde{w}_{2} \}$, by means of the decoding regions of Figure \ref{fig:PairDecoding}, defined in \eqref{eq:PairRegionsB}:
\begin{align}   \label{eq:PairRegionsB}
\mathcal{R}_{w,w'} &= \left\{ z^{\lambda_2 n}: || u^{\lambda_2 n}(q=\{w,w'\}) - z^{\lambda_2 n}|| \right. \left. \leq || u^{\lambda_2 n}(q=\{w,w''\}) - z^{\lambda_2 n}||  \right. \text{ and, }\\ \nonumber
&\left. || u^{\lambda_2 n}(q=\{w,w'\}) - z^{\lambda_2 n}|| \right. \left. \leq|| u^{\lambda_2 n}(q=\{w',w''\}) - z^{\lambda_2 n}||  \right\}. \nonumber
\end{align}
\begin{figure}[H]
	\centering
	\includegraphics[width=0.35\textwidth]{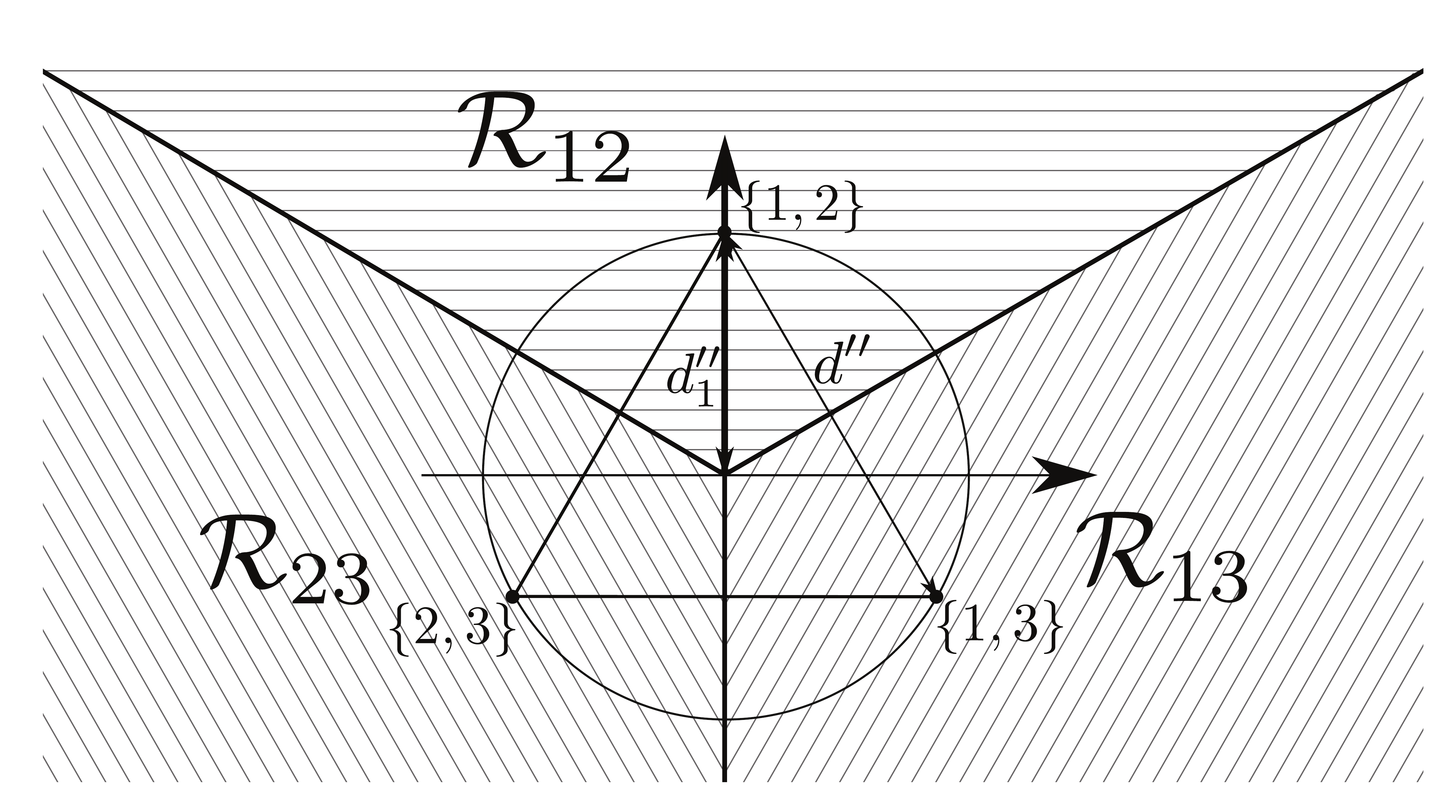}
	\caption{Decoding regions for active feedback. Here, $d''= \sqrt{3 n P_{\text{FB}}}$ and $d''_1 = \sqrt{nP_{\text{FB}}}$. }
	\label{fig:PairDecoding}
\end{figure}
At the end of the feedback stage an error occurs if the most likely pair determined at the receiver $q$ is different from that decoded at the transmitter $\hat{q}$. Recalling that $W=1$ was sent, the active feedback leads to an error if the transmission stage led to a signal received in either $A'_{12} $ or $ A'_{13}$, and one of the following two events occur:
\begin{align} 
\mathcal{E}_{\text{AFB}_1} &= \left\{ \left( Z^{\lambda_2 n} \in  \mathcal{R}_{13} \cup \mathcal{R}_{23} \right) \cap \left( Y^{\lambda_1 n} \in  A'_{12} \right)  \right\} \label{eq:EAFB001}  \\
\mathcal{E}_{\text{AFB}_2} &= \left\{ \left(Z^{\lambda_2 n} \in  \mathcal{R}_{12} \cup \mathcal{R}_{23}  \right) \cap \left( Y^{\lambda_1 n} \in  A'_{13}  \right) \right\}. \label{eq:EAFB002}
\end{align}
{In the events above, note that although $\hat{q}$ may contain the true codeword (i.e. if $W=2$, $q=\{1,2\}$ and $\hat{q}=\{2,3\}$), if $\hat{q}$ does not match $q$ exactly, it is counted as an error.  We do this as in the upcoming retransmission stage, antipodal signaling is used to correct potential decoding errors: a negative symbol indicates the true message is the largest (number)  in the pair, a positive symbol the smallest (number) in the pair.  As such, while in some cases an erroneous $\hat{q}$ could still lead to the correct re-transmission, we choose to simply count them all as errors and require $\hat{q}=q$.  
Let $\mathsf{P}(\mathcal{E}_{\text{AFB}})$ denote the probability that either of the two events happen. Then $\mathsf{P}(\mathcal{E}_{\text{AFB}}) \leq \mathsf{P}(\mathcal{E}_{\text{AFB}_1})+ \mathsf{P}(\mathcal{E}_{\text{AFB}_2}) = 2 \mathsf{P}(\mathcal{E}_{\text{AFB}_1})$, and the achievable error exponent results from \eqref{Eq:NoFBErrExp} and the use of a simplex code:
\begin{equation} \label{eq:ErrExpAFB}
\lim_{n\rightarrow\infty} -\frac{1}{n}\log \left(\mathsf{P}({\cal E}_{\text{AFB}})\right) :=  		E_{\mathcal{E}_{\text{AFB}}} \geq \frac{3}{8}  \frac{P_{\text{FB}}}{\sigma_{\text{FB}}^2}.
\end{equation}

	\textbf{3. Retransmission}: 
	The last stage lasts for the remaining $\lambda_3 n$ channel uses. At this point, transmitter and receiver have estimates of the two most likely codewords, $q=\{\hat{w}_{1}, \hat{w}_{2}\}$, and $\hat{q} = \{\tilde{w}_{1}, \tilde{w}_{2} \}$, which we assume are the same otherwise this is counted as an error in the feedback stage. Then, based on $\hat{q}$, the transmitter uses antipodal signaling to send the true codeword among these two candidates using its  remaining power:
\begin{align} \label{eq:AntipodalRTX}
&X^{\lambda_3 n} = 
\begin{cases*}
\sqrt{(1-\lambda_1) n P} \cdot (+1,0,...,0), &\text{if } $w = \min\{ \tilde{w}_{1}, \tilde{w}_{2} \}$\\ 
\sqrt{(1-\lambda_1) n P} \cdot (-1,0,...,0), &\text{if } $w = \max\{ \tilde{w}_{1}, \tilde{w}_{2} \}$\\
(0,0,...,0), &\text{otherwise}.
\end{cases*}
\end{align}	
Note that the transmission of this signal employs one channel use only,  and that parameters $\lambda_1$, $\lambda_2$ and $\lambda_3$ are used for proper power allocation. 
Message decoding follows a similar approach as that proposed in \cite{Kim:ThreeCWPeak2013}, where the receiver decodes message $W$ immediately at the end of the \textit{transmission stage} if $y^{\lambda_1 n}$ is received within a protection region and ignores all other signals received in other channel uses. 
If $y^{\lambda_1 n}$ is not in any $B_w$, the decoder determines the most likely codeword pair based on minimum distance decoding and sends it to the transmitter using active feedback. Once the second stage is finished, the receiver waits for the retransmission signal, and uses it along with the first stage transmission to determine $\hat{W}$ using the decoding rule in \eqref{eq:DecRuleS}, as with passive feedback \cite{Kim:ThreeCWPeak2013}: 
\begin{align} \label{eq:DecRuleS}
	\hat{w} &= \text{arg} \min_{w \in \{ \hat{w}_{1}, \hat{w}_{2} \}}{ || x^{n}(w) - y^{n} ||} \\ \notag
	&= \text{arg} \min_{w \in \{ \hat{w}_{1}, \hat{w}_{2} \}} { \left( || x^{\lambda_1 n}(w) - y^{\lambda_1 n} ||^2 + || x^{\lambda_3 n}(w) - y^{\lambda_3 n} ||^2  \right)^{\frac{1}{2}}	}.
	\end{align}
	The decoder errs  
	if given that $y^{\lambda_1 n}$ is outside the protection regions and the feedback stage led to $q=\hat{q}$, the following event occurs: 
	\begin{align} \label{eq:ERTp}
	\mathcal{E}_{\text{RT}} &= \left\{\left( w  \in q = \{\hat{w}_{1_1},\hat{w}_{1_2} \}\right) \cap \left( w \in \hat{q} = \{\tilde{w}_{1_1},\tilde{w}_{1_2} \} \right)\cap \left(\hat{w} \neq w \right) \right\}.
	\end{align}
	Then, the error exponent \eqref{eq:ErrExpRT} can be derived in a similar way
	as in  \cite[$\mathsf{P}(\mathcal{E}_2)$ in Section II-A]{Kim:ThreeCWPeak2013} yielding:
		\begin{equation} \label{eq:ErrExpRT}
	\lim_{n\rightarrow\infty} -\frac{1}{n}\log \left(\mathsf{P}({\cal E}_{\text{RT}})\right)	:= E_{\mathcal{E}_{\text{RT}}} \geq \left( 1 - \frac{\lambda_1}{4}\right) \frac{P}{2\sigma^2}.
	\end{equation}
	
	The overall error exponent expression is derived from \eqref{eq:ErrExpT}, \eqref{eq:ErrExpAFB} and \eqref{eq:ErrExpRT}, as the minimum of the three stages' 
	\begin{align} \label{eq:ErrExpM3ActiveDef3}
	&E_{12}\left(M=3,\frac{P}{\sigma^2}, \frac{P_{\text{FB}}}{\sigma_{\text{FB}}^2} ,s \right)  \geq \min{ \left\{ \frac{\lambda_1 P}{8\sigma^2}(s^2-2s+4) ,  \frac{3}{8}  \frac{P_{\text{FB}}}{\sigma_{\text{FB}}^2} ,\left( 1 - \frac{\lambda_1}{4}\right) \frac{P}{2\sigma^2} \right\} }.
	\end{align}
Equating the first and third terms, we obtain $\lambda_1$ as: $\lambda_1 = \frac{4}{s^2-2s+5}$. This reduces the number of arguments of \eqref{eq:ErrExpM3ActiveDef3} to two, which leads to \eqref{eq:ASactiveM} evaluated at $M=3$.
	
\medskip
In the above, we have sketched the overall scheme and left out the details of the probability of error analysis. We present this next in greater detail for $M=3$, with the goal of making the extension to arbitrary $M$ more transparent. \\ 
		
\textbf{Probability of error analysis:}
The error exponent expression of Equation \eqref{eq:ErrExpM3ActiveDef3} corresponds to the minimum of the  contribution to the probability of error of three events.  
By symmetry and without loss of generality, we assume that message $W=1$ is transmitted, and denote its transmission probability of error as  $ \prob(\text{error} \mid W =1) =  \prob_{1}(W \neq \hat{W})$. Note that $\prob_{1}(W \neq \hat{W}) = \prob_{2}(W \neq \hat{W}) = P_{3}(W \neq \hat{W})$, and hence, since all three messages are equally likely, the overall probability of error is
$\prob(\text{error}) = \frac{1}{3} \sum_{i=1}^3 \prob_{i}(W \neq \hat{W})  = \prob_{1}(W \neq \hat{W})$. 
\begin{align} \label{eq:PerrorM3Active}
P_1(W \neq \hat{W}) &= \mathsf{P}(\text{Err.Tx}) \\ \notag
&+ \mathsf{P}\left(\text{Err.FB} \mid \overbrace{Y^{\lambda_1n} \in \{ A_{12}' \cup A_{13}' \} }^{\text{Not in Prot.Reg.}}  \right) \cdot \mathsf{P}(Y^{\lambda_1n} \in \{ A_{12}' \cup A_{13}' \})	 \notag\\
		&+ \mathsf{P}\left(\text{Err.Rtx} \mid  \overbrace{ Y^{\lambda_1n} \in \{ A_{12}' \cup A_{13}' \} }^{\text{Not in Prot.Reg.}}  \cap  \overbrace{\left\{ W \in q \right\} \cap \left\{ W \in \hat{q} \right\}}^{\text{No mis-coord. error}} \right)  \notag\\
		& \cdot \mathsf{P}\left(   \left\{Y_2^{\lambda_1n} \in \{ A_{12}' \cup A_{13}' \} \right\}  \cap  \left\{ W \in q \right\} \cap \left\{ W \in \hat{q} \right\} \right),  \notag
\end{align}
where, Err.Tx, Err.FB and Err. Rtx correspond to error events for each corresponding stage described as:
\begin{itemize}
	\item Error event in transmission stage: $\text{Err.Tx} :=Y^{\lambda_1n} \in \{ B_2 \cup B_3 \cup A'_{23} \}$.
	\item Error events in FB stage given that $Y^{\lambda_1n}$ was not received within a protection region:
	\begin{align}
	  \text{Err.FB} &:= \left\{ \left( Z^{\lambda_2n} \in \mathcal{R}_{12}\cup \mathcal{R}_{23} \right)\cap \left(Y^{\lambda_1n} \in A'_{13} \right) \right\} \cup \left\{\left(Z^{\lambda_2n} \in \mathcal{R}_{13} \cup \mathcal{R}_{23}  \right)\cap \left( Y^{\lambda_1n} \in A'_{12}\right)\right\}. \notag
	\end{align}
	\item Error event in retransmission stage: $\text{Err.Rtx} := \{ w \neq \hat{w}\}$.
\end{itemize}
Then, \eqref{eq:PerrorM3Active} can be expressed as:  
\begin{align} \notag
\prob_1(W \neq \hat{W}) &= \mathsf{P}(Y^{\lambda_1n} \in \{ B_2 \cup B_3 \cup A'_{23} \} ) \\ \notag
		&+ \mathsf{P}\left( \left\{ \left(Z^{\lambda_2n} \in \mathcal{R}_{12} \cup \mathcal{R}_{23} \right)\cap \left(Y^{\lambda_1n} \in A'_{13} \right) \right\} \cup \left\{ \left(Z^{\lambda_2n} \in \mathcal{R}_{13} \cup \mathcal{R}_{23} \right)\cap \left( Y^{\lambda_1n} \in A'_{12} \right) \right\} \right) \\ \notag
		&\cdot \mathsf{P}\left(Y^{\lambda_1n} \in \{A_{12}' \cup A_{13}' \} \right)	\\ \notag
		&+ \mathsf{P}\left(\{ w \neq \hat{w}\} \cap \left\{ Y^{\lambda_1n} \in \{ A'_{12} \cup A'_{13} \cup A'_{23}\} \right\} \cap \{ W \in q \} \cap \{ W \in \hat{q} \} \right) \notag \\
 &\leq \mathsf{P}(Y^{\lambda_1n} \in \{ B_2 \cup B_3 \cup A'_{23} \} ) \\ \notag
		&+\mathsf{P}\left( \left\{ \left(Z^{\lambda_2n} \in \mathcal{R}_{12} \cap \mathcal{R}_{23} \right)\cap \left(Y^{\lambda_1n} \in A'_{13} \right) \right\} \right) \cdot \mathsf{P}\left( Y^{\lambda_1n} \in \{A_{12}' \cup A_{13}' \}  \right)	\\ \notag
		&+\mathsf{P}\left( \left\{ \left( Z^{\lambda_2n} \in \mathcal{R}_{13} \cap \mathcal{R}_{23} \right)\cap \left( Y^{\lambda_1n} \in A'_{12} \right)\right\}  \right) \cdot \mathsf{P}\left( Y^{\lambda_1n} \in \{A'_{12} \cup A'_{13} \} \right) \\ \notag
		&+P \left(\{ w \neq \hat{w}\} \cap \left\{ Y^{\lambda_1n} \in \{ A'_{12} \cup A'_{13} \} \right\}  \cap \{ W \in q \} \cap \{ W \in \hat{q} \} \right) \notag \\
 &\leq \mathsf{P}(Y^{\lambda_1n} \in \{ B_2 \cup B_3 \cup A'_{23} \} ) +\mathsf{P}\left( \left\{Z^{\lambda_2n} \in \mathcal{R}_{12} \cap \mathcal{R}_{23}  \right\} \cap \left\{ Y^{\lambda_1n} \in A'_{13} \right\} \right) \notag   \\ 
		&+\mathsf{P}\left( \left\{ Z^{\lambda_2n} \in \mathcal{R}_{13} \cap \mathcal{R}_{23} \right\} \cap \left\{ Y^{\lambda_1n} \in A'_{12}\right\}  \right) +\mathsf{P}\left(\{ w \neq \hat{w}\} \cap \{ W \in q \} \cap \{ W \in \hat{q} \} \right)  \label{eq:EventSum}\\
		&= \prob(\mathcal{E}_{T}) +  \prob(\mathcal{E}_{AFB}) + \prob(\mathcal{E}_{RT}) \label{eq:PEventSUm}
\end{align}
Observe that the terms in the sum of Equation \eqref{eq:EventSum} correspond to the error events defined for each stage: 
the first term is $\mathcal{E}_{T}$ from Equation \eqref{eq:ET}, the second and third terms correspond to $\mathcal{E}_{AFB}$ \eqref{eq:EAFB001} and \eqref{eq:EAFB002}, and the fourth is related to $\mathcal{E}_{RT}$, Equation \eqref{eq:ERTp}. Next, we show that the summation in \eqref{eq:PEventSUm} leads to the error exponent of Equation \eqref{eq:ErrExpM3ActiveDef3}, and how the probabilities of these events can be upper bounded in a way that can be generalized to larger $M$. 

\textbf{Probability of error in the transmission stage:}
Probability of error event $	\mathcal{E}_T= \{ y^{\lambda_1n} \in \{ B_2 \cup B_3 \cup A'_{23} \}\} $, can be upper bounded similarly as in \cite[see Figure \ref{fig:ProtRegionB}]{Kim:ThreeCWPeak2013}, yielding error exponent \eqref{eq:ErrExpT} as\footnote{Where $ Q(x) = \frac{1}{\sqrt{2\pi}} \int_{x}^{\infty}  \exp\left( -\frac{y^2}{2} \right) dy $.}:
	\begin{align}	\label{eq:PET}
			\mathsf{P}(\mathcal{E}_{T})  &\leq 2Q\left(\frac{d_5}{\sigma}\right) \\ \notag
						&\leq \exp{\left(\ -n \frac{\lambda_1 P}{8\sigma^2}(s^2-2s+4) \right)},
	\end{align}

\textbf{Probability of error in the active feedback stage:}
If the transmission of $W=1$ is received in regions $A_{12} \cup A_{13}$, the feedback transmission may cause a mis-coordination error $(q \neq \hat{q})$, whenever the events  in Equations \eqref{eq:EAFB001} and \eqref{eq:EAFB002} occur.
Since 
$			\mathsf{P}(\mathcal{E}_{\text{AFB}}) \leq \mathsf{P}(\mathcal{E}_{\text{AFB}_1})+ \mathsf{P}(\mathcal{E}_{\text{AFB}_2}) = 2 \mathsf{P}(\mathcal{E}_{\text{AFB}_1})$, we can upper bound the $\mathsf{P}(\mathcal{E}_{\text{AFB}_1})$ probability to obtain the error exponent expression of Equation \eqref{eq:ErrExpAFB} (active feedback stage) as:
\begin{align} \label{eq:PAFB}
		\mathsf{P}(\mathcal{E}_{\text{AFB}_1}) &= \mathsf{P}\left( Z^{\lambda_2 n} \in  \mathcal{R}_{13} \cup \mathcal{R}_{23} \mid Y^{\lambda_1 n} \in  A'_{12} \right) \cdot \mathsf{P}(Y^{\lambda_1 n} \in  A'_{12})\\ \notag
					&\leq \mathsf{P}\left( Z^{\lambda_2 n} \in  \mathcal{R}_{13} \cup \mathcal{R}_{23} \mid Y^{\lambda_1 n} \in  A'_{12} \right) \\ \notag
					&\leq 2Q\left(\frac{d''}{2\sigma_{\text{FB}}} \right) = 2Q\left( \frac{\sqrt{3  n P_{\text{FB}} } }{2\sigma_1} \right) \\ \notag
					&\leq \exp{\left( - n \frac{ 3 P_{\text{FB}} }{8\sigma_{\text{FB}}^2}  \right)}.
\end{align}

\textbf{Probability of error for the retransmission stage:}
	The probability of error of this stage is linked to the occurrence of the event $\mathcal{E}_{\text{RT}}$, and shown in \eqref{eq:ERT}.
The errors produced in the previous stages are captured by the corresponding error events defined above. Therefore, this transmission assumes that the two previous stages are correct, which is equivalent to the noiseless passive feedback case analyzed in \cite{Kim:ThreeCWPeak2013}, in which transmitter and receiver agree on the most likely pair of codewords (for the active feedback setting of this scheme, this is equivalent to $q=\hat{q}$). Thus, the probability of error leading to the error exponent of \eqref{eq:ErrExpRT} (via a Chernoff upper-bound) is derived in a similar manner and given by: 
\begin{align}  \label{eq:ERT}
		\mathsf{P}(\mathcal{E}_{\text{RT}}) &= Q\left( \sqrt{\left( 1 - \frac{\lambda_1}{4}\right) \frac{P}{\sigma^2} n  }  \right).
\end{align}

It follows that Equation \eqref{eq:ErrExpM3ActiveDef3} results from plugging in Equations \eqref{eq:PET}, \eqref{eq:PAFB} and \eqref{eq:ERT} into \eqref{eq:PEventSUm} and noting that the error exponent is dominated by the minimum exponential decay of the three terms. \qed


\subsection{Generalization to the transmission of $M$ messages}
In \cite{Kim:ThreeCWPeak2013}, the authors present a geometric approach that allows generalizing the result they found for $M=3$ to any arbitrary $M$. Here, we utilize a similar generalization for the result we presented for active feedback. In this approach, all distances utilized for three messages are generalized to any $M$ using the relation $d_j^{(3)} = d_j^{(M)} \sqrt{3(M-1)/(2M)}$. Since the first and third stages follow in the same way as that for passive feedback, we use the results in \cite[Appendix]{Kim:ThreeCWPeak2013}. For the transmission stage:
\begin{align} \label{eq:EPET}
	\prob(\mathcal{E}_{T}) &\leq M^2 Q\left(d^{(M)}_5\right)\\
												 &\leq \frac{M^2}{2} \exp\left( -n \frac{M\lambda_1}{12(M-1)} \frac{P}{\sigma^2} (s^2-2s+4)  \right),
\end{align}
and for the retransmission stage:
\begin{align} \label{eq:EPERT}
	\prob(\mathcal{E}_{RT}) &= Q\left( -\sqrt{ \left( 1- \lambda_1 \frac{M-2}{2(M-1)}   \right) \frac{P_{\text{FB}}}{\sigma^2_{\text{FB}}}  } \right)\\
													&\leq \frac{1}{2} \exp\left[ -n \frac{P}{2\sigma^2}  \left( 1- \lambda_1 \frac{M-2}{2(M-1)} \right)\right].	
\end{align}


In the feedback stage, the most likely codeword pair $q$ is returned to the transmitter.  
Since the antipodal signaling of the retransmission stage sends a positive signal to indicate the true message is the smallest of the pair, and a negative signal to indicate the true message is the largest. The order of the elements in each pair does not matter as long as the pair is correctly identified. As such, since there are ${M \choose 2}$ such unordered pairs the size of the simplex code used in the active feedback stage is ${M \choose 2}$, and the probability of error of this stage can be upper bounded as in \eqref{Eq:NoFBErrExp}:
\begin{align} \label{eq:EPEAFB}
	\prob(\mathcal{E}_{AFB}) &\leq (M-1) \cdot Q\left( \sqrt{ n \frac{P_{\text{FB}}}{\sigma^2_{\text{FB}}}  \frac{{M \choose 2}}{2\left({M \choose 2}-1\right)}}   \right)\\
												   &\leq \frac{M-1}{2} \exp\left( - n  \frac{P_{\text{FB}}}{\sigma^2_{\text{FB}}} \frac{{M \choose 2}}{4\left({M \choose 2}-1\right)} \right).
\end{align}



Finally, the overall error exponent is:
\begin{align}
	E_{12}^{AS} &= \limsup_{n\to \infty} - \frac{1}{n} \ln \prob(\text{error})\\
							&\geq \limsup_{n\to \infty} - \frac{1}{n} \max \left\{  \ln \prob(\mathcal{E}_T), \ln \prob(\mathcal{E}_{AFB}) ,\ln \prob(\mathcal{E}_{RT})   \right\} \label{eq:ErrExpASM}\\
							&\geq \min \left\{ \lambda_1 \frac{MP}{12\sigma^2(M-1)} (s^2-2s+4),  \frac{P_{\text{FB}}}{\sigma^2_{\text{FB}}} \frac{{M \choose 2}}{4\left({M \choose 2}-1\right)} , \frac{P}{2\sigma^2}  \left( 1- \lambda_1 \frac{M-2}{2(M-1)} \right)  \right\} \label{eq:ThreeTermsAS_M}.
\end{align}
By equating the first and third arguments, $\lambda_1 = \frac{6(M-1)}{M(s^2-2s+4)+3(M-2)} $, thus yielding \eqref{eq:ASactiveM}. \qed

\bigskip
\section{Achievable error exponents for the One-Way AWGN channel under the expected block power constraint and the transmission of $M$ messages: Proof of Theorem \ref{th:OneWayEXPM2}.}
\label{sec:ProofOneWayEXPM}

This section presents the proof of Theorem \ref{th:OneWayEXPM2}, as an extension of the achievability result derived by Kim, Lapidoth and Weissman in \cite{Kim:ActiveNoisyFB} under the EXP power constraint for the transmission of two messages for the channel of Figure \ref{eq:OneWayGEnModelY}. We show how this technique can be extended to any finite number of messages $M$. The case of $M=3$ is first introduced using a geometric approach that can be easily extended to larger $M$. Our generalization is based on a modification of the  building block (BB) originally designed for the transmission of two messages and proposed in \cite{Kim:ActiveNoisyFB}. The building block for general $M$ uses a different retransmission encoding function, that generates retransmission codewords of length $M$, in contrast to that of length one employed for the transmission of two messages. 

Next, we summarize the operation of the original building block, which comprises three stages and achieves an error exponent of $\frac{2P}{\sigma^2}$ for a block length $n$ (if used in the forward direction, or $\frac{2P_{\text{FB}}}{\sigma_{\text{FB}}^2}$ for messages in the backwards direction). 
The  first $n-1$ channel uses are used to transmit message $w \in \{0,1\}$ using the following signaling:
\begin{equation} \label{eq:CodeM2}
		X_{k}(w) = 		
		\begin{cases*}
      \sqrt{P},						& if $w = 0$ \\
			-\sqrt{P},					& if $w = 1$ \\
    \end{cases*}.
\end{equation}
At the end of this transmission the receiver computes $ S = \sum_{i=1}^{n-1} Y_i $ and declares the transmission to be either non-valid (NACK) if $\left| \frac{1}{n-1} \cdot S \right| \leq (1-\delta) \sqrt{P}$, or otherwise, valid (ACK). The distribution of $\frac{S}{n-1}$ is
  $  \mathcal{N}(\sqrt{P}, \frac{\sigma^2}{n-1} )$, if $w = 0$, and $\mathcal{N}(-\sqrt{P}, \frac{\sigma^2}{n-1} )$, if $w = 1$, as illustrated in Figure \ref{fig:OldNackBandsM2}.
\begin{figure}[H]
	\centering
		\includegraphics[width=0.5\textwidth]{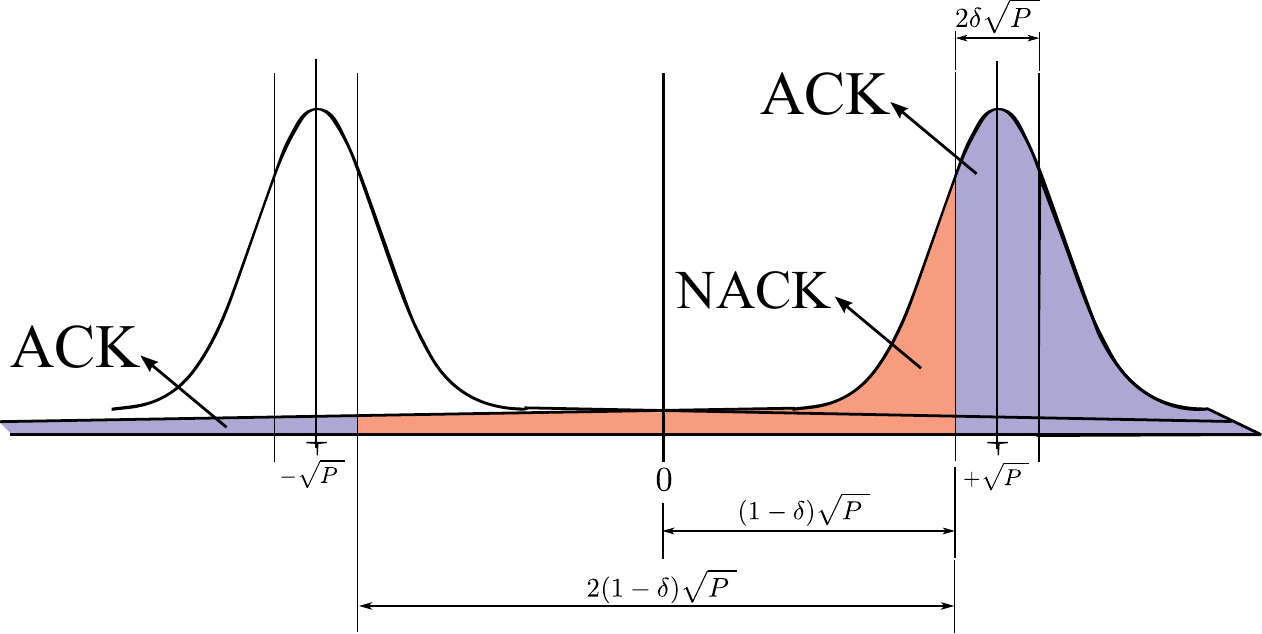}
		\caption{Distribution of random variable $\frac{S}{n-1} $.  The NACK region in shown in red, whereas the ACK regions are shown in light blue. }
	\label{fig:OldNackBandsM2}
\end{figure}
We term the region in which a NACK is declared as the ``NACK band'' (to be re-used again later, shown in Figure \ref{fig:OldNackBandsM2}). 
In the feedback stage, the ACK/NACK decision is sent to the transmitter using a single zero or a single very high amplitude signal respectively.  Since the NACK event occurs with exponentially small probability (as demonstrated in \cite[Equation (126)]{Kim:ActiveNoisyFB}), this transmission satisfies the EXP power constraint. The transmitter correspondingly decodes these transmissions as ``ACK''/``NACK''. To this end, it compares the single channel use received signal with a very large threshold, which achieves an asymptotically infinite error exponent (see \cite[ Equations (128a), (128b)]{Kim:ActiveNoisyFB}). {The retransmission stage depends on feedback stage result. Thus,} if a ``NACK'' is declared at the transmitter, a very high  amplitude antipodal signaling is used to retransmit the original message, again achieving an infinite error exponent (see \cite[Equations (130a), (130b)]{Kim:ActiveNoisyFB}). The only source of error in this BB is caused by incorrect decoding when the ACK event is declared at the receiver.
This scheme yields error exponent $\frac{2P}{\sigma^2}$, as shown in \cite[Equation (138)]{Kim:ActiveNoisyFB}. 


The BB is used as part of a three stage scheme, the first two lasting for $\frac{n-1}{2}$ channel uses and the last one for a single channel use: in the first stage, the BB is used to transmit the original message $W$, which is decoded by the receiver as $W' = \hat{W}$. In the second stage, the BB is used to feed back $W'$, thus the transmitter can estimate the symbol received at the destination as $W''=\hat{W'}$, and compare it with the true message $W$. Finally, a single channel use retransmission stage is used to encode the true message in the polarity of a very high amplitude antipodal signaling when $W \neq W''$, or a silent transmission when the preliminary decision is correct ($W=W''$). The key feature of this scheme is the use of very reliable high amplitude transmissions occurring with exponentially small probability.

 Next, we present a modification of the BB for $M=3$ messages and explain how the modified version may be used in an analogous three stage communication scheme. 

\subsection{A communication building block for $M=3$ messages}
\label{sec:BuildingBlockM3}

Consider the AWGN channel with active feedback of Figure \ref{fig:OneWayGaussianActive} and  described by Equations \eqref{eq:OneWayGEnModelY} and \eqref{eq:OneWayGEnModelZ}, where both channel inputs are subject to EXP power constraints: $\mathsf{E}\left[ \sum^{n}_{k=1} X_{k}^2 \right] \leq n P$ and $\mathsf{E}\left[ \sum^{n}_{k=1} U_{k}^2 \right] \leq n P_{\text{FB}}$. The building block for the transmission of $M=3$ messages chosen equally likely from $\mathcal{W} \in \{1,2,3\}$ {and block length $n$}, consists of three stages: transmission, feedback and retransmission, which we present in the following section.

\subsubsection{Building block operation}
\label{sec:OperationDescription}
The first stage uses the simplex code $\mathcal{C}\left(\mathcal{W}, \sqrt{nP} \right)$ and lasts for $j=n-4$ channel uses. During this transmission the receiver remains silent. Figure \ref{fig:NackBands} shows the constellation resulting from encoded messages $x^j(w)$ as well as their corresponding decoding regions. {Similar to the BB for two messages, once the first stage is complete, the receiver uses the sequence $y^j$ to determine whether the received signal is a valid transmission (ACK), or non-valid (NACK). The latter occurs for signals received within the gray ``NACK bands'' regions shown below. The width of these bands is proportional to the distance between any two symbols and parametrized by $t \in (0,1)$}. Each region $A_i$ for $i \in \{1,2,3 \}$ corresponds to the decoding region for messages $ w_i \in \mathcal{W}$. 
\begin{figure}[htb]
	\centering
		\includegraphics[width=0.4\textwidth]{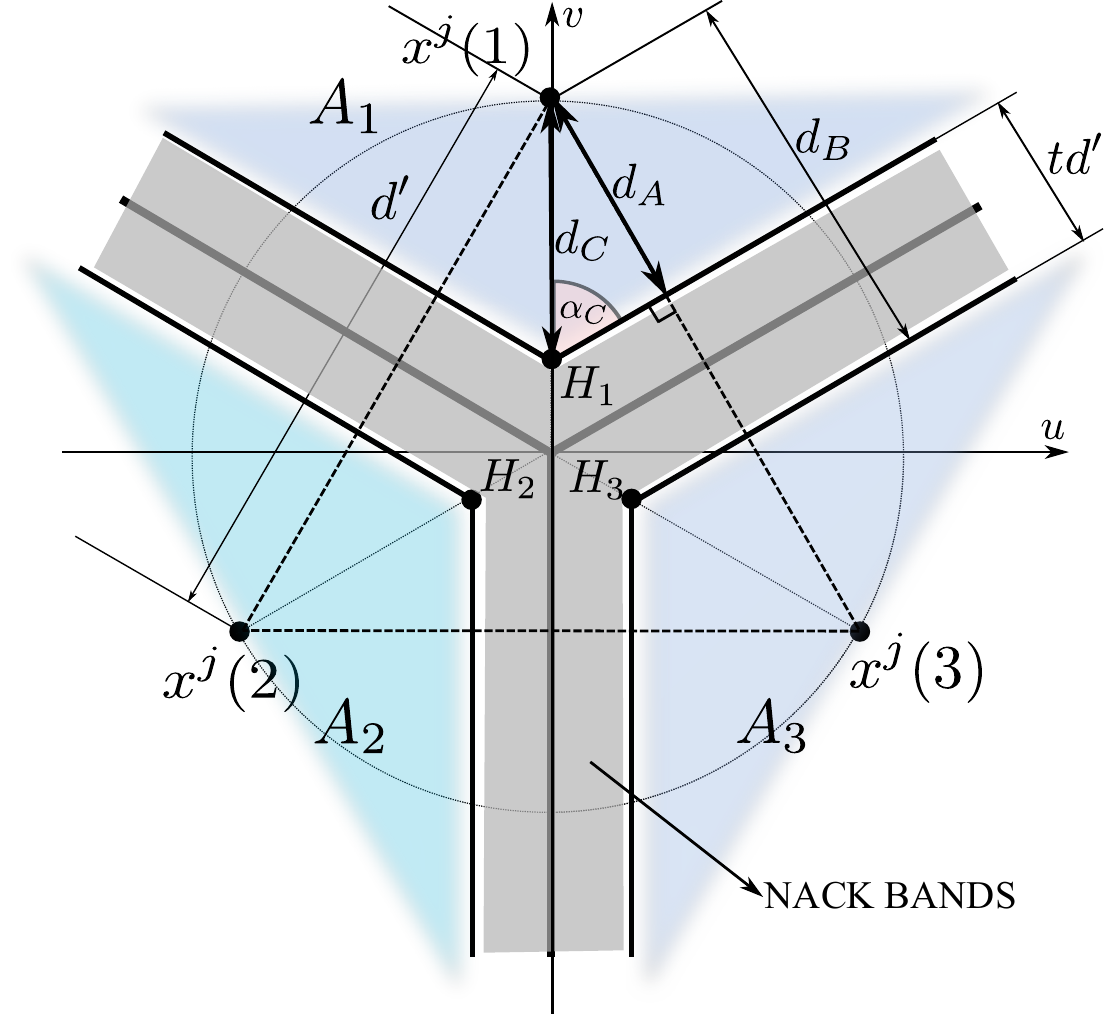}
		\caption{Nack-Bands Regions, here $d' = \sqrt{3nP}$.}
	\label{fig:NackBands}
\end{figure}
An ACK event is declared if: $ y^j \in \bigcup_{i=1}^3 A_i $, otherwise,  the receiver declares a NACK. In general, each ACK region $A_i$ corresponds to a space determined by $M-1$ hyperplanes defined in $M-1$ dimensions. {Specifically, ACK regions are determined by the hyperplanes between symbol $i$ and each of the other $M-1$ symbols, each plane being perpendicular to the line connecting them and located at a distance $d_A$ from symbol $i$. To illustrate this for $M=3$, consider the case of $A_1$:} there exists two hyperplanes (two lines), one between symbols $1-2$ and another between $1-3$. Each line can be represented using point $H_1:(H_{1_u},H_{1_v})$ (which belongs to both lines) and the vectors in the direction of the lines connecting connecting symbols $1-2$ and $1-3$ respectively.  Note that the coordinates of points $H_i$ can be determined using distance $d_C$, that depends on $d_A$ and the angle $\alpha_C$ as $d_C = \frac{d_A}{\sin{\alpha_C}}$ .
Next, we present the ACK region definition for $A_1$ only, since those for $A_2$ and $A_3$ follow in a similar way: 
\begin{align} \label{eq:ACKregionsM3EXP2}
	A_1 &= \left\{ y^j:=(y_1,y_2) : y_2 \geq \frac{ - \left( x^j_1(3)-x^j_1(1) \right)}{x^j_2(3)-x^j_2(1) } ( y_1 - H_{1_u} ) + H_{1_v} \right.\notag \\
	& \text{ and }  \left. y_2 \geq \frac{ - \left( x^j_1(2)-x^j_1(1) \right)}{x^j_2(2)-x^j_2(1) } ( y_1 - H_{1_u} ) + H_{1_v} \right\}.
\end{align}

Assume without loss of generality that $W=1$ was sent. The transmitter is informed of the receiver's decoding decision (ACK/NACK event declaration) using the single channel use feedback codeword in \eqref{eq:FBcode} that yields codeword $U_{j+1}$.  If a NACK occurs, a very high amplitude signal is transmitted since $\prob_1(\text{NACK})$ is exponentially small, otherwise a zero is transmitted. 
 \begin{equation} \label{eq:FBcode}
		U_{j+1} = 		
		\begin{cases*}
     			0,																	& if ACK \\
	   			\sqrt{\frac{\Delta}{ \prob_1(\text{NACK})} },	& if NACK
    \end{cases*}.
\end{equation}
By symmetry, note that $\prob_1(\text{NACK})=\prob_2(\text{NACK})=\prob_3(\text{NACK})$. To verify that NACK events are very rare, observe first that: $\prob_1(\text{NACK}) = 1 - \prob_1(\text{ACK})$. An upper bound on $\prob_1(\text{NACK})$ follows from lower bounding $\prob_1(\text{ACK})$ as 
\begin{align}
	\prob_1(\text{ACK}) &\geq  \prob_1( y^j \in A_1 ) \geq 1 - 2 Q \left(\frac{d_A}{\sigma}\right),
\end{align}
such that
\begin{align} \label{eq:PNACK2}
	\prob_1(\text{NACK}) 	&\leq 2 Q \left(\frac{d_A}{\sigma}\right) \leq \exp{ \left(- \frac{d_A^2}{2\sigma^2} \right)}.
\end{align}
The last inequality comes from applying the Chernoff upper bound on the $Q$-function, valid for $x>0 $, as 
$ Q(x) \leq \frac{1}{2} \exp{\left( -x^2/2 \right)}$.
Plugging in $d_A = \frac{1}{2} (d' - t d') $ and since $d' = \sqrt{3nP}$,  the probability of the NACK event is upper bounded as \eqref{eq:PNACK3}, which shows that a NACK event occurs with exponentially small probability:
 \begin{align} 
 \prob_1(\text{NACK}) &\leq \exp{ \left(- n \frac{3P}{8\sigma^2} (1-t)^2 \right)}. \label{eq:PNACK3} 
 \end{align}
 
Signal $u_{j+1}$ is sent through the noisy feedback channel and decoded by the transmitter by comparing the received $z_{j+1}$ with a large threshold $\Upsilon$. A ``NACK'' is declared if $ z_{j+1} > \Upsilon $, otherwise, an ``ACK'' is declared.  As in \cite{Kim:ActiveNoisyFB}, $\Upsilon$ is chosen to be $n$, and one can verify that with this choice,  $\prob_{w}(\text{``NACK''} \mid \text{ACK} )$ and $\prob _{w}(\text{``ACK''} \mid \text{NACK} )$ decay to zero in $n$ faster than any exponential, for all ${w} \in \{1,2,3 \}$, i.e.
\begin{align}
	-\lim_{n\to \infty} \frac{1}{n} \ln \prob_{w}(\text{``NACK''}\mid \text{ACK}) &= +\infty  \label{eq:NACKACK}\\
-\lim_{n\to \infty} \frac{1}{n} \ln \prob_{w}(\text{``ACK''}\mid \text{NACK}) &= +\infty \label{eq:ACKNACK}.
\end{align}

The retransmission stage transmits a length $M=3$ codeword that is generated depending on the true message and whether a ``NACK'' or an ``ACK'' is declared by the transmitter at the end of the feedback stage. This codeword is generated using the code $\mathcal{G}\left(\mathcal{W},  \sqrt{\frac{P}{\prob_w({\text{``NACK''}})} }\right)$ for the transmission of $|\mathcal{W}|+1 = M+1$ messages: $w'\in \{ \mathcal{W} \cup \{0\} \} $ yielding retransmission codewords $X^{M}(w')$, defined in Equation \eqref{eq:ReTransCODEM3} for $M=3$. The code consists of an all zeros codeword associated to the ``ACK'' event regardless of what true message $W$ is, and an index location based signaling in which the codeword is only non-zero in the location indexed by the true message $W = w \in \{1,..,M\}$, i.e. transmits a very high amplitude $ \sqrt{\frac{P}{\prob_w({\text{``NACK''}})} }$ in position $w$ (since $\prob_w(\text{``NACK''})$ is exponentially small) and zero in all the remaining $M-1$ positions. 
Thus, if the true message is $W=1$ and a ``NACK'' is declared at the transmitter, only the first slot of the retransmission codeword sequence contains a high amplitude signal, while the other two will be zero. 
\begin{equation} \label{eq:ReTransCODEM3}
		X^{M=3}(w') = 		
		\begin{cases*}
      \sqrt{\frac{P}{\prob_w({\text{``NACK''}})} }\cdot (0,0,0),								&if ``ACK'' \& $ \forall w'$\\
	 \sqrt{\frac{P}{\prob_w({\text{``NACK''}})} }\cdot\left(1,0,0 \right),								&if ``NACK'' \& $w'=1$ \\
	 \sqrt{\frac{P}{\prob_w({\text{``NACK''}})} }\cdot\left(0,1,0 \right),								&if ``NACK'' \& $w'=2$ \\
	 \sqrt{\frac{P}{\prob_w({\text{``NACK''}})} }\cdot\left(0,0,1 \right),								&if ``NACK'' \& $w'=3$\\
    \end{cases*} .
\end{equation}

At the end of the retransmission stage, the receiver makes a decision by comparing the signals received in the last $M=3$ time slots, $y_{i}$ for $i=\{n-2,n-1,n\}$ with a very large threshold $\Upsilon =n$, such that:
\begin{align}
	&-\lim_{n\to \infty} \frac{1}{n} \ln \prob_{1}\left( Y_{i} < \Upsilon \mid X_i = \sqrt{\frac{P}{\prob_1{(\text{``NACK''})}} } \right) = +\infty  \label{eq:ReTx1}\\
	&-\lim_{n\to \infty} \frac{1}{n} \ln \prob_{1}\left( Y_{i} > \Upsilon \mid X_i = 0 \right) = +\infty . \label{eq:ReTx2}
\end{align}

To provide some insight into the decoding rule operation, note from \eqref{eq:ReTransCODEM3} and the assumption $W=1$ is sent, that the received signal at the $i$-th position during the last $M$ channel uses, $Y_i$, is a random variable which is distributed in the case of ``ACK'' as:
\begin{equation} \label{eq:YidistrACK}
Y_{i} \sim \mathcal{N}\left(0,\sigma^2 \right),	\;\; \text{ for } i = n-2 \text{ and } i=n-1 \text{ and } i=n,
\end{equation}
whereas for ``NACK'' as:
\begin{equation} \label{eq:XidistrNACK}
Y_{i} \sim 		
\begin{cases*}
\mathcal{N}\left(\sqrt{ \frac{P}{ \prob_1(\text{``NACK''}) } } ,\sigma^2 \right),	& if $i=n-2$ \\
\mathcal{N}\left(0,\sigma^2 \right),								& if $i = n-1$ and $i=n$.
\end{cases*}
\end{equation}

Observe that at the end of the three stages, \eqref{eq:ReTx1} and \eqref{eq:ReTx2} guarantee that once the retransmission sequence of length $M$ is received, the decoder can use the following rule to determine $\hat{W}$ as: 
\begin{equation} \label{eq:DecRuleM3}
		\hat{W} =		
		\begin{cases*}
      \tilde{W},		&if ACK\\
			1        ,		&if NACK and $Y_{n-2} > \Upsilon$ , $Y_{n-1} < \Upsilon$ , $Y_{n} < \Upsilon$ \\
			2        ,		&if NACK and $Y_{n-2} < \Upsilon$ , $Y_{n-1} > \Upsilon$ , $Y_{n} < \Upsilon$ \\
			3        ,		&if NACK and $Y_{n-2} < \Upsilon$ , $Y_{n-1} < \Upsilon$ , $Y_{n} > \Upsilon$ \\
    \end{cases*} ,
\end{equation}
where $\tilde{W}$ corresponds to the codeword decoded at the receiver based on the sequence received in the first $j=n-(M+1)$ channel uses using the minimum distance decoding rule 
\begin{align} \label{eq:DecRuleM3Fin}
	\tilde{W} &= \text{arg} \min_{w \in \{ 1, 2 , 3 \}}{ || x^{j}(w) - y^j ||} .
\end{align}

\subsubsection{Probability of error analysis}
 
 
Assuming $W=1$ was sent, note from the geometry of the problem in Figure  \ref{fig:NackBands}, 
that it suffices to upper bound $\prob(\text{error})  =\prob_1(W \neq \hat{W})$ as 
\begin{align}
	\prob_1(\text{error}) &= \overbrace{\prob_1(\text{NACK}) \prob_1(\text{``NACK''}\mid \text{NACK})}^{\leq1} \cdot \overbrace{\prob_1(\text{error} \mid \text{NACK , ``NACK''})}^{\text{exceedingly small by \eqref{eq:ReTx1} and \eqref{eq:ReTx2} }} \notag \\
		&+ \overbrace{\prob_1(\text{NACK})}^{\leq1} \overbrace{\prob_1(\text{``ACK''}\mid \text{NACK})}^{\text{exceedingly small by \eqref{eq:ACKNACK}}} \cdot \overbrace{\prob_1(\text{error} \mid \text{NACK , ``ACK''})}^{\leq1} \notag \\
		&+ \overbrace{\prob_1(\text{ACK})}^{\leq1} \overbrace{\prob_1(\text{``NACK''}\mid \text{ACK})}^{\text{exceedingly small by \eqref{eq:NACKACK}}} \cdot \overbrace{\prob_1(\text{error} \mid \text{ACK , ``NACK''})}^{\leq1} \notag \\
		&+ \prob_1(\text{ACK}) \prob_1(\text{``ACK''}\mid \text{ACK}) \cdot \prob_1(\text{error} \mid \text{ACK , ``ACK''}) \notag  \\
	&\doteq \prob_1(\text{ACK}) \prob_1(\text{``ACK''}\mid \text{ACK}) \cdot \prob_1(\text{error} \mid \text{ACK , ``ACK''}) \notag  \\	
	&\leq \prob_1(\text{error} \mid \text{ACK , ``ACK''}) \notag\\
	&= \prob_1(\text{error} \mid \text{ACK}). \label{eq:PerrACKBBM}
\end{align} 

 Equation \eqref{eq:PerrACKBBM} is similar to \cite[Equation (132)]{Kim:ActiveNoisyFB}. 
To upper bound $\prob_1(\text{error} \mid \text{ACK})$ note first that an ACK implies the received signal is within one of the three codeword regions $A_w$ (shown in Figure  \ref{fig:NackBands}). Since $W=1$, an error occurs only if the received signal is in one of the two codeword regions $A_2$ or $A_3$, and so
\begin{align}
	\prob_1(\text{error} \mid \text{ACK}) &= \frac{ \prob_1(\text{error} , \text{ACK})  } {\prob_1(\text{ACK}) } \notag\\
	&= \frac{ \prob_1\left( y^j \in  A_2 \cup A_3 \right) }{ \prob_1\left( y^j \in A_1 \cup A_2 \cup A_3 \right) }. \label{eq:Perror}
\end{align}
By symmetry, $\prob_1( y^j \in A_2 ) = \prob_1( y^j \in A_3 ) $, and hence
\begin{align}
	\prob_1\left( y^j \in  A_2 \cup A_3 \right) &\leq \prob_1\left( y^j \in  A_2 \right) +   \prob_1\left( y^j \in  A_3 \right) \notag \\
	    &= 2 \prob_1\left( y^j \in  A_2 \right) \notag \\
			&\leq 2 Q\left( \frac{ d_B }{ \sigma } \right) \notag \\
			&\leq \exp{\left( - n \frac{3P}{8\sigma^2}(1+t)^2 \right)}
\end{align}
where, $d_B = \frac{1}{2} (d' + t d') $ is as shown in Figure  \eqref{fig:NackBands}. 
We can further upper bound \eqref{eq:Perror} by lower bounding the denominator as: 
\begin{align}
\prob_1\left( y^j \in A_1 \cup A_2 \cup A_3 \right) &\geq  \prob_1\left( y^j \in A_1 \right) \notag \\
   &\geq 1 - 2 Q \left(\frac{d_A}{\sigma}\right) \notag \\
	 &\geq 1 - \exp{\left( -n \frac{3P}{8\sigma^2} (1-t)^2  \right)}. 
\end{align}
Therefore, \eqref{eq:Perror} can be written as:
\begin{align}
		\prob_1(\text{error} \mid \text{ACK}) &\leq  \frac{\exp{\left( - n \frac{3P}{8\sigma^2}(1+t)^2 \right)}}{1 - \exp{\left( -n \frac{3P}{8\sigma^2} (1-t)^2  \right)}  	} \doteq \exp{\left( - n \frac{3P}{8\sigma^2}(1+t)^2 \right)} \label{eq:PerrACKTOT}. 
\end{align}
Finally, the achievable error exponent for the building block, since parameter $t$ can be chosen sufficiently close to 1, is:
\begin{equation} \label{eq:ErrExpBB}
		E^{\text{EXP}}_{\text{BB}} \geq \frac{3P}{8\sigma^2}(1+t)^2 = \frac{3P}{2\sigma^2}.
\end{equation}
Equation \eqref{eq:ErrExpBB} shows that the building block leads to a four-fold gain over non-feedback transmission under the AS power constraint in \eqref{Eq:NoFBErrExp}: $E^{AS} = \frac{3P}{8\sigma^2}$. This also illustrates how a more flexible power constraint may lead to higher achievable error exponents. Next, following \cite{Kim:ActiveNoisyFB}, we utilize the building block for the transmission of $M=3$ messages in a three-stage communication scheme.

\subsection{A transmission scheme based on the building block for the transmission of $M=3$ messages under the EXP power constraint}
\label{sec:SchemeBasedOnBB}


This scheme comprises three stages: transmission, feedback and retransmission, lasting $\lambda(n-M)$, $(1-\lambda)(n-M)$  and $M$ channel uses respectively for $\lambda \in (0,1)$. 
In the transmission stage, the message $W$ is transmitted using the building block, using power $\frac{1}{\lambda} P - \eta$ on the forward link and $\eta$ in the feedback link. The transmitter reserves power $ 0 < \eta < \min\{P, P_{\text{FB}} \}$ to provide feedback to the receiver's transmission in the next stage. Denoting the estimation of the true message at the end of this stage by $W'$, 
using \eqref{eq:ErrExpBB}, the probability of error of this stage is:
\begin{align}
	\prob_1(W'\neq W) &= \exp{\left( -\lambda (n-M) \frac{3 (\frac{1}{\lambda}P - \eta) }{2\sigma^2}   \right)}  \notag\\
	 &\doteq \exp{\left( - n \frac{3 (P-\lambda \eta)}{2\sigma^2} \right)}  \label{eq:PerrM3BBOW}.
\end{align}
In the feedback stage, $W'$ is sent to the transmitter using the building block in $(1-\lambda)(n-M)$ channel uses. The receiver uses power $\frac{1}{1-\lambda}P_{\text{FB}}-\eta$ while the transmitter uses $\eta$. The estimate $W''= \hat{W}'$ leads to the probability of error,  
given by \eqref{eq:ErrExpBB}:
\begin{align}
	\prob_1(W''\neq W' \mid W') &\doteq \exp{\left( - n \frac{3 (P_{\text{FB}}- (1-\lambda) \eta)}{2\sigma_{\text{FB}}^2}   \right)}  \label{eq:PerrorM3BBOWfb}.
\end{align}
 

The retransmission stage follows similarly as in the building block, but using code $\mathcal{G}\left(\mathcal{W}, \sqrt{\frac{P}{\prob_w(W''\neq W)}}\right)$. 
Thus, the transmitter compares $W''$ with $W=w$, and generates a length $M$ retransmission codeword using \eqref{eq:ReTransCODEM32} {(shown for $M=3$)}.
\begin{equation} \label{eq:ReTransCODEM32}
		X^{M}(w) = 		
		\begin{cases*}
      \sqrt{\frac{P}{\prob_w(W''\neq W)}}\cdot(0,0,0),								&if $W''=W$ \\
			\sqrt{\frac{P}{\prob_w(W''\neq W)}}\cdot(1,0,0),								&if $W''\neq W$ \& $w=1$ \\
			\sqrt{\frac{P}{\prob_w(W''\neq W)}}\cdot(0,1,0),								&if $W''\neq W$ \& $w=2$ \\
			\sqrt{\frac{P}{\prob_w(W''\neq W)}}\cdot(0,0,1),								&if $W''\neq W$ \& $w=3$\\
    \end{cases*}.
\end{equation}
Finally, the receiver uses the decoding rule of Equation \eqref{eq:DecRuleM32} (setting $\tilde{W} = W'$) to estimate message $W$, based on the length $M=3$ codeword received during the retransmission stage:
\begin{equation} \label{eq:DecRuleM32}
		\hat{W} =		
		\begin{cases*}
      W',		&if $Y_{n-2} < \Upsilon$ , $Y_{n-1} < \Upsilon$ , $Y_{n} < \Upsilon$ \\
			1        ,		&if $Y_{n-2} > \Upsilon$ , $Y_{n-1} < \Upsilon$ , $Y_{n} < \Upsilon$ \\
			2        ,		&if $Y_{n-2} < \Upsilon$ , $Y_{n-1} > \Upsilon$ , $Y_{n} < \Upsilon$ \\
			3        ,		&if $Y_{n-2} < \Upsilon$ , $Y_{n-1} < \Upsilon$ , $Y_{n} > \Upsilon$ \\
    \end{cases*}.
\end{equation}
As in the building block decoding rule, each of the $M$ signals $Y_i^M$ is compared with a threshold $\Upsilon =n$, which has been chosen to be very large such that:
 \begin{align}
	&-\lim_{n\to \infty} \frac{1}{n} \ln \prob_{w}\left( Y_{i} < \Upsilon \mid X_i^M = \sqrt{\frac{P}{\prob_w{(W''\neq W)}} } \right) = +\infty  \label{eq:ReTx1B}\\
	&-\lim_{n\to \infty} \frac{1}{n} \ln \prob_{w}\left( Y_{i} > \Upsilon \mid X_i^M = 0 \right) = +\infty  \label{eq:ReTx2B}.
\end{align}
Note that $\prob_w{(W''\neq W)} $ can be shown to be exponentially small too. 

The probability of error of this communication scheme, considering decoding rule of Equation \eqref{eq:DecRuleM32}, along with \eqref{eq:ReTx1B} and \eqref{eq:ReTx2B}  is:
\begin{align}
	\prob_1(\text{error}) &= \overbrace{\prob_1(\text{error} \mid W'' = 1, W' = 1 )}^{\text{negligible by \eqref{eq:ReTx2B}}} \cdot  \overbrace{\prob_1( W''=1,  W'=1 )}^{\leq1} \notag \\
	&+ \prob_1(\text{error} \mid W'' =1 , W' = 2 ) \cdot  \prob_1( W'' = 1 , W' = 2 ) \notag \\
	&+ \prob_1(\text{error} \mid W'' =1 , W' = 3 ) \cdot  \prob_1( W'' = 1 , W' = 3 ) \notag \\
&+ \overbrace{\prob_1(\text{error} \mid W'' = 2, W' = 1 )}^{\text{negligible by \eqref{eq:ReTx1B} and  \eqref{eq:ReTx2B}  }} \cdot  \overbrace{\prob_1( W''=2,  W'=1 ) }^{\leq 1}\notag \\
	&+ \overbrace{\prob_1(\text{error} \mid W'' =2 , W' = 2 )}^{\text{negligible by \eqref{eq:ReTx1B} and  \eqref{eq:ReTx2B}  }} \cdot  \overbrace{\prob_1( W'' = 2 , W' = 2 )}^{\leq 1} \notag \\
	&+ \overbrace{\prob_1(\text{error} \mid W'' =2 , W' = 3 )}^{\text{negligible by \eqref{eq:ReTx1B} and  \eqref{eq:ReTx2B}  }} \cdot  \overbrace{\prob_1( W'' = 2 , W' = 3 )}^{\leq 1} \notag \\
&+ \overbrace{\prob_1(\text{error} \mid W'' = 3, W' = 1 )}^{\text{negligible by \eqref{eq:ReTx1B} and  \eqref{eq:ReTx2B}  }} \cdot \overbrace{ \prob_1( W''=3,  W'=1 )}^{\leq 1} \notag \\
	&+ \overbrace{\prob_1(\text{error} \mid W'' =3 , W' = 2 )}^{\text{negligible by \eqref{eq:ReTx1B} and  \eqref{eq:ReTx2B}  }} \cdot  \overbrace{\prob_1( W'' = 3 , W' = 2 )}^{\leq 1} \notag \\
	&+ \overbrace{\prob_1(\text{error} \mid W'' =3 , W' = 3 )}^{\text{negligible by \eqref{eq:ReTx1B} and  \eqref{eq:ReTx2B}  }} \cdot  \overbrace{\prob_1( W'' = 3 , W' = 3 )}^{\leq 1} \notag \\	
	&\doteq \prob_1(\text{error} \mid W'' =1 , W' = 2 ) \cdot  \prob_1( W'' = 1 , W' = 2 ) \notag \\
	&+ \prob_1(\text{error} \mid W'' =1 , W' = 3 ) \cdot  \prob_1( W'' = 1 , W' = 3 ) \notag \\
 &\leq \prob_1(W''= 1, W'= 2) + \prob_1(W''= 1, W'= 3).
\end{align}
Since $\prob_1(W''= 1, W'= 2) = \prob_1(W''= 1, W'= 3) $,
\begin{align}
 \prob_1(\text{error})  &\leq 2 \prob_1(W''= 1, W'\neq W) .  \label{eq:PerrorM3}
\end{align}
Using \eqref{eq:PerrM3BBOW} and \eqref{eq:PerrorM3BBOWfb}, \eqref{eq:PerrorM3} can be written as:
\begin{align}
	\prob_1(\text{error})  &\leq 2\prob(W'' = 1 \mid W'\neq W, W=1 )\cdot \prob( W'\neq W \mid W=1) \notag\\												
												 &= 2 \left( \frac{1}{2} \exp{\left( - n \frac{3 (P_{\text{FB}}- (1-\lambda) \eta)}{2\sigma_{\text{FB}}^2}   \right)}   \right) \cdot \exp{\left( - n \frac{3 (P-\lambda \eta)}{2\sigma^2} \right)}. \label{eq:PerrSchM3}
\end{align}
Above, the $\frac{1}{2}$ factor next to the first exponential function results from the fact that when $W'\neq 1$ is sent back to the transmitter (meaning that $W'= 2$ or $W'=3$), {a retransmission error will occur only for the case a $W''= W = 1'$ ($\prob_1(W''\neq W' \mid W') $), implying that for any $W'\neq W$ only half of the possibilities may lead to $W''=1$.}
Then, from \eqref{eq:PerrSchM3}, it follows that:
\begin{align}
	\prob_1(\text{error}) &\doteq \exp{\left( -n  \left( \frac{3 (P-\lambda \eta)}{2\sigma^2}   +  \frac{3 (P_{\text{FB}}- (1-\lambda) \eta)}{2\sigma_{\text{FB}}^2}    \right)  \right)}. \label{eq:Perr2stagesB}
\end{align}
This result implies that the following error exponent is achievable since $\eta$ can be chosen sufficiently small:
%
\begin{equation}
E^{\text{EXP}} \geq \frac{3}{2} \left( \frac{P}{\sigma^2}   +  \frac{ P_{\text{FB}}}{\sigma_{\text{FB}}^2}    \right).
\end{equation}

Next, we generalize this result for any finite number of messages $M$. First, we show the generalization of the building block operation, followed by its use in the three stage communication scheme.

\subsection{Building block for general finite $M$}
\label{sec:GeneralM}

This generalization is based on the use of a simplex code of $M$ messages for the transmission and feedback stages, with symbols having energy $nP$ and $nP_{\text{FB}}$ respectively. First, we transform the geometry for $M=3$ into the higher dimensional space required for a constellation of size $M$ using the mapping introduced in \cite{Kim:ThreeCWPeak2013}. For notational convenience, let the distance between two codewords shown in Figure \ref{fig:NackBands} and denoted as $d'$ for $M=3$, be denoted as $d'^{(3)}$. 
Then, the equivalent distance for general $M$ is denoted as $d'^{(M)}$. 
The following relation holds for all $M$, i.e. using  \eqref{eq:GeometryMapping} for a simplex code of size $M=4$  yields $d'^{(4)} = \sqrt{nP}  \sqrt{ \frac{8}{3} }$,
\begin{align} \label{eq:GeometryMapping}
		d'^{(M)} = { d'^{(3)} }{ \sqrt{ \frac{2M } { 3(M-1)}  } }.
\end{align}
 The building block operation for $M>3$ remains unchanged. Thus, it suffices 
 to prove the exponentially small probability of the NACK event in this new space. 
A simplex code for $M$ symbols requires $(M-1)$ dimensions, in which the NACK bands region idea still hold. 
Next, we analyze the key elements and differences in the scheme for the transmission of $M$ messages. 

\subsubsection{NACK $(M-1)$-volume and probability of NACK event}
\label{sec:PNACKgenM}
For $M$ symbols, the NACK bands region becomes an $M-1$ dimensional space defined as the complement of the union of the ACK volumes: $\bigcup_{i=1}^M A_i$, where each $A_i$ is defined in a similar way as \eqref{eq:ACKregionsM3EXP2}.
Note that for $M>3$, the $M-1$ hyperplanes defining each $A_i$ region can be obtained using points $H_i$ and the $M-1$ vectors in the direction of the lines that connect the $M-1$ symbols to symbol $i$. Then, each $A_i$ region corresponds to the space in the intersection of the regions above or below the hyperplanes that contain symbol $i$.
	%
	%

\begin{figure}[H]
	\centering
		\includegraphics[width=0.7\textwidth]{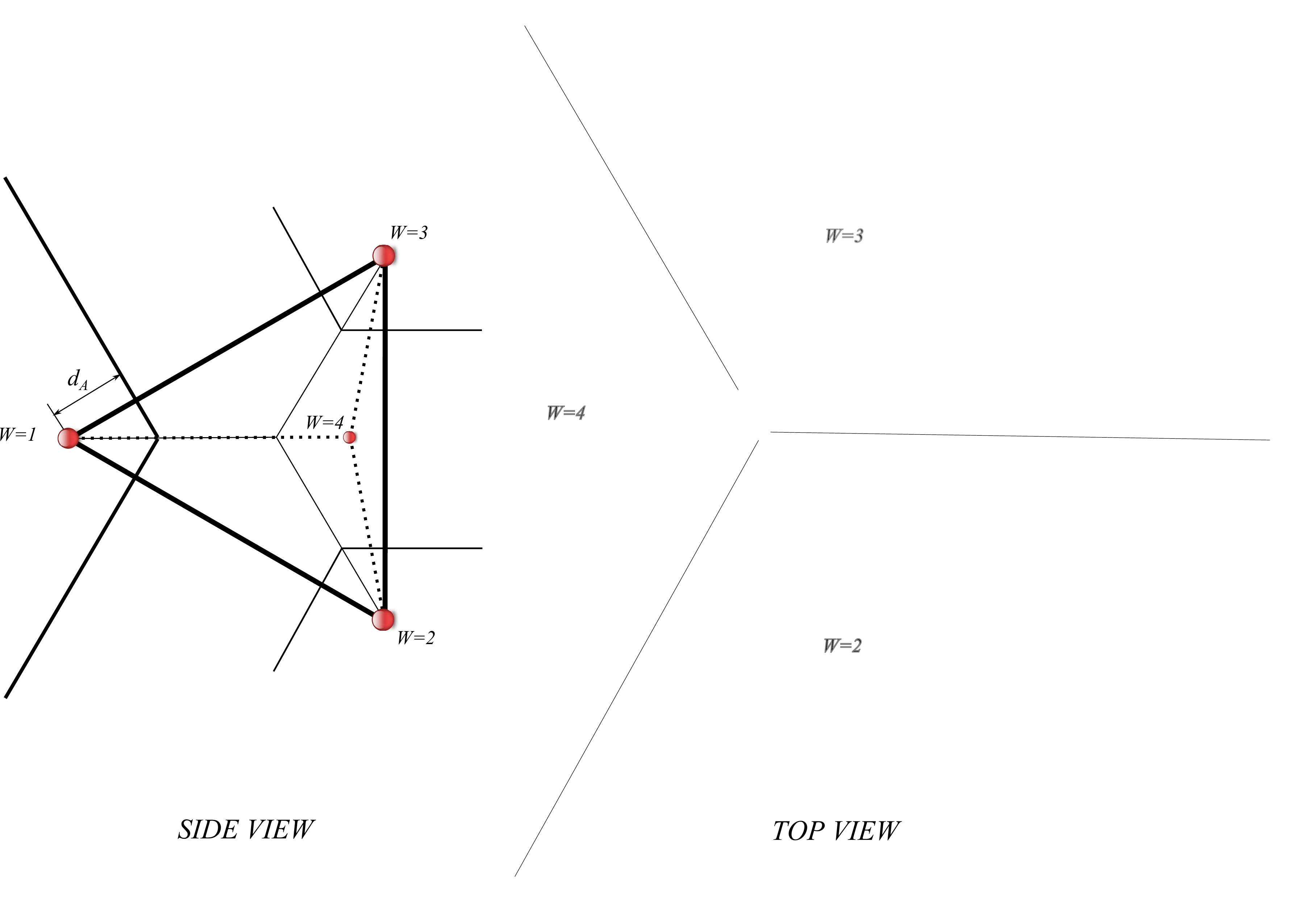}
		\caption{Top view of a simplex code constellation for $M=4$ and the hyperplanes defining $A_1$. Distance $d_A$ corresponds to the distance along the perpendicular line from symbol $W=1$ to each of the three hyperplanes.}
	\label{fig:piramidTOPii}
\end{figure}

Assume without loss of generality that symbol $W=1$ is transmitted. We first prove that the NACK event occurs with exponentially small probability. As in the case of $M=3$, since $	\prob_1(\text{NACK} ) = 1 - \prob_1(\text{ACK} )$, we obtain a lower bound on $\prob_1(\text{ACK} )$ to upper bound $\prob_1(\text{NACK} )$.  
For $M$ symbols, an ACK event is declared whenever the received signal, $y^j$, lies within one of the $A_i$ regions associated to each symbol:
\begin{align} \label{eq:PACKM}
	\prob_1(\text{ACK} ) &= \prob_1(y^j \in \bigcup_{i=1}^M A_i ) \\ \notag
											 &\geq \prob_1(y^j \in A_1 ) \\ \notag
											 &\geq 1 - \left[ (M-1 ) Q\left( \frac{d_A^{(M)}}{\sigma} \right) \right],\notag
\end{align} 
where $d_A^{(M)}$ denotes the distance between symbol $W=1$ and the NACK volume closest boundary, i.e. $d_A^{(M)} = (1-t) \frac{d'^{(M)}}{2}$, see Figure \ref{fig:piramidTOPii}. Then,
\begin{align} \label{eq:PyinA1}
	\prob_1(y^j \in A_1) \geq 1 -  \frac{(M-1)}{2} \exp{\left\{ -n \frac{P}{4\sigma^2} \frac{M}{M-1} (1-t)^2  \right\} }.
\end{align}
Therefore, the upper bound on $\prob_1(\text{NACK} ) \leq 1 - \prob_1(\text{ACK} )$ can be written as:
\begin{align}
		\prob_1(\text{NACK} ) &\leq (M-1 ) Q\left( \frac{d_A^{(M)}}{\sigma} \right) \\ \notag
					&\leq \frac{M-1}{2} \exp{ \left(-\frac{(d_A^{(M)} )^2}{2\sigma^2} \right)  }. \notag
\end{align}
Using  the mapping in \eqref{eq:GeometryMapping}, and the equivalent distance $d_A^{(M)} = (1-t)\sqrt{3nP} \sqrt{\frac{M}{6(M-1)}} $, we conclude that $\prob_1(\text{NACK} )$ occurs with exponentially small probability 
\begin{align} \label{eq:PNACKMexp}
		\prob_1(\text{NACK} ) &\leq \frac{M-1}{2} \exp{ \left(-n \frac{3P}{2\sigma^2} \left(  \frac{M}{6(M-1)} \right) (1-t)^2  \right)  }. 
\end{align}

\subsubsection{Probability of error analysis for the building block for $M$ messages}
\label{sec:PerrorGenM}

As in the case of $M=3$, once the transmitter has sent   codeword $x^j(w)$, the the receiver determines whether $y^j$ lies inside the NACK volume or in one of the ACK regions $A_i$. Then, the signaling of \eqref{eq:FBcode} is used to report this result to the transmitter. The feedback signaling decoding result may lead to a ``NACK'' or an ``ACK'' event decision that is used to generate the length $M$ retransmission, which exactly follows the $M=3$ scheme.  Therefore, a decoding error may occur only when the receiver declares an ACK, and the decoding decision is solely based on $y^j$. Since we assumed $W=1$:
\begin{align}
	\prob_1(\text{error}|\text{ACK}) &= \frac{ \prob_1(\text{error},\text{ACK}) }{ \prob_1(\text{ACK}) } \\ \notag
					&= \frac{ \prob_1\left( y^j \in \bigcup_{i=2}^M A_i  \right)}{ \prob_1\left( y^j \in \bigcup_{i= 1}^M A_i  \right) } \\ \notag 
					&\leq \frac{ \sum_{i=2}^M  \prob_1\left( y^j \in A_i  \right)}{ \prob_1\left( y^j \in  A_1  \right) },\notag 
\end{align}		
where the last inequality comes from the lower bound $\prob_1\left( y^j \in \bigcup_{i= 1}^M A_i  \right) \geq \prob_1\left( y^j \in A_1  \right)  $ for the denominator and the union bound on the numerator. Next, noting that for all $i\geq 2$, (see Figure  \ref{fig:NackBands}):
\begin{equation} \label{eq:NumBound}
	\prob_1(y^j \in A_i) \leq Q\left(\frac{d_B^{(M)}}{\sigma}\right),
\end{equation}
where $d_B^{(M)}$ is the distance between symbol $W=1$ and the furthest NACK region boundary: $d_B^{(M)} = (1+t) \frac{d'^{(M)}}{2}$. By the mapping \eqref{eq:GeometryMapping}, the equivalent distance $d_B^{(M)} = (1+t)\sqrt{3nP} \sqrt{\frac{M}{6(M-1)}} $.
Then, by \eqref{eq:PyinA1} and \eqref{eq:NumBound}, and the Chernoff bound:
\begin{align} \label{eq:PerrACKsimple}
	\prob_1(\text{error}|\text{ACK}) &\leq \frac {  \frac{M-1}{2} \exp{\left\{ -n \frac{P}{4\sigma^2} \frac{M}{M-1} (1+t)^2  \right\}  } } { 1 -  \frac{M-1}{2} \exp{\left\{ -n \frac{P}{4\sigma^2} \frac{M}{M-1} (1-t)^2  \right\}  } }.
\end{align}		
The denominator in \eqref{eq:PerrACKsimple} can be factored as:
\begin{align}
	 1 -  \frac{M-1}{2} \exp{\left\{ -n \frac{P}{4\sigma^2} \frac{M}{M-1} (1-t)^2  \right\}  }  &=\frac{M-1}{2} \exp{\left\{ -n \frac{P}{4\sigma^2} \frac{M}{M-1} (1-t)^2  \right\}  } \cdot \\ \notag
	&\left[ \frac{M-1}{2} \exp{\left\{ +n \frac{P}{4\sigma^2} \frac{M}{M-1} (1-t)^2  \right\}  } -1     \right],
\end{align}
such that for very large $n$, the expression in the brackets tends to $\frac{M-1}{2} \exp{\left\{ +n \frac{P}{4\sigma^2} \frac{M}{M-1} (1-t)^2  \right\}  }$ and consequently, the denominator of \eqref{eq:PerrACKsimple} tends to $1$ for very large $n$. 
From this, we obtain: 
\begin{align} \label{eq:PerrACKMfin}
	\prob_1(\text{error}|\text{ACK}) &\doteq   \frac{M-1}{2} \exp{\left\{ -n \frac{P}{4\sigma^2} \frac{M}{M-1} (1+t)^2  \right\}  }.
\end{align}		
Since parameter $t$ can be chosen to be very close to $1$, the achievable error exponent is:
\begin{equation} \label{eq:ErrExpMfin}
		E^{\text{EXP}}_{\text{BB}} \geq \frac{P}{\sigma^2} \frac{M}{M-1},
\end{equation}
where, the subscript BB stands for building block error exponent.

Note that for $M=2$, \eqref{eq:ErrExpMfin} leads to the result of \cite{Kim:ActiveNoisyFB}: $		E^{\text{EXP}} \geq \frac{2P}{\sigma^2}$; and for $M=3$, it leads to \eqref{eq:ErrExpBB}: $		E^{\text{EXP}} \geq \frac{3P}{2\sigma^2}$.

\subsection{A three stage transmission scheme based on the building block for $M$ messages}
\label{sec:SchemeGenM}

In general, the three stage operation remains unchanged. In the transmission stage, $W$ is transmitted using the building block. Note that $\prob_1(W'\neq W) = \prob_2(W'\neq W)  =...= \prob_w(W'\neq W) $.  The receiver estimates $W'$ with a probability of error given by Equation \eqref{eq:PerrACKMfin}:
\begin{align}
	\prob_1(W'\neq W) &\leq\frac{M-1}{2} \exp{\left\{ -\lambda (n-M) \frac{(\frac{1}{\lambda}P - \eta) }{4\sigma^2} \frac{M}{M-1} (1+t)^2  \right\}  }  \\ \notag
	&\doteq \frac{M-1}{2} \exp{\left\{ - n \frac{(P - \lambda\eta) }{4\sigma^2} \frac{M}{M-1} (1+t)^2  \right\}  } \label{eq:PerrMBBOW}.
\end{align}
For the feedback stage, $W'$ is sent to the transmitter using the building block in $(1-\lambda)(n-M)$ channel uses. By the simplex code geometry, $\prob_1(W''\neq W' \mid W') = \prob_2(W''\neq W' \mid W')  = ... =  \prob_w(W''\neq W' \mid W')$. The feedback transmission yields a probability of error given by \eqref{eq:PerrACKMfin}:
\begin{align}
	\prob_1(W''\neq W' \mid W') &\leq \frac{M-1}{2} \exp{\left\{ -(1-\lambda)(n-M) \frac{(\frac{1}{1-\lambda}P_2 - \eta) }{4\sigma_{\text{FB}}^2} \frac{M}{M-1} (1+t)^2  \right\}  }  \\ \notag
	&\doteq \frac{M-1}{2} \exp{\left\{ -n \frac{(P_2 - (1-\lambda)\eta)}{4\sigma_{\text{FB}}^2} \frac{M}{M-1} (1+t)^2  \right\}  }. \label{eq:PerrMBBOWFB}
\end{align}


In the retransmission stage the transmitter compares $W''$ with $W$, and generates a retransmission codeword of length $M$ based on the index location code $\mathcal{G}\left(\mathcal{W},\sqrt{\frac{P}{\prob_w(W''\neq W)}}\right)$ of size $M$. The final decoding rule follows directly from extending \eqref{eq:DecRuleM3} to $M>3$, noting that \eqref{eq:ReTx1B} and \eqref{eq:ReTx2B} still hold.

%
%
%

The probability of error of this scheme is dominated by the events where an incorrect symbol is decoded after the transmission stage, and the feedback stage leads to incorrectly decoding $W''$ as the true message.
\begin{align}
	\prob_1(\text{error}) &\doteq  \overbrace{\prob_1(\text{error} \mid W'' = 1, W' = 1 )}^{\text{negligible by \eqref{eq:ReTx2B}}} \cdot  \overbrace{\prob_1( W''=1,  W'=1 )}^{\leq1} \notag \\
	&+ \prob_1(\text{error} \mid W'' =1 , W' = 2 ) \cdot  \prob_1( W'' = 1 , W' = 2 ) \notag \\
	&+ \prob_1(\text{error} \mid W'' =1 , W' = 3 ) \cdot  \prob_1( W'' = 1 , W' = 3 ) \notag \\
	&+ ... \notag \\
	&+ \prob_1(\text{error} \mid W'' =1 , W' = M ) \cdot  \prob_1( W'' = 1 , W' = M ) \notag \\
&+ \overbrace{\prob_1(\text{error} \mid W'' = 2, W' = 1 )}^{\text{negligible by \eqref{eq:ReTx1B} and  \eqref{eq:ReTx2B}  }} \cdot  \overbrace{\prob_1( W''=2,  W'=1 ) }^{\leq 1}\notag \\
	&+ \overbrace{\prob_1(\text{error} \mid W'' =2 , W' = 2 )}^{\text{negligible by \eqref{eq:ReTx1B} and  \eqref{eq:ReTx2B}  }} \cdot  \overbrace{\prob_1( W'' = 2 , W' = 2 )}^{\leq 1} \notag \\
	&+ \overbrace{\prob_1(\text{error} \mid W'' =2 , W' = 3 )}^{\text{negligible by \eqref{eq:ReTx1B} and  \eqref{eq:ReTx2B}  }} \cdot  \overbrace{\prob_1( W'' = 2 , W' = 3 )}^{\leq 1} \notag \\
&+ ... \notag \\	
&+ \overbrace{\prob_1(\text{error} \mid W'' =2 , W' = M )}^{\text{negligible by \eqref{eq:ReTx1B} and  \eqref{eq:ReTx2B}  }} \cdot  \overbrace{\prob_1( W'' = 2 , W' = M )}^{\leq 1} \notag \\
&+ \overbrace{\prob_1(\text{error} \mid W'' = 3, W' = 1 )}^{\text{negligible by \eqref{eq:ReTx1B} and  \eqref{eq:ReTx2B}  }} \cdot \overbrace{ \prob_1( W''=3,  W'=1 )}^{\leq 1} \notag \\
	&+ \overbrace{\prob_1(\text{error} \mid W'' =3 , W' = 2 )}^{\text{negligible by \eqref{eq:ReTx1B} and  \eqref{eq:ReTx2B}  }} \cdot  \overbrace{\prob_1( W'' = 3 , W' = 2 )}^{\leq 1} \notag \\
	&+ \overbrace{\prob_1(\text{error} \mid W'' =3 , W' = 3 )}^{\text{negligible by \eqref{eq:ReTx1B} and  \eqref{eq:ReTx2B}  }} \cdot  \overbrace{\prob_1( W'' = 3 , W' = 3 )}^{\leq 1} \notag \\	
	&+ ... \notag \\	
	&+ \overbrace{\prob_1(\text{error} \mid W'' =3 , W' = M )}^{\text{negligible by \eqref{eq:ReTx1B} and  \eqref{eq:ReTx2B}  }} \cdot  \overbrace{\prob_1( W'' = 3 , W' = M )}^{\leq 1} \notag \\		
	&+ .. \notag \\	
	&+ . \notag \\	
	&+ \overbrace{\prob_1(\text{error} \mid W'' = M, W' = 1 )}^{\text{negligible by \eqref{eq:ReTx1B} and  \eqref{eq:ReTx2B}  }} \cdot \overbrace{ \prob_1( W''=M,  W'=1 )}^{\leq 1} \notag \\
	&+ \overbrace{\prob_1(\text{error} \mid W'' =M , W' = 2 )}^{\text{negligible by \eqref{eq:ReTx1B} and  \eqref{eq:ReTx2B}  }} \cdot  \overbrace{\prob_1( W'' = M , W' = 2 )}^{\leq 1} \notag \\
	&+ \overbrace{\prob_1(\text{error} \mid W'' =M , W' = 3 )}^{\text{negligible by \eqref{eq:ReTx1B} and  \eqref{eq:ReTx2B}  }} \cdot  \overbrace{\prob_1( W'' = M , W' = 3 )}^{\leq 1} \notag \\	
	&+ ... \notag \\	
	&+ \overbrace{\prob_1(\text{error} \mid W'' =M , W' = M )}^{\text{negligible by \eqref{eq:ReTx1B} and  \eqref{eq:ReTx2B}  }} \cdot  \overbrace{\prob_1( W'' = M , W' = M )}^{\leq 1} \notag \\		
	&\doteq \prob_1(\text{error} \mid W'' =1 , W' = 2 ) \cdot  \prob_1( W'' = 1 , W' = 2 ) \notag \\
	&+ \prob_1(\text{error} \mid W'' =1 , W' = 3 ) \cdot  \prob_1( W'' = 1 , W' = 3 ) \notag \\
	&+ ... \notag \\	
	&+ \prob_1(\text{error} \mid W'' =1 , W' = M ) \cdot  \prob_1( W'' = 1 , W' = M ) \notag \\
 &\leq \prob_1(W''= 1, W'= 2) + \prob_1(W''= 1, W'= 3) + ... \notag \\
 & ... + \prob_1(W''= 1, W'= M) \notag\\
 &\leq \sum_{k=2}^M \prob_1(W''= 1, W'= k)
\end{align}
Since $\prob_1(W''= 1, W' =2) = \prob_1(W''= 1, W'= 3) = ... =  \prob_1(W''= 1, W'= M)$,
\begin{align}
 \prob_1(\text{error})  &\leq (M-1) \prob_1(W''= 1, W'\neq W).  \label{eq:PerrorM}
\end{align}
Using \eqref{eq:PerrMBBOW} and \eqref{eq:PerrMBBOWFB}, \eqref{eq:PerrorM} can be written as:
\begin{align}
	\prob_1(\text{error})  &\leq (M-1) \cdot \prob( W'\neq W \mid W=1) \cdot \prob(W'' = 1 \mid W'\neq W, W=1 ) \notag\\												
												 &= (M-1) \cdot   \frac{M-1}{2} \exp{\left\{ - n \frac{(P - \lambda\eta) }{4\sigma^2} \frac{M}{M-1} (1+t)^2  \right\} }  \cdot  \frac{1}{2} \exp{\left\{ -n \frac{(P_{\text{FB}} - (1-\lambda)\eta)}{4\sigma_{\text{FB}}^2} \frac{M}{M-1} (1+t)^2  \right\}  }  \label{eq:PerrSchM} \\
%
	& \leq \frac{(M-1)^2}{4}  \cdot \exp{\left[ -n \frac{M (1+t)^2}{M-1}  \left( \frac{(P - \lambda\eta) }{4\sigma^2} + \frac{(P_{\text{FB}} - (1-\lambda)\eta)}{4\sigma_{\text{FB}}^2}   \right)  \right]}. \label{eq:Perr2stagesBG}
\end{align}

This result concludes the proof of Theorem \ref{th:OneWayEXPM2}, since the following error exponent is achievable, given $\eta$ can be sufficiently small and $t$ very close to $1$: 
\begin{align}
		E^{\text{EXP}}_{\text{FB}^{\text{EXP}}}\left(M,P,\sigma^2,P_{\text{FB}},\sigma^2_{\text{FB}}\right)  &\geq \frac{M}{M-1} \left( \frac{P }{\sigma^2} + \frac{P_2}{\sigma_{\text{FB}}^2}   \right).
\end{align}

 \qed

\section{Achievable error exponents region for the Two-Way AWGN channel and the transmission of $M$ messages under the EXP power constraint: Proof of Theorem \ref{th:EXPMachievTW}}
\label{sec:M2EXP}

In this section we present how the result of Theorem \ref{th:EXPMachievTW} can be obtained from the achievable error exponent for the one-way AWGN channel presented in Section \ref{sec:ProofOneWayEXPM}.  An initial result for the transmission of $M=2$ messages can be found in \cite{PalacioDevroyeISIT2018}; here we extend this to general $M$.


\subsubsection{A Two-way Communications Building Block}
\label{sec:BuildingBlock}

A length $n$ building block based on that in \cite{Kim:ActiveNoisyFB} is proposed for two-way communications under the EXP power constraint for both directions. Messages $W_1$ and $W_2$ are transmitted simultaneously following the channel model described by Equation \eqref{eq:TWChModel1}. The probability of error given that  $W_i=w_i$ is sent in each direction follows directly from that obtained for the one-way channel building block, in particular, as Equation \eqref{eq:PerrACKMfin}. Then, by denoting the estimates of each transmitted message by $W'_1$ and $W'_2$ respectively, the probabilities of error on decoding messages $W_i$ for each direction, conditioned on $W_i=w_i$ sent are:
\begin{align}
  \prob_{i(3-i)}(W_i \neq W'_i \mid W_i =w_i ) &= \mathsf{P}_{w_i}(W_i \neq W'_i )
\end{align}
 Assume that messages $W_1 = W_2 =1$ are transmitted, since by the geometry of the simplex code, the remaining results follow in a similar way. Then, by \eqref{eq:PerrACKMfin}, the above probabilities can be easily obtained as: 
\begin{align} \label{eq:PerrACKMfinTW}
	\prob_{w_i=1}(\text{error}) &\leq  \frac{M-1}{2} \exp{\left\{ -n \frac{P_i}{4\sigma_{3-i}^2} \frac{M}{M-1} (1+t)^2  \right\}  }
\end{align}	
Similarly, as $t$ can be chosen very close to $1$, the following pair is achievable for the two-way building block:
%
\begin{align} \label{eq:EEexpTW}
		E_{i(3-i)}^{\text{EXP}} \geq \frac{M}{M-1} \frac{P_i}{\sigma_{3-i}^2}.
\end{align}

\subsubsection{A three phase communication scheme for the two-way AWGN channel based on the use of the building block}
\label{sec:ThreePhases}

Similar to the one-way approach, we present a scheme based on the two-way building block comprised of three stages: 
Transmission, Feedback and Retransmission, whose durations are respectively given by $\lambda(n-M)$, $(1-\lambda)(n-M)$ and $M$ channel uses, for  $\lambda \in (0,1)$ .

\textbf{1. Transmission}:	The two-way building block is used for the simultaneous transmission of messages $W_1$ and $W_2$ using $\lambda (n-M)$ channel uses. These messages are estimated at their corresponding destinations as ${W}'_1$ and  ${W}'_2$. We allocate power $K_1 P_1$ for the $1\to2$ direction transmission, and power $K_2 P_2$ for the $2\to1$ direction. Parameters $K_1, K_2 \in [0,1/\lambda]$ indicate the fraction of the available power that is allocated for the transmission phase in each direction. Then, assuming the transmission of $W_1=1$ and $W_2=1$ and considering the building block probability of error analysis of \eqref{eq:PerrACKMfin}, we obtain the following for the $i \to (3-i)$ directions respectively for $i \in \{ 1,2\}$:
\begin{align} \label{eq:PerrACKMfinTWB}
	\prob_{w_i=1}^{i(3-i)}(W'_i \neq W_i) &\leq   \frac{M-1}{2} \exp{\left\{ -\lambda(n-M) \frac{K_iP_i}{4\sigma_{3-i}^2} \frac{M}{M-1} (1+t)^2  \right\}  } ,
\end{align}	
 leading to the achievability of the following error exponent pair:
\begin{align} \label{eq:EEexpTWM}
	E_{i(3-i)}^{\text{EXP}}  &\geq \lambda \frac{M}{M-1} K_i \frac{P_i}{\sigma_{3-i}^2}
\end{align}	
	\textbf{2. Feedback}: 	In this stage, the $W'_i$ messages are sent back to the terminals they originated from, using $(1-\lambda) (n-M)$ channel uses. 
	We parametrize the fraction of power allocated for this transmission by means of $J_1 , J_2 \in [0,1/(1-\lambda)]$ respectively for each terminal.  At the end of this stage, each terminal has an estimate of the messages that were fedback, denoted by $W''_i$. 
	Since these stage transmissions are performed using the basic two-way building block, the probability of error for each direction is upper bounded by:
\begin{align} \label{eq:PerrACKMfinTWBFB}
	\prob_{w_i=1}^{i(3-i)}(W''_i \neq W'_i) &\leq   \frac{M-1}{2} \exp{\left\{ -(1-\lambda)(n-M) \frac{J_iP_i}{4\sigma_{3-i}^2} \frac{M}{M-1} (1+t)^2  \right\}  }.
\end{align}		
	

	\textbf{3. Retransmission}: At the end of the feedback stage, each terminal compares the true message $W_i$ with that estimated from the feedback received $W''_i$. As a result of this comparison, a length $M$ codeword based on the index location code $ \mathcal{G}\left( \mathcal{W}, \sqrt{\frac{P_i}{\prob_1(W''_i\neq W_i)} } \right)$ is transmitted from each terminal, which is later decoded following the same approach used for one-way AWGN channel for $M$ messages. 
	Note as well that the probability of error in each direction follows seamlessly as the one-way case but using the appropriate power allocation factors:
\begin{align}
 \prob_{w_i=1}^{i(3-i)}(\text{error})  &\leq (M-1) \prob_{w_i=1}(W''_i= 1, W'_i\neq W_i).  \label{eq:PerrorM2W}
\end{align}	

Using \eqref{eq:PerrMBBOW} and \eqref{eq:PerrMBBOWFB}, \eqref{eq:PerrorM2W} can be written as:
\begin{align}
	\prob_{w_i=1}^{i(3-i)}(\text{error})  &\leq (M-1) \cdot \prob( W'_i\neq W_i \mid W_i=1) \cdot \prob(W''_i = 1 \mid W'\neq W, W=1 ) \notag\\												
												 &= (M-1)  \frac{M-1}{2} \exp{\left\{ -\lambda(n-M) \frac{K_iP_i}{4\sigma_{3-i}^2} \frac{M}{M-1} (1+t)^2  \right\}  } \notag \\
												 & \;\;\;\;\;\; \cdot \frac{1}{2} \exp{\left\{ -(1-\lambda)(n-M) \frac{J_{3-i}P_{3-i}}{4\sigma_{i}^2} \frac{M}{M-1} (1+t)^2  \right\}  }, 
\end{align}
hence, 
\begin{align*}
	\prob_{w_i=1}^{i(3-i)}(\text{error})  &\leq \frac{(M-1)^2}{4} \exp \left\{ -\lambda(n-M) \frac{K_iP_i}{4\sigma_{3-i}^2} \frac{M}{M-1} (1+t)^2  \right. \left. -(1-\lambda)(n-M) \frac{J_{3-i}P_{3-i}}{4\sigma_{i}^2} \frac{M}{M-1} (1+t)^2  \right\}.   
\end{align*}
From this, the achievable error exponent, leading to \eqref{eq:E12expMTW} and \eqref{eq:E21expMTW}, is for $i \in \{1,2 \}$:
\begin{align} \label{eq:EmTWG}
	E^{i(3-i)} &\geq  \frac{M}{M-1} \left( \lambda \frac{K_iP_i}{\sigma_{3-i}^2} + (1-\lambda) \frac{J_{3-i}P_{3-i}}{\sigma_{i}^2} \right)  .
\end{align}
\normalsize
Since during the retransmission stage there is only a high amplitude transmission occurring with exponentially small probability, the power constraint holds
\begin{equation} 
\overbrace{\lambda (n-M) K_i P_i}^{\text{transmission phase}} + \overbrace{(1-\lambda) (n-M) J_i P_i}^{\text{feedback phase}} \leq nP_i.
\end{equation}
Since $n$ is much larger than any finite $M$, the above equation can be written as:
\begin{equation} \label{eqn:PwrUsr1}
\left[ \lambda (n) K_i + (1-\lambda) (n) J_i \right] P_i   \leq nP_i ,
\end{equation}
and hence $ \lambda K_i + (1-\lambda) J_i  \leq 1  $ for $\lambda\in [0,1]$, $K_1,K_2 \in [0, 1/\lambda)$ and $J_1,J_2 \in [0, 1/(1-\lambda))$.


\section{On the largest number of transmitted messages $M$}
\label{sec:HowLargeIsM}

We now explore how large $M$ may be while still achieving the presented error exponent regions. 
The main restriction on $M$ comes from our use of simplex codes
for both the one-way and two-way AWGN channels. These codes require all symbols to have the same energy and the same pairwise distance, as in an $n$-dimensional tetrahedron. It is known \cite{BalakrishnanContribution,LandauSlepianOptSimplex} \cite[pp. 65--67, Proposition 4.1]{MathNmaterials} that the unique solution to placing $M \leq n+1$ equally likely points on the surface of a unitary sphere in $\mathbb{R}^{n}$ such that the distance between any two points is maximized corresponds to a regular simplex. Thus, codewords of length $n$ may be used to transmit a point from a regular simplex code with $M$ messages as long as $M\leq n+1$.

\subsection{Bounds on $M$ for the AS constraint}
\label{sec:LargestMasOW}

In Theorem \ref{th:OneWayActiveASM3} 
the active feedback transmits the set of codeword pairs that is most likely to have occurred based on the first stage of transmission, as described in 
Section \ref{sec:ActiveM3OneWay}, and this stage becomes the bottleneck, as this number is 
%
%
$L={M \choose 2}= \frac{M(M-1)}{2}$.
Figure \ref{fig:BLKDGMcombinedOWayACTIVEonlyONE} shows that the number of dimensions required by the simplex code in the transmission stage is $\lambda_1 (n-1)$ whereas for the active feedback stage is $\lambda_2 (n-1)$. Only one channel use must be used for the antipodal signaling of the retransmission stage, see Equation \eqref{eq:AntipodalRTX}.
\begin{figure}[H]
	\centering
		\includegraphics[width=0.5\textwidth]{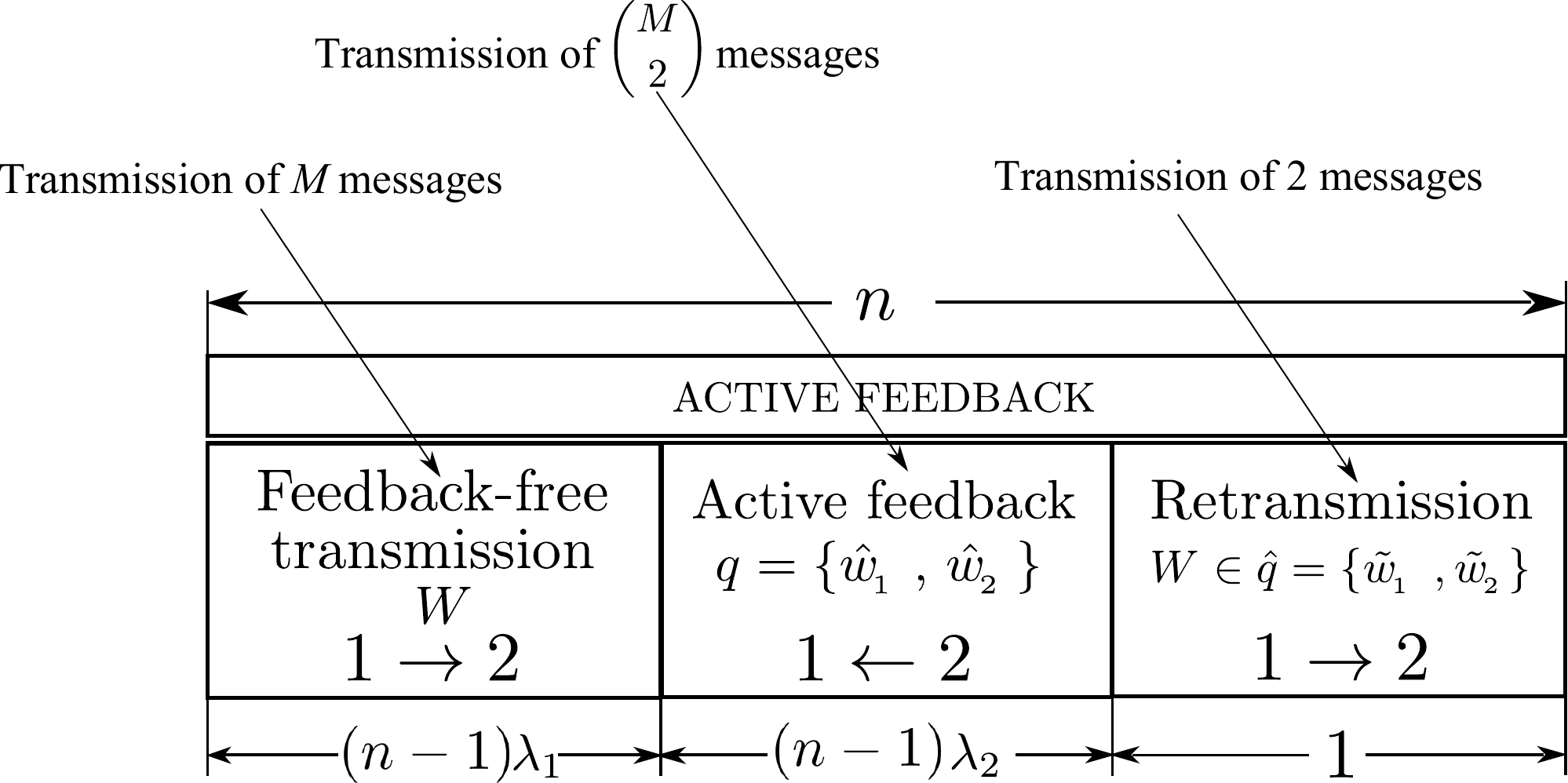}
		\caption{Achievability scheme block diagram for One-Way channel with AS constraint.}
	\label{fig:BLKDGMcombinedOWayACTIVEonlyONE}
\end{figure} 

The number of dimensions a simplex code requires to support $L$ symbols is thus
\begin{align} 
		\lambda_2 (n-1) &= L-1 \notag \\
								&= \frac{M^2-M}{2}-1. \label{eq:lambda2}
\end{align}
In the first stage, the number of dimension of the simplex code is:
\begin{align}
		\lambda_1 (n-1) &= M -1 .	 \label{eq:lambda1}					
\end{align}
Next, note from Figure \ref{fig:BLKDGMcombinedOWayACTIVEonlyONE}, that the total block length $n$ is:
\begin{align}
		\lambda_1 (n-1) + \lambda_2 (n-1) + 1 &= n \\
		\left(M-1\right) + \left(	\frac{M^2-M}{2}-1\right) +1		&= n
\end{align}
The above result can be represented as the quadratic equation $M^2 + M - (2+2n) = 0$
whose positive root is:
\begin{align} 
		M &= \frac{\sqrt{8n+9}-1}{2} \notag\\
			&= \sqrt{2n+\frac{9}{4}}-\frac{1}{2} \label{eq:FiniteMasN}
\end{align}
Note that as $n$ becomes large, $M\approx \sqrt{2n}$ may thus be supported. Then, parameters $\lambda_2$ and $\lambda_1$ can be derived from \eqref{eq:FiniteMasN}, considering \eqref{eq:lambda2} and \eqref{eq:lambda1}:
\begin{align} 
		\lambda_2  &= \frac{1}{(n-1)}\left( \frac{M^2-M}{2}-1\right) \notag \\
							 &= \frac{n}{(n-1)} + \frac{1}{2(n-1)} - \frac{\sqrt{8n+9}-1}{2(n-1)} \\
							 &  \notag \\
		\lambda_1  &= \frac{1}{(n-1)}\left( M-1\right) \notag \\
							 &=  \frac{\sqrt{8n+9}}{2(n-1)} - \frac{3}{2(n-1)}
\end{align}
In the above results, note that for large $n$, these two parameters tend to:
\begin{equation}  \label{eq:lambdasAS}
\lambda_2 \approx 1 - \frac{2}{\sqrt{2n}}; \;\;\;\;\; \lambda_1\approx \frac{2}{\sqrt{2n}} ,
\end{equation}
{In the case of passive feedback, \cite{Kim:ThreeCWPeak2013}, while it was not mentioned, observe that $M$ can be as large as $n$ by setting the duration of the retransmission stage to one channel use, leaving $n-1$ channel uses for the simplex code of the transmission/passive-feedback stage,  thus leaving room for exactly $n$ symbols.}

\medskip
{To analyze the relationship between $M$ and $n$ in the two-way channel, refer to Theorems \ref{Prop2} and \ref{Prop3}, whose block diagrams are depicted in Figure \ref{fig:BLOCK2wayAS} in Section \ref{sec:ResultsTwoWayII}. Recall that both terminals transmit the same number of messages $M$. First, we consider the case of passive feedback, Theorem \ref{Prop2}, that allocates the first $\lambda n$ channel uses to the transmission of message $W_2$ over the $1\leftarrow 2$ channel without feedback. The remaining channel uses are used for the transmission of $W_1$ in the opposite (weaker) direction with the help of passive feedback, which comprises two stages: transmission/passive-feedback lasting for $(1-\lambda)n-1$ channel uses, and retransmission lasting for a single channel use (see \cite[Section II-A]{Kim:ThreeCWPeak2013}). Thus, the total block length $n$ may be expressed as in Equation \eqref{eq:BlockLenghtPassive}. Recalling that the simplex code for $M$ symbols transmitted in each direction requires $M-1$ channel uses, it follows that the number of messages to be transmitted can be as large as $M \approx \frac{n}{2}$, since:}
\begin{align} 
		\lambda n + [(1-\lambda) n -1 ]  + 1	&= n  \label{eq:BlockLenghtPassive} \\
						(M-1) + (M-1) + 1 &= n  \notag \\
														M &= \frac{n+1}{2}  \notag
\end{align}
 
\medskip
{To study the case of active feedback, consider the block diagram of Figure \ref{fig:BLKDGMcombinedOWay} (right), and the stage duration labeling shown in Figure \ref{fig:BLKDGMcombinedOWayACTIVEonlyTWO}. We use parameters $\lambda_1$ and $\lambda_2$ (satisfying $\lambda_1 + \lambda_2 = 1$) to characterize the stage durations supporting the largest number of messages $M$.} 
\begin{figure}[H]
	\centering
		\includegraphics[width=0.5\textwidth]{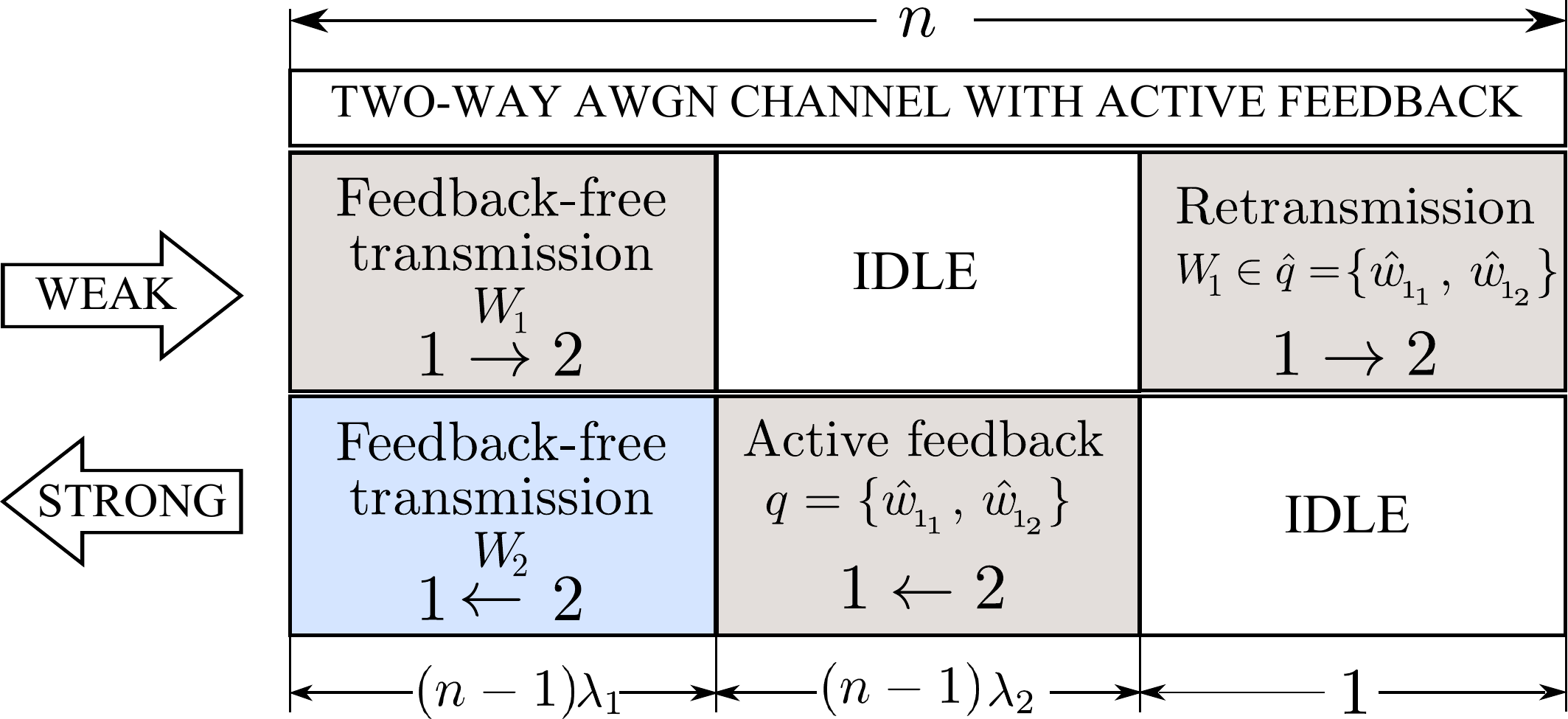}
		\caption{Two-way AWGN channel achievability scheme for Theorem \ref{Prop2} and large $M$.}
	\label{fig:BLKDGMcombinedOWayACTIVEonlyTWO}
\end{figure}
{In the block diagram shown above, the feedback free transmission of the stronger direction (transmission of message $W_2$ in the $1\leftarrow 2$ direction), and the feedback-free transmission of the first stage of the active feedback-aided scheme employed in the weaker direction (transmission of $W_1$ in the $1\to 2$  direction) occur simultaneously, as both channels are independent. These transmissions are based on a simplex code of $M$ symbols and therefore
$\lambda_1 (n-1) = M-1 $.  The active feedback stage occurs in the next $\lambda_2(n-1)$ channel uses, followed by a single channel use for the retransmission, see Equation \eqref{eq:AntipodalRTX}. Note that $\lambda_1$ and $\lambda_2$ are the same as those shown in Equation \eqref{eq:lambdasAS}. Also, note that the number of messages is mainly determined by the active feedback stage, since the largest fraction of the block length $n$ is used for this stage. Thus, again, $M\approx \sqrt{2n}$.}


\subsection{Bounds on $M$ for the  EXP constraint}
\label{sec:LargestMexpOW}

In this case, for the scheme presented in Theorem \ref{th:OneWayEXPM2}, $M$ is constrained by the 
third stage. This stage corresponds to the index location code we have proposed and lasts for $M$ channel uses. This implies that only $n-M$ channel uses are available for the transmission and feedback stages. Each of these two stages transmits  one out of $M$ possible messages from $\mathcal{M} =\{1,2,...,M\}$ using a simplex code, and requiring $M-1$ channel uses. 
Thus, $M$ must satisfy $(M - 1) + (M - 1) + M \leq n$, and hence  $M \approx \frac{n}{3}$ for large $n$.

In the case of the two-way AWGN channel, the transmission of $M = \frac{n}{3}$ in each direction determines the duration of transmission and feedback stages as $\frac{n}{3}$, therefore, Theorem \ref{th:EXPMachievTW} is slightly modified by setting parameter $\lambda = \frac{1}{3}$, and may be restated as: 

\textit{An achievable error exponent region for the two-way AWGN channel and the transmission of a finite number of messages $M$ under the EXP power constraint for both directions is given by the union over all error exponent pairs $(E_{12},E_{21})^{\text{EXP}}$  over $K_1, K_2 \in [0, 3]$ and $ J_1, J_2 \in [0,3]$  such that $\frac{1}{3} K_i + \frac{1}{3}J_i \leq 1$ for $i = 1,2$.  for which: }
\begin{align} \label{eq:E12expMTWnw}
	E_{12}^{\text{EXP}} &\geq \left(\frac{1}{3}\right)  \left( K_1 \frac{ P_1}{\sigma_2^2} +  J_2 \frac{P_2}{\sigma_1^2} \right) \notag\\
	E_{21}^{\text{EXP}} &\geq \left(\frac{1}{3}\right)  \left( K_2\frac{ P_2}{\sigma_1^2} +  J_1 \frac{ P_1}{\sigma_2^2} \right). \notag
\end{align}	
The achievable error exponent region achieved by the above equations is the same as that presented in Theorem \ref{th:EXPMachievTW}, Figure \ref{fig:M2allsimpleAnalytic} and presented later in Figure \ref{fig:M2allsimple} of the next section. Since no extra power is used for the retransmission codeword (since it occurs with exponentially small probability), fixing $\lambda = 1/3$ does not affect the shape of the time-sharing like triangular region as it can be achieved by proper power allocation characterized by parameters $K_i$ and $J_i$. 

{Results derived in this section are summarized in Table \ref{tab:largeMtable}, where we note that all corresponding theorems hold with $M$ as large as shown below (as $n\rightarrow \infty$).}


\begin{table}[H]
\centering
\caption{Summary of results for the largest $M$ in each achievability scheme}
	\begin{tabular}{c||c|c|c}
	Power Contraint	& FB-type			&  One-Way 				&  Two-Way 					\\ \hline \hline
				AS				& Passive 		&  $n$ 						&	 $\frac{n}{2}$   \\ 
									& Active  		& $\sqrt{2n}$ 		&  $\sqrt{2n}$  	 \\ \hline								 
			EXP					&	Active  		& $\frac{n}{3}$ 	&  $\frac{n}{3}$  	
	\end{tabular}
	\label{tab:largeMtable}
\end{table}


\section{Numerical simulations}
\label{sec:Simul}


This section presents numerical simulations that evaluate the newly derived inner bounds for different SNR conditions and number of messages.
We begin with the one-way channel and illustrate how active feedback leads to larger achievable error exponents than passive feedback for the same SNR observed in the feedback link. Later, we show error exponent regions for the two-way AWGN channel under the AS power constraint and observe the effect of active and passive feedback for two SNR scenarios and the transmission of $M=3$ messages. Next, we present the achievable region under the EXP power constraint for different number of transmitted messages $M$. 
Simulations also show the feedback-free and perfect feedback achievable error exponents (or regions in case of the two-way channel) as references. From these, one can easily see when noisy feedback or interaction  / adaptation may be beneficial over a non-feedback scheme. The perfect feedback scenario provides an upper bound on what can be achieved with noisy feedback (or the two-way channel). 

\subsection{Simulations for the One-Way AWGN Channel}


Figure \ref{fig:OWactive} presents a numerical evaluation that illustrates how under the AS power constraint, the active feedback strategy proposed improves the error exponent over both existing passive and non-feedback schemes. This error exponent ($E_{\text{FB}}^{\text{AS}}$ on the vertical axis) is characterized by Equation \eqref{eq:ASactiveM} --Theorem \ref{th:OneWayActiveASM3}-- for active feedback (shown in blue solid line), and by Equation \eqref{eq:PassiveFBM3simplified} for passive feedback --derived from Section II-B of \cite{Kim:ThreeCWPeak2013}-- (shown in red dashed line). The non-feedback transmission error exponent provided by Equation \eqref{Eq:NoFBErrExp}, is also shown with a solid black line as the benchmark to improve upon using noisy feedback. The three exponents are plotted for all possible choices of parameter $s \in (0,1)$ on the horizontal axis. 
\begin{figure}[htb]
	\centering
		\includegraphics[width=1\textwidth]{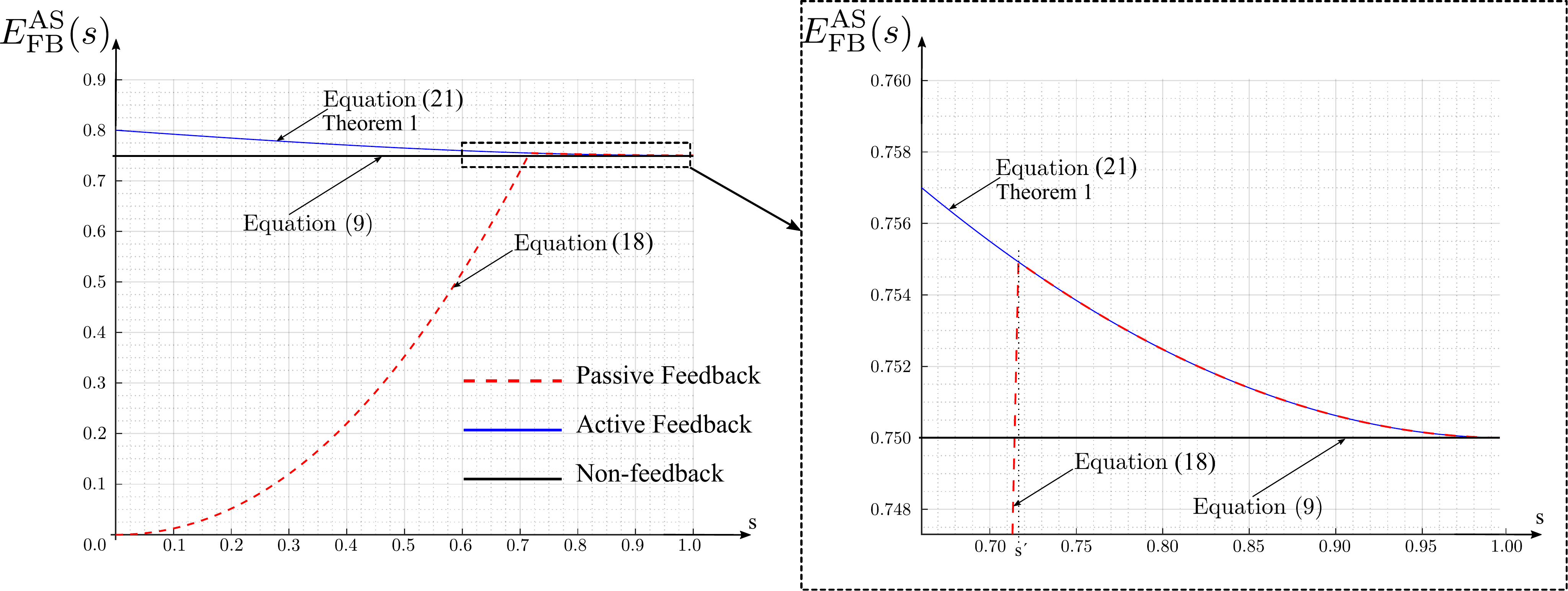}
		\caption{Error exponent for the One-Way AWGN channel under the AS constraint using noisy passive and active feedback --Equations (18) and (21)--, and without feedback --Equation (9)--. The forward channel SNR is $\frac{P}{\sigma^2}=2$ and feedback channel SNR is $\frac{P_{\text{FB}}}{\sigma_{\text{FB}}^2}=16$ for $M=3$.} 
	\label{fig:OWactive}
\end{figure}


Given the SNR pair we evaluated, observe that active and passive feedback provide a similar error exponent improvement over the feedback-free exponent boundary ($E^{\text{AS}} = 3/4$) over the interval $s\in (s',1)$ for $s'=\frac{\sqrt{21}-1}{5} \approx 0.71651$. This occurs since Equations \eqref{eq:PassiveFBM3simplified} and \eqref{eq:ASactiveM} are dominated by the first argument in the $\min$ operator (similar for both equations). Outside this interval, passive feedback error exponent is dominated by the second argument, and decays dramatically as $s$ approaches zero, falling below the feedback-free error exponent. In contrast, active feedback remains dominated by the first argument since the second one is larger than the first for all $s\in(0,1)$. Thus, it is easy to see that the a small $s>0$ is the best choice for active feedback, since attains the largest exponent, whereas for passive feedback, this is given for $s'$. The highest error exponent of the passive feedback scheme is $E^{\text{AS}}_{\text{FB}}\approx 0.7549238<0.755$, that is less than $0.67\%$ over the non-feedback transmission ($0.75$). Active feedback can achieve an error exponent of up to $0.8$, or $6.67\%$ over the non-feedback error exponent. 

In the case of ideal feedback, the second argument of the active and passive feedback error exponent expressions will tend to infinity, and therefore to be dominated by the first argument. Thus, it can concluded that even with perfect feedback, the largest gain that these two schemes may achieve is given by evaluating the first argument in the $\min$ operator at $s$ very close to zero, for example with $M=3$, this leads to $E_{\text{FB}}^{\text{AS}} = \frac{2}{5} \frac{P}{\sigma^2}$.


\subsection{Simulations for the two-way AWGN Channel}

This section shows  two-way achievable error exponent regions under different power constraints. 

\medskip

\subsubsection{Numerical simulation for $M=3$ under AS power constraint}
\label{sec:ASactiveM3}

This simulation illustrates how under the AS power constraint, and a two-way channel with one direction's SNR much larger than the other, our schemes lead to an achievable error exponent region that describes may improve over the feedback-free error exponent region of the weak direction, at the price of a reduction of the one of the strong direction.

Consider Figure \ref{fig:M3soloSIMPLE}:  the left plot depicts a two-way channel with SNRs $\frac{P_1}{\sigma_2^2} = 2$ and $\frac{P_1}{\sigma_2^2} = 16$, whereas on the right, the SNRs are  $\frac{P_1}{\sigma_2^2} = 2$ and $\frac{P_1}{\sigma_2^2} = 160$.  Note that both plots include the non-feedback inner bound provided by Proposition \ref{Prop1} depicted with a black solid line,  the dotted green line corresponding to passive feedback based interaction (Theorem \ref{Prop2}), the dashed red line corresponding to  active feedback based interaction (Theorem \ref{Prop3}), and the dashed orange line corresponding to the outer bound obtained from Pinsker's noiseless feedback bound for the one-way AWGN channel used for each direction \cite{pinsker1968probability}. 
Note that for the evaluated SNR conditions, active feedback achieves a higher error exponent in the $1\to 2$ direction at a smaller reduction in the $1\leftarrow 2$ direction, than the scheme based on passive feedback. This reduction is caused by the amount of power conceded by terminal 2 to provide feedback and improve the error exponent of terminal 1.
\begin{figure}[htb]
	\centering
		\includegraphics[width=0.7\textwidth]{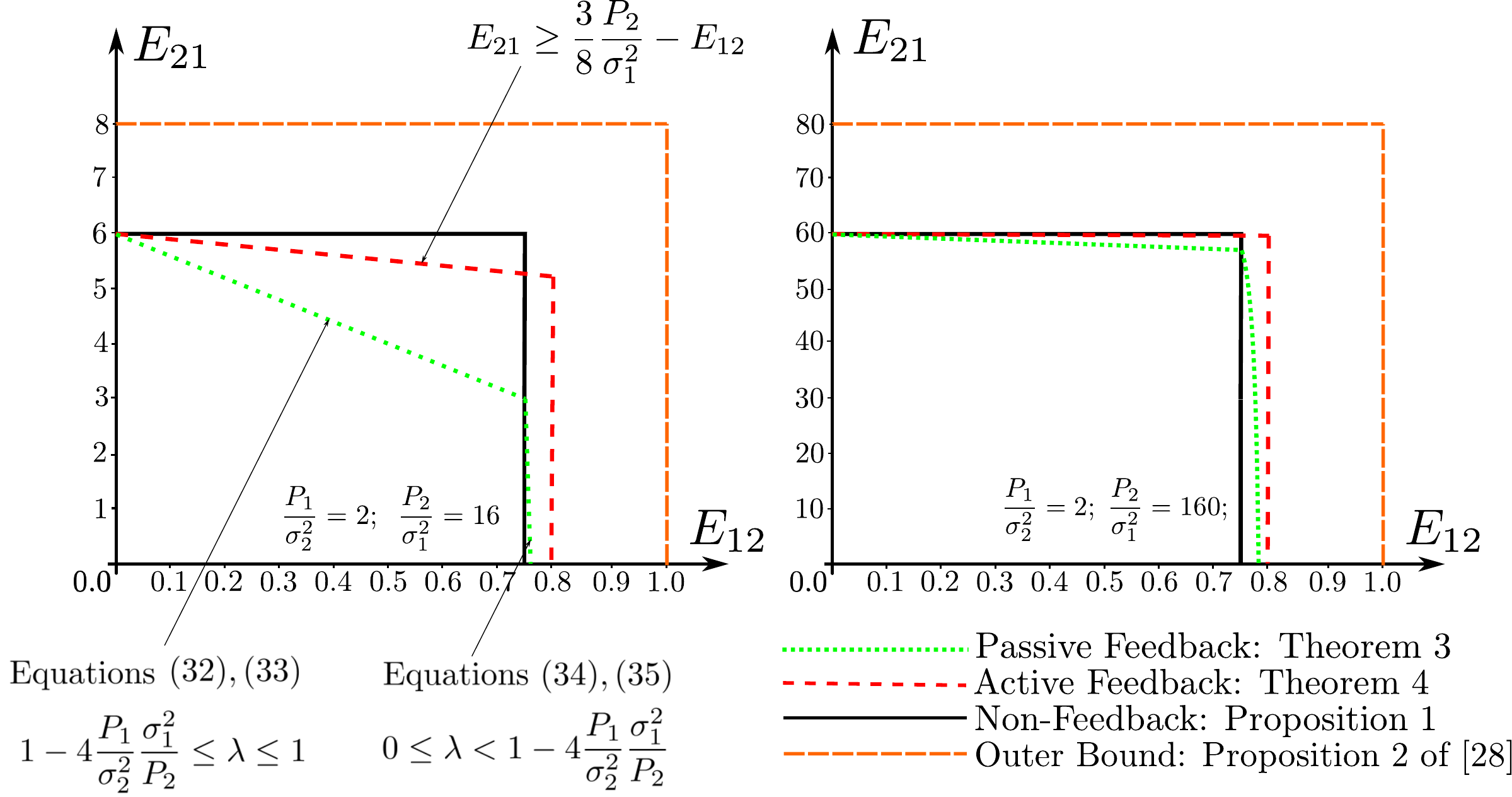}
\caption{Error exponent region for the two-way AWGN channel under the AS constraint and the transmission of three messages.}
\label{fig:M3soloSIMPLE}
\end{figure}

Figure \ref{fig:M3soloSIMPLE2SNR160ALL} shows the achievable error exponent region obtained by Theorems \ref{Prop2} and \ref{Prop3} for different values of $M$. Note that as $M$ increases, the achievable feedback-free error exponent in both directions decrease until $1/4$ of the corresponding channel SNR. This numerical evaluation allows us to observe that active feedback always attains error exponents above those attained with passive feedback.
\begin{figure}[htb]
	\centering
		\includegraphics[width=0.4\textwidth]{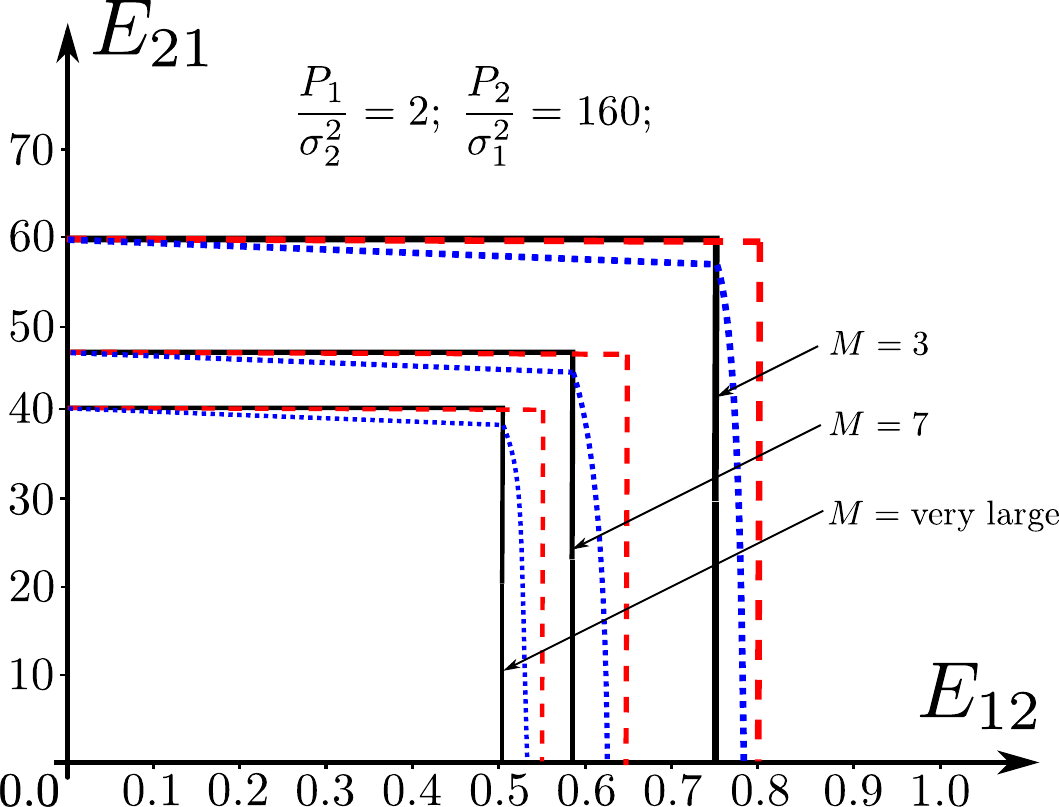}
		\caption{Achievable error exponent regions provided by Theorems \ref{Prop2} and \ref{Prop3} for different values of $M$. } 
	\label{fig:M3soloSIMPLE2SNR160ALL}
\end{figure}

\bigskip

\subsubsection{Numerical simulation for $M$ under EXP power constraint}

Figure  \ref{fig:M2allsimple} shows the achievable error exponent region for the two-way AWGN channel with active noisy feedback under the EXP power constraint for an increasing number of transmitted messages $M$. The largest region is attained for the transmission of two messages, as studied in \cite{PalacioDevroyeISIT2018}. It is interesting that the triangular regions are bounded by a line of similar slope but decreasing in sum-value as $M$ increases.  
Note that the red line corresponds to a very large $M$ and that even for this case, the EXP constraints yield an achievable region that completely contains what is achievable under the AS and shown with a dotted line at the bottom left.

\begin{figure}[H]
	\centering
		\includegraphics[width=0.5\textwidth]{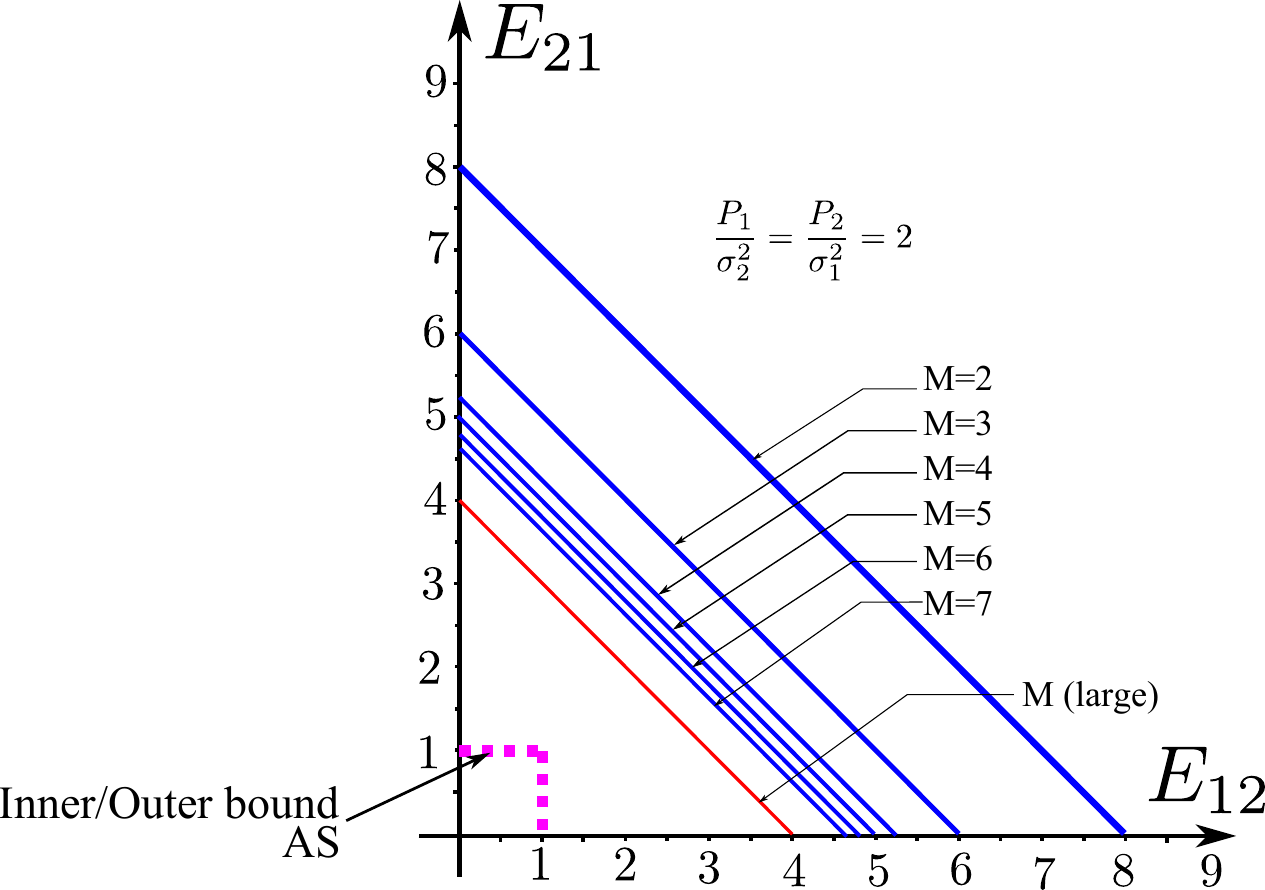}
		\caption{Achievable error exponent region for the two-way AWGN channel with active noisy feedback under the EXP power constraint and symmetric SNR.}
	\label{fig:M2allsimple}
\end{figure}

\section{Conclusions}
\label{sec:Conclu}
Achievable error exponents for the one-way and two-way AWGN channel under the AS and EXP power constraints were derived for the transmission of an  arbitrary but finite number of messages. The main conclusions drawn are that, under the AS power constraint, 
noisy feedback schemes may achieve error exponents between the feedback-free and perfect feedback, and are only beneficial when the feedback channel SNR is (considerably) larger than the forward channel SNR. 
We considered both passive noisy feedback and encoded / active feedback that make use of simplex codes. 
In the two-way setting, we demonstrated achievable error exponent pairs that illustrate a tradeoff between allocating resources towards one channel's transmission versus aiding in feedback for the other.

 

In contrast to the AS, the EXP power constraint allows more flexibility in coding  and power allocation strategies. In particular, it allows transmitters to generate very high amplitude transmissions (that allow for very high error exponents) with exponentially small probability while still meeting the power constraints. 
By extending the error exponents for the one-way AWGN channel with active noisy feedback for two messages by Kim, Lapidoth and Weissmann to any arbitrary but finite number of messages $M$, we obtained a general expression for achievable one-way error exponents and showed how they may be incorporated into error exponent regions for the two-way AWGN channel.

\section{Acknowledgements}

This work was partially supported by NSF under awards 1053933 and 1645381. The contents of this article are solely the responsibility of the authors and do not necessarily represent the official views of the NSF.

\bibliographystyle{IEEEtran}
\bibliography{Palacio-Baus-Devroye-TIT-submission-11-2018}

\newpage
\appendices

\end{document}